\documentclass[10pt]{article}    
\usepackage{amssymb,latexsym,amsmath} 
\usepackage{amsthm}                   

\usepackage[mathscr]{eucal}  

\theoremstyle{plain}

\newtheorem{theorem}{Theorem} 
\newtheorem{lemma}{Lemma}     

\newfont{\sffl}{msbm10 at 16pt} 
\newfont{\sff}{msbm10 at 10pt}

\begin{document}           
\title{\vskip -.75in
Dirichlet integral point-source harmonic interpolation\\
       over ${\mathbb{R}}^3$ spherical interiors: DIDACKS II
\thanks{\small{Approved for public  release.}}
}

\author{Alan Rufty\\         
\\
P.O. Box 711\\
Dahlgren, VA. 22448}
\maketitle                 

\newcommand{\KD}{K_{\text{D}}}
\begin{abstract}
 This article addresses the interpolation of harmonic functions over the interior of a $\mathbb{R}^3$ unit sphere by linear combinations of fundamental-solution point-source basis functions, where all the sources are assumed to be outside the sphere.  While it is natural to formulate approaches to harmonic approximation, interpolation and/or boundary value problems for spherical interiors in terms of minimizing some standard Dirichlet integral, there is no established approach along these lines that yields interpolating solutions for $\mathbb{R}^3$ point-source basis functions.  Here it is shown that by introducing a simple weighting function, exact closed-form inner-products result for the expressions needed; hence, minimizing the appropriate weighted Dirichlet integral yields exact closed-form linear equation sets for the source strengths and the fits resulting from these equation sets match all of the prescribed values at the interpolation points.  Further, the formalism can be extended in a natural fashion to handle interpolation fits using higher-order point-multipole basis functions (such as point dipoles and point quadrupoles), so that interpolations for higher-order partials of harmonic functions can be easily implemented.  Since the source and field points are in different domains, the fundamental-solution basis functions are bounded and can be regarded as defining a new type of kernel space that is related to, but distinct from, a reproducing kernel Hilbert space (RKHS).  This new space is labeled a Dirichlet integral dual-access collocation-kernel space (DIDACKS).  While the corresponding DIDACKS formalisms for $\mathbb{R}^3$ spherical exteriors was developed elsewhere and thus is only summarized here, a full discussion of the $\mathbb{R}^3$ spherical interior formalism is given here, including all the relevant derivations.  This presentation includes numerical tests of the DIDACKS point source fitting method.  Finally, although DIDACKS fits to data obtained from actual point measurements are possible, all of the required interpolation point data is assumed to be exact here and neither the treatment nor the consequences of measurement errors are considered in this paper, but ill-conditioning and condition number issues are addressed. 

 Like RKHS (or least-squares collocation) fits, DIDACKS fits simultaneously satisfy two distinct norm minimization conditions: (1) The norm of the difference between the function to be fit and the interpolating fit is minimized (the least-squares norm property). (2) Of all those functions that match the values of the function to be fit at the specified interpolation points, the interpolating fit is the one with the smallest norm (the collocation minimum norm property).  
The least-squares norm property follows directly from the DIDACKS formalism itself.  A derivation of the collocation minimum norm property is given here, which is a generalization of the standard proofs encountered in a RKHS setting.  In concert with these minimum norm properties, DIDACKS theory has implications for other integral and kernel based approaches---especially the method of fundamental solutions (MFS).  Specifically, it is shown that natural connections between DIDACKS and MFS approaches imply that two distinct previously unrecognized minimum norm conditions hold for MFS techniques as commonly implemented; furthermore, these conditions imply that MFS approaches are on a more solid theoretical footing than has been commonly recognized.  Finally, the existence of obvious connections between DIDACKS and the fast multipole method (FMM), as well as the boundary element method (BEM), are also pointed out, but they are not explored in any detail here. 
\end{abstract}

\newcommand{\SubSec}[1]{

\vskip .18in
\noindent
\underline{{#1}}
\vskip .08in

}

\newcommand{\ls}{\vphantom{\big)}} 
\newcommand{\lsm}{\!\vphantom{\big)}} 

\newcommand{\mLarge}[1]{\text{\begin{Large} $#1$ \end{Large}}}
\newcommand{\mSmall}[1]{\text{\begin{footnotesize} $#1$ \end{footnotesize}}}

\newcommand{\Blbrac}{\rlap{\bigg{\lceil}}\bigg{\lfloor}} 
\newcommand{\Brbrac}{\bigg{\rbrack}} 

\newcommand{\R}[1]{${\mathbb{R}}^{#1}$}  
\newcommand{\mR}[1]{{\mathbb{R}}^{#1}}
\newcommand{\RR}{${\mathbb{R}}$}
\newcommand{\mRR}{{\mathbb{R}}}
\newcommand{\C}{${\mathbb{C}}$}
\newcommand{\mC}{{\mathbb{C}}}
\newcommand{\riii}{${\mathbb{R}}^3$} 
\newcommand{\mriii}{{\mathbb{R}}^3}

\newcommand{\Dr}{\mathscr{D}_r}

\newcommand{\sps}{0} 
\newcommand{\spss}{0}

\newcommand{\smallindex}[1]{\text{\raisebox {1.5pt} {${}_{#1}$}}}

\newcommand{\eq}{:=}  
\hyphenation{DIDACKS}  

\vskip .05in
\noindent
\begin{itemize}
\item[\ \ ] \small{\textbf{Key words:} {Laplace's equation, inverse problem, Dirichlet form, point collocation, reproducing\\ \phantom{Key words. L} kernels,} 
fundamental solutions, point sources, multipole, potential theory}
\item[\ \ ] \small{\textbf{AMS subject classification (2000):} {Primary 65Nxx. Secondary 31Bxx, 35J99, 65D05}}
\end{itemize}

\renewcommand {\baselinestretch}{1.35} 

 \section{Introduction}\label{S:intro}

  Given a real valued function $f$ satisfying Laplace's equation over the interior and on the boundary of an \riii\ unit sphere, this article shows that an interpolating function $\varphi$ can be constructed from linear combinations of inverse-distance point-source basis functions, where the sources are located outside of the unit sphere.  Further, these interpolating solutions are obtained by directly minimizing $\|f - \varphi\|{\ls}_{\text{II}}^2$\,, where $\|\,\cdot\,\|{\ls}_{\text{II}}$ denotes the interior integral (II) norm---a positive definite norm defined here in a simple but nonintuitive way.  It is shown that this norm is proportional to the Dirichlet integral of the field difference $(f - \varphi)$ with an inverse radial distance weighting function.  The developed formalism prescribes a one-to-one correspondence between each pair of interpolation and point-source locations so that if there are $N_k$ interpolation points then there must be $N_k$ point-source basis functions.  (It is assumed throughout that $N_k$ is a fixed finite integer.)  The basis functions have the form  $1/\ell_k$, where ${\ell}_k$ is the distance between source point $k$ and some field point that is in the interior of the sphere.  These basis functions are often called ``fundamental solutions'' since they satisfy Laplace's equation for interior points, ${\nabla}^2 \ell^{-1}_k = 0$, as well as Poisson's equation over the entire field and source region; moreover, other solutions can be built up from them by  means of the integral form of Poisson's equation.   Here $\varphi \eq \sum^{N_k}_{k=1}q_k/\ell_k$, which is the discrete analog of the integral form of Poisson's equation, specifies the basic form of the interpolating function.  The source strengths $q_k \in \mRR$ are determined by a closed-form linear equation set that yields a $\varphi$ with the desired interpolating properties.  This equation set follows directly from the standard approximation procedure of setting the partials of 
\begin{equation}\label{E:PhiDef}
\Phi \eq \|f - \varphi\|{\ls}_{\text{II}}^2 =  (f,\,f){\ls}_{\text{II}} - 2(f,\,\varphi){\ls}_{\text{II}} + (\varphi,\,\varphi){\ls}_{\text{II}}
\end{equation}
 with respect to the source strengths equal to zero and diving by two:
\begin{equation}\label{E:LLSQ}
\boxed{\,\mathbf{T}\,\mathbf{q} = \mathbf{A}\,}\ , 
\end{equation}
which obviously minimizes $\Phi$.
Here $\mathbf{T}$ denotes the matrix whose elements are $T_{k'\,k} \eq ({\ell}^{-1}_{k'},\,{\ell}^{-1}_{k}){\ls}_{\text{II}}$ for $k$ and $k' = 1,\,2,\,3,\,\ldots\,N_k$; $\mathbf{A}$ the vector whose elements are  $A_{k'} \eq (f,\,{\ell}^{-1}_{k'}){\ls}_{\text{II}}$; and finally, $\mathbf{q}$ the vector whose elements are ${q}_k$.  This interpolation procedure is made possible by the existence of closed-form inner products of the form  $({\ell}^{-1}_{k'},\,{\ell}^{-1}_{k}){\ls}_{\text{II}}$ and $(W,\,{\ell}^{-1}_k){\ls}_{\text{II}}$\,, where $(\,\cdot\,,\,\cdot\,){\ls}_{\text{II}}$ is the II inner-product associated with the II norm.  (Subsequently, since the definition of an inner product and norm are linked in an obvious fashion, when no confusion can result the descriptive adjective ``norm'' will be used to imply both the inner-product and norm setting.)

   Besides the fact that the II norm yields closed form expressions for inner products of  ${\ell}^{-1}_k$ with any harmonic function [i.e., (\ref{E:IIntRep})], it also has various other interesting properties.  One is that the II norm of any suitable function can be expressed as the sum of a weighted Dirichlet integral term and of function evaluations at the origin (\ref{E:IInorm2d}), while a second is that the II norm of any function can be written as the sum of an unweighted Dirichlet integral term and a surface integral term (\ref{E:IIrel2}).  Yet another interesting property emerges from the question: Does a point replication and inner product structure exist that allows one to reinterpret Poisson's integral formula for $f$ in a natural way as the inner product of some fundamental solution kernel and $f$ so that the end result is $f$ evaluated at some specified point?  As the derivation leading up to (\ref{E:IIe}) reveals, the II norm with kernel ${\ell}^{-1}_k$ provides an affirmative answer to this question.  (It is also worth noting that the Poisson kernel itself can be extended in a natural way that allows for a reinterpretation of Poisson's integral formula as an inner-product evaluation \cite[p. 122]{AxlerEtAll}).  Other interesting aspects of the II norm should be apparent in the sequel.

   While various procedures for \riii\ harmonic interpolation have been proposed over the years that are largely unrelated to the approach presented here, the only direct precedent appears to be an analogous formalism for \riii\ spherical exteriors developed by the author \cite{Ruf1}.   As one would expect, for this exterior problem the source points are inside the sphere and the harmonic field region is in the exterior.  There corresponding relationships yielding closed form inner products and relationships to weighted and standard unweighted Dirichlet integrals are derived.
 It was also shown there that a similar theory can be developed for \riii\ half-space.  This exterior and half-space theory are briefly summarized in Section~\ref{S:Recap}.  
As elaborated on below, the label used for this basic interpolation approach and the associated  space for these and other geometric settings is a Dirichlet integral dual-access collocation-kernel space (DIDACKS).

  Although the goal here is to minimize the overlap between this and the companion article just mentioned, in order to make the present article self-contained there are three areas of modest overlap.  First, given that there is a great deal of overall mathematical similarity between the interior and exterior \riii\ DIDACKS formalisms, in order to facilitate comparisons it is useful to have a ready summary of the relationships for both settings available and thus Section~\ref{S:Recap} fills this role.  The interior relationships summarized in Section~\ref{S:Recap} are proved in Section~\ref{S:InteriorRelationships}, but the derivation of the exterior relationships are not duplicated.  Second, just as in the companion article, it is also pointed out in Section~\ref{S:Recap} that the basic DIDACKS interpolation formalism can be directly generalized in an obvious way to allow for the matching of higher-order partial derivatives of $f$.  In this case, the interpolating basis functions that must be used are corresponding components of higher-order point multipoles as summarized in Table~\ref{Ta:MeasSour}.  Hence from a fitting perspective, various DIDACKS multipole fits, including combinations of point dipoles and point quadrupoles, can easily be performed.  These possibilities are discussed at the end of Section~\ref{S:Recap}.  Third and finally, just as in the companion article normalized basis functions are often used here, where for scalar point sources they are given by $\hat{\varphi}_k \eq {{\ell}^{-1}_k}/{\|{\ell}^{-1}_k\|}$.  (Generally the basis functions $\hat{\varphi}_k$ will be highly non-orthogonal.)  As previously noted in the companion article, the use of normalized basis functions is strongly recommended for cases where large differences of physical scale may occur and produce unwanted effects (i.e., large condition numbers) in the solution process of (\ref{E:LLSQ}).  This occurs, for example, when scalar point sources and dipoles are both present.

  The DIDACKS interpolation procedures studied here are analogous to reproducing kernel Hilbert space (RKHS) interpolation techniques where an appropriate linear combination of reproducing kernel basis functions are used---Bergman kernel interpolation being one notable example.  It is thus useful to briefly compare and contrast these two approaches.  (The reader who is unfamiliar with RKHS theory may wish to consult a standard reference, such as \cite{Aronszajn}, \cite{Mate} or \cite{Moritz}, but aside from some of the material in Appendix~A these connections are not actually an integral part of what follows and they need not be completely mastered so a detailed knowledge of RKHS theory is not really mandatory.)  For some region $\Omega \subset \mathbb{R}^3$, associated with each RKHS and its inner product form $(\,\cdot\,,\,\cdot\,)$ is a unique symmetric reproducing kernel (SRK): $K(\vec{P},\,\vec{Q}) = K(\vec{Q},\,\vec{P})$, where $\vec{P} \in \mriii$ and $\vec{Q} \in \mriii$ .  Here $K(\vec{X},\,\vec{P})$ satisfies the reproducing property
\begin{equation}\label{E:RKHS}
(K,\,f) = f(\vec{P})\ ,
\end{equation}
where $\vec{X} \in \mriii$ is a dummy argument tied to the inner-product structure.

The inner-product evaluation property of kernels like ${\ell}_k^{-1}$, which have one argument in one domain and the other argument in the complimentary domain, is called a replication property in order to distinguish it from the standard reproducing kernel property (\ref{E:RKHS}),
where both of the arguments involved are in the same domain.  The term collocation property will be used to indicate either the possibility of a replication or a reproducing property; moreover, it will also be used in its more commonly used sense to indicate a simple forced matching of a linear combination of $N_k$ basis functions to $N_k$ specified point values.  Kernels, of course, will be labeled by their properties so that, for example, ${\ell}_k^{-1}$ is called a replication or collocation kernel.  Although it cannot be considered a reproducing kernel, since its two arguments are in different domains, ${\ell}_k^{-1}$ is a bounded kernel like reproducing kernels due to the restrictions on these arguments.  In recognition of the dual argument nature of ${\ell}_k^{-1}$, as well as the fact that it has a replication property, it and like kernels will be given the generic label dual-access collocation kernel (DACK).  The main focus of this article, of course, is the more interesting, but specialized, case where the normed space arises from a (weighted) Dirichlet integral (i.e., a DIDACKS), so that the kernel form employed here, ${\ell}_k^{-1}$, is regarded as a Dirichlet integral dual-access collocation kernel (DIDACK).

  Before considering the differences between DIDACK and SRK approaches, consider the similarities.  Although they will not be explored in detail here, the existence of connections between DACKs and reproducing kernels are obvious for the DIDACK kernel form ${\ell}_k^{-1}$.  As it should be obvious from the sequel, the point is that the DIDACKS kernel form ${\ell}_k^{-1}$ can always be transformed into a symmetric reproducing kernel form and then the associated replication property can be transformed into a reproducing property.   For example, consider the II norm replication property given by (\ref{E:IIntRep}).  The DIDACK ${\ell}_k^{-1}$ can be transformed into a SRK by first using (\ref{E:PkXk}) to eliminate the exterior source point in favor of a corresponding interior point and then substituting the result into (\ref{E:IIntRep}).  Dividing the resulting expression for (\ref{E:IIntRep}) by ${P}_k$ then yields the SRK form (\ref{E:RKHS}).  As discussed at length in \cite{Ruf1}, for the exterior setting a similar reformulation is possible and the resulting kernel and associated weighted Dirichlet norm was first studied by Krarup \cite{Krarup}.  The various differences between DIDACK and SRK based approaches were addressed in some detail in \cite{Ruf1} and thus need not be reconsidered here, but it is worth noting that vast differences are readily apparent when dipole or other multipole like fits are contemplated.  These effects arises from the differences in action of a linear differential operator in the two space settings.  Thus suppose that $\mathscr{L}_k$ is a differential operator that is, for example, some combination of partials with respect to the components of $\vec{P}_k$ (or of $\vec{P}$), then it is easy to see that the action of  $\mathscr{L}_k$ on (\ref{E:IIntRep}) and its RKHS analog (\ref{E:RKHS}) are quite different and correspond to entirely different fits.  Moreover, all of the operators considered here act with respect to source components rather than components of $\vec{P}_k$, so that these differences are even more pronounced.  There are also other general differences between SRKs and the kernels studied here, but questions of an abstract nature regarding kernel structure are not addressed in this article. 

  Next briefly consider the general function space setting for DIDACKS.  Observe that all the solutions to (\ref{E:LLSQ}) are contained in the finite linear span framed by the set of basis functions used and that Cauchy sequences thus need not be admitted.  These Cauchy sequences can be potentially problematic here; hence, for the main interior setting of interest the family of admissible functions is assumed to be the linear space of bounded harmonic functions with bounded partials in the interior of a unit sphere and on its boundary and Cauchy sequences of such functions are not admitted, which means an underlying pre-Hilbert space setting is assumed.

  Next, consider the overall structure of the paper.  Earlier it was pointed out that Section~\ref{S:Recap} contains a summary of DIDACKS spherical exterior as well as a preview of the DIDACKS interior relationships.  The derivations of the DIDACKS interior relationships given in Section~\ref{S:InteriorRelationships} are followed up in Section~\ref{S:Analysis} with a brief synopsis  of the collocation minimum norm result for the DIDACK $1/\ell_k$.  This result is the analog of a like result from RKHS theory.  A more general discussion of this result and related theorems is given in  Appendix~A.  It is worth noting that the basic proof of the minimum norm collocation principle presented in Section~\ref{S:Analysis} (which is the same line of argument that is generalized in Appendix~A to include the RKHS setting) is very concise; moreover, the surrounding context in Section~\ref{S:Analysis} and Appendix~A contains various conditions and restrictions that should normally be stated for these RKHS proofs, but are not.  Further, since there does not seem to be a complete discussion of all of the associated mathematical consequences and pitfalls, an attempt to raise most of the major points is made in Appendix~A.  Including the possible generalities and associated side issues results in a somewhat circuitous and lengthy discussion in Appendix~A, but as mentioned in Section~\ref{S:Analysis} significant theoretical and practical consequences follow from this minimum norm collocation principle so a thorough understanding of it is useful.  In this connection, one of the primary tools introduced in Section~\ref{S:Analysis} and Appendix~A is the concept of pointwise independence and Appendix~B gives a attempt to directly prove that harmonic functions are pointwise independent.   Finally, although it is obvious that the minimum norm properties described in Section~\ref{S:Analysis} can be used to derive various interesting energy norm (and weighted energy norm) inequalities, these possibilities are only briefly touched on at the end of Section~\ref{S:Analysis}.  

Since the basic DIDACKS interpolation formalism developed here is new, for completeness and in order to give some level of assurance in this basic formalism itself numerical test results are given in Section~\ref{S:Tests}.  While all of the numerical tests in Section~\ref{S:Tests} are with respect to elementary basis functions of the form  $1/\ell_k$, as noted in \cite{Ruf1} the author has had extensive favorable experience with combined point mass and point dipole fits (and even some with point quadrupole fits) for the exterior setting where the formalism is quite analogous; consequently, a duplication of these types of tests for the interior setting was not considered particularly relevant here.  While the numerical tests were not chosen  to be representative of typical applications, they do serve as a vehicle for the consideration of various points of interest; thus a partial list of relevant points suggested by the numerical results in Section~\ref{S:Tests} is given at the end of Section~\ref{S:Tests}.  As an aside, in conjunction with these numerical tests, a simple algorithm is described, which may be useful elsewhere, that sets up a semi-regular grid on a sphere (i.e., a grid where the points are more-or-less equal distances apart).

  Finally, after a little reflection, it is apparent that the DIDACKS interpolation algorithms described here can be used as a boundary value problem (BVP) solution technique for \riii\ harmonic problems over not only spherical interiors, but, as elaborated on below, over general bounded connected regions that have suitably shaped bounding surfaces since all that is really required for the present interpolation formalism is that the sources be in the exterior of some sphere and that the harmonic region be contained inside the same sphere---in fact many of the numerical trials given in Section~\ref{S:Tests} can be reinterpreted as the method of fundamental solutions (MFS) results for spherical BVPs. (Using the exterior relationships summarized in Section~\ref{S:Recap}, it is clear that a completely analogous treatment of harmonic BVP exterior problems using interior sources is possible.)  Section~\ref{S:MFS} addresses these important  MFS/BVP connections to the DIDACKS approach, including the significance of the above mentioned minimum norm collocation principle and the standard DIDACKS least-squares minimum norm principle for MFS.  These minimum norm conditions imply that MFS approaches are on a more solid theoretical footing than has been commonly recognized and that this footing implies, among other things, that a greater numerical stability results than might otherwise be expected for deep source placements when using MFS for exterior BVP problems---which has been frequently observed and often commented on in the MFS literature (where it is usually noted as the confluence of good solutions and large condition numbers).  Of course, the same source placement effect occurs for the interior setting and it has the same explanation.  It is thus worth noting that exploring this source placement issue in the interior setting is one of the design criteria behind the test scenarios presented in Section~\ref{S:Tests}.

  Finally, it may be useful to briefly touch on other significant topics that are not addressed elsewhere in the article.  While there are many such topics, only three are considered below.  
Corresponding to each of these three topics is an active  area of applied numerical research.  As a guide to the uninitiated in any of these applied areas, at the end of the discussion of each of these three topics the name of journal(s) where relevant research articles often currently occur is given in curly braces.

First, given the well-known historical connections of MFS approaches to the boundary element method (BEM) and the closely related boundary integral method (BIM) [also called boundary integral equation (BIE) method] techniques \cite[p.286]{BEMhistory}, it would seem to be obvious that there are analogous connections between BEM (or BIM) and DIDACKS approaches for Laplacian problems.   Moreover, for Laplacian based problems with BVP surfaces that have appropriate geometries, the resulting connections have significant consequences, including, but not limited to, implied implicit minimum norm conditions for BIM and BEM techniques that have previously gone unnoticed.  This family of integral equation approaches is most commonly presently referred to as BEM approaches and they have a truly remarkable range of applicability.  Some hint of these BEM connections is given at the beginning of Section~\ref{S:MFS} where the case of BVPs with spherical boundary surfaces is discussed; however, even a preliminary discussion of BEM/DIDACKS connections is outside the scope of the present article. Nevertheless, before proceeding it is interesting to consider several follow-on questions that may, in some sense or other, prove to be suggestive: Do these DIDACKS/BEM connections imply that certain applications exist where a BEM-like integral approach can be developed that introduces a (small) layer-like region separating the sources and the boundary surface so that numerical singularities can be avoided?  Is it possible to reformulate some BEM-like approaches to certain problems as a Dirichlet integral norm based approach somewhat along the lines of the DIDACKS approach (this question is raised by the fact that it is possible to reformulate the DIDACKS technique as a surface distribution approach).  Above, it was noted that BEM-like approaches have a broad range of applicability---how broad is the area of applicability where some corresponding implicit norm criteria can be retained.  Although it would seem to be clear that something of interest can be done for MFS or BEM problems with spherical boundaries, what is the most general boundary surface geometry that can be considered, which still has some relevant implicit norm condition?  It is not clear how much can actually be done here, but these questions are raised in the hopes that something can indeed be done and that others will explore the possibilities that are relevant. 
\{For BEM related articles see, for example, the journal \emph{Engineering Analysis with Boundary Elements}.\}  \{For MFS related articles see, for example, the journals \emph{Engineering Analysis with Boundary Elements}, \emph{International Journal for Numerical Methods in Engineering} or \emph{Numerical Methods in Partial Differential Equations}.\}

 Second, elaborating on and adapting the ideas just noted in connection with MFS/BVP approaches, it is readily apparent that the formalism developed here can be more or less directly adapted to problems of inverse source estimation for bounded regions satisfying Laplace's equation provided that the geometry of the bounded region under consideration is fairly benign (i.e., somewhat spherical in shape) since all that is really required in the present formalism is that the source region be in the exterior of some sphere and that the harmonic region be contained inside the same sphere.  Although the region associated with the norm may not match the shape of this boundary, its norm minimization criteria still should generally provide solutions that have a significant physical grounding and it is thus not clear to what extent these shapes need to actually match.  This issue is addressed further in Section~\ref{S:MFS}.  Given the general ambiguities inherent in inverse source estimation problems; it is, however, clear that even for spherical regions there are significant implementation issues and, as such, a separate article is planned to deal with them.  Since inverse source estimation is more commonly associated with problem settings for exterior geometries, this subsequent article will addresses only exterior settings.  Also it is clear in general that relevant points made in conjunction with the exterior setting also hold for the interior setting and conversely; consequently, discussions of relevant points will not be duplicated from one setting to the next unless clarity of exposition dictates otherwise, since it is assumed that these points can be carried over from one setting (or article) to another setting (or article).  (The same comment also holds for different geometries, such as half-space settings, as well as for different underlying spaces.)  There are nevertheless several unique problem areas of interest where sources are in the exterior region and harmonic interpolating solutions for spherical interiors are needed so they are noted here in passing for future reference.  One example of where this might occur is in a physics laboratory or other experimental setting where a bounded static electrical (or magnetic) field test domain is of interest (which can be taken to be spherical or enclosed in a spherical region) and has unwanted field inhomogeneities.  The goal, then, might be to find a set of external sources to counter these unwanted effects.  Various other scenarios obviously exist.  [Although it is outside the scope of the present article, it is worth noting that various ways of performing dipole I norm or II norm DIDACKS fits without assuming explicit point potential values exist and that one of these techniques was briefly described in \cite{Ruf1}.]  \{For inverse source estimation related articles see, for example, the journal \emph{Inverse Problems} or the journal \emph{Geoexploration}.\}

  Third, a common set of problems where sources are in the exterior region and point source interpolating functions are desired in the interior of a bounded region are problems of a type  usually addressed by the Fast Multipole Method (FMM).  Specifically,
in many engineering settings there may be numerous known sources in some exterior region that produce fields whose effects need to be computed in some interior region, which may be taken to be the interior of a sphere (here often there are also sources in the interior region of interest, but their effects can easily be computed separately).  The inherent difficulty is that there are generally so many sources in the exterior region that a direct computation of the resulting interior field is overly burdensome.  FMMs are designed to overcome this computational difficulty and they have found broad usage in diverse areas such as computational field engineering and astronomy.  Generally the various sets of sources are broken down into different regimes depending on how far away they are from the interior field region, so that, for example, the nearer regimes produce greater variations over the field region of interest, but correspond to fewer given sources.  DIDACKS interpolation might profitably be used to speed up this computational process since a properly implemented DIDACKS fit with a reasonable number of sources might well be expected to replicate each of these different field variations over the entire spherical interior in an accurate and efficient way.  To achieve this accuracy and efficiency, DIDACKS fits here clearly must be done for each of the needed regimes separately.   Generally when FMM approaches are used the vector field is the quantity of interest rather than the potential itself and from Table~\ref{Ta:MeasSour} a combined scalar and dipole DIDACKS interpolation scheme is most appropriate for use in these cases.  Thus in effect, aside from the additional burden of computing the potential, the only significant computational load would be the needed FMM computations at the few interpolation points initially required to generate the (dipole/point mass) fits themselves.  A major point here is that there is generally a natural decomposition into near, intermediate and far source terms for most FMM areas of application and from the author's experience with gravity field modeling \cite{Ruf1}, a decomposition into these various field parts is very desirable from both a field modeling and computational efficiency perspective.  It is clearly generally advantageous to treat parts of a field with different characteristics separately when using a DIDACKS based treatment.  This is obviously due to very different spectral properties and associated attenuation attributes that each of the different parts of the field might have; moreover, field parts with different behaviors require quite different associated source spacing in order to obtain good modeling.  For example, for a single source layer, positioning is important not only with regards to the tangential (or horizontal) source interspacing, but also with regards to the overall radial (or vertical) source positioning (for multiple source layers, spacing issues are even more involved, but analogous considerations enter).  In the literature, although somewhat different proposals for simultaneously using both MFS and FMM approaches to related application areas have recently been proposed \cite{MFS6,MFS2}, this is clearly an area warranting further study and in this endeavor DIDACKS theory could prove to be of general relevance.  Although there are other pertinent points that could have been made, the above DIDACKS based approach just outlined was given since it is conceptually simple and may eventually prove to be of some interest for certain applications.  \{For FMM related articles see, for example, \emph{International Journal for Numerical Methods in Engineering}.\}

\section{Summary of Spherical Exterior and Interior Relationships}\label{S:Recap} 

  This section introduces the standard notation to be used and it also outlines the basic inner-product relations of interest for both the $\mathbb{R}^3$ spherical interior and exterior harmonic field regions.  For completeness the basic $\mathbb{R}^3$ point replicating inner-product equation needed to handle half-space interpolation and approximation is also presented without proof.

  Although it is slightly pedestrian, for clarity, consistency and in order to avoid various notational conflicts and ambiguities (such as double subscripts), in the main body of this article three dimensional Cartesian vectors are used and denoted by overset vectors:  $\vec{X} = (x, y, z)^T \in  \mriii$ (for superscript $T \eq$ transpose), while for $n \neq 3$ $n$-dimensional vectors and matrices are denoted by lower and uppercase bold letters, respectively.  (This also serves to make the \riii\ nature of the relationships here transparent and serves to emphasize the fact that only vector analysis is required for their derivation.)  For vectors, slightly different conventions are adopted in Appendix~A as the need arises.  As in \cite{Ruf1}, the general exterior field region is:  $\Omega_{\sps} \eq \{\vec{X} \in  \mriii\mid |\vec{X}| \geqq {R}_{\sps} \}$.  For the half-space case the origin is chosen in the plane $\partial \Omega_1 \eq \{\vec{X} \in  \mriii\mid z = 0 \}$ itself and the field region of interest is $\Omega_1  \eq \{\vec{X} \in  \mriii \mid z \geqq 0 \}$.  The general spherical interior field region is denoted by $\Omega_{[\sps]} \eq \{\vec{X} \in  \mriii\mid |\vec{X}| \leqq {R}_{\sps} \}$, where the subscript ${\sps}$ has been surrounded by a set of square brackets that has been introduced here to denote the closed interior spherical region versus the subscript ${\sps}$ without brackets that is used to denote the corresponding exterior region.  This subscript convention has been introduced here in analogy to the standard notation for open and closed intervals  $(a,\,b)$ and $[a,\,b]$:  thus, for example, $\Omega_{(\sps)} \eq \{\vec{X} \in  \mriii\mid |\vec{X}| < {R}_{\sps} \}$.

   Due to the occurrence of extraneous factors of ${R}_{\sps}$, for the exterior and interior of unit spheres the mathematical formalism developed here assumes a slightly more elegant form that it does for spheres of general radius ${R}_{\sps}$.  Furthermore, since it is fairly simple matter to rescale any given spherical problem to a corresponding unit sphere problem and then transcribe the resulting solution back, letting ${R}_{\sps}$ be arbritrary adds little of significant value.  Thus instead of dealing directly with the regions $\Omega_{\sps}$, $\Omega_{(\sps)}$ and $\Omega_{[\sps]}$, the formalism will be presented for the following unit-sphere analogs:
\begin{align}
 \widehat{\Omega}_{\sps}\  \eq \{\vec{X} &\in \, \mriii\mid \,\, |\vec{X}| \geqq 1\,\}\notag\\ 
 \widehat{\Omega}_{(\sps)} \eq \{\vec{X} &\in \, \mriii\mid \,\, |\vec{X}| < 1\,\}
\\  
 \widehat{\Omega}_{[\sps]}\, \eq \{\vec{X} &\in \, \mriii\mid \,\, |\vec{X}| \leqq 1\,\}\notag 
\end{align}
where the commonly used hat notation for unit vectors has been mimicked.  The region $\widehat{\Omega}_{[\sps]}$ is of particular interest here.

  For each of these three geometries the compliment of the field region will be denoted using a superscript prime and locations in this complimentary region will be denoted by primed vectors.  Thus a typical source point will be denoted $\vec{X}'_k = (x'_k,\,y'_k,\,z'_k)^T$ for $k = 1,\,2,\,3,\,\ldots,\,N_k $, so that ${\ell}_k \eq |\vec{X} - \vec{X'_k}|$.  In all cases it is assumed that the fixed source region is bounded, closed and an interior subset of the compliment of the field region and this compliment is open so that $1/{\ell}_k$ is bounded.  It is also assumed that $\vec{X}'_{k'} \neq \vec{X}'_k$ for all $k' \neq k$.

   For each of these three geometries, let $f(\vec{X})$ and $g(\vec{X})$ denote typical bounded harmonic functions for the region of interest $\Omega$, which can be either $\Omega_1$, $\widehat{\Omega}_{\sps}$ or $\widehat{\Omega}_{[\sps]}$.   For the unbounded domains $\widehat{\Omega}_{\sps}$ and $\Omega_1$ it is also necessary to stipulate that $f$ and $g$ fall off at least as fast as $1/r$ as $r \to \infty$, where $r \eq |\vec{X}|$, so that $|rg| < \infty$ and $|rf| < \infty$ as $r \to \infty$.  Then the inner-product relations of interest for spherical geometries can be compactly stated in terms of weighted Dirichlet integrals over $\Omega$ by introducing an appropriate notation for it.  Recall first that the standard Dirichlet integral over $\Omega$ is generally denoted  $\text{D}[f,\,g] = \iiint_\Omega \vec{\nabla} f\cdot\vec{\nabla} g \,\,dV$.  Generalizing this notation in a natural way to indicate not only the weighting function $\mu = \mu(\vec{X})$, but the domain $\Omega$ as well gives
\begin{equation}
\text{D}[f,\,g,\,\mu,\,\Omega] \eq  \iiint\limits_\Omega \mu\,\vec{\nabla} f\cdot\vec{\nabla} g \,\,dV\ . 
\end{equation}
 Clearly $\text{D}[f,\,g,\,1,\,\Omega] = \text{D}[f,\,g]$.

  Consider the half-space case where $\Omega = \Omega_1$ so that $z \geqq 0$ and $z'_k < 0$.   Then the $ \Omega_1$ point replication (or generalized collocation) property for the DIDACKS kernel ${\ell}_k^{-1}$ is simply
\begin{equation}\label{E:rep1}
 \text{D}[f,\,{\ell}_k^{-1},\,1,\,\Omega_1]\, =\, 2\pi\,f(x'_k,\,y'_k,-z'_k)\ .
\end{equation}

  For the spherical exterior, where $\widehat{\Omega}_{\sps}$ denotes the field domain of interest, $|\vec{X}| \geqq  1$ and $|{\vec{X}}'_k| <  1$.  As an intermediate step it is useful to first introduce the integral (I) norm discussed in \cite{Ruf1}:
\begin{equation}\label{E:Inorm}
(f,\,g){\ls}_{\text{I}} \eq \,-\,\frac1{4\pi}\negthinspace\iint\limits_{\sigma} \Dr (rf\,g)\,\, d\,\sigma,\ \text{ with }\ \Dr \eq \frac{\partial\ }{\partial r}\ ,
\end{equation}
and where $\sigma$ and $d\,\sigma$ have the following meaning when associated with the integral of $f(\vec{X})$
\begin{equation}\label{E:sigeqn}
 \iint\limits_{\sigma} f(r, \theta,\,\phi)\,d\,\sigma\, \eq \int\limits_{\theta=0\ \ }^{\pi}\negthickspace\negthickspace\!\!\int\limits_{\ \phi=0}^{\ \ 2\pi} \negmedspace\left[f(r,\,\theta,\,\phi)\right]{\Big|}_{r=1}\negmedspace\,\, \text{sin}\,\theta\,\,d\,\theta\,d\,\phi
\end{equation}
for standard spherical coordinates $r,\, \theta$ and $\phi$, so that an evaluation on the surface of a unit sphere (i.e., at $r = 1$) is understood for the integrand of (\ref{E:Inorm}).   The integral evaluation convention  introduced by (\ref{E:sigeqn}) is useful in deriving new results since, among other things, it decouples radial and angular dependencies from each other in surface integral expressions.

  Here the label ``integral norm'' derives form the fact that closed-form expressions of integrals can be obtained for the resulting exterior inner-product expressions that involve $f$ and $1/{\ell}_k$.  Specifically
\begin{equation}\label{E:IntRep}
({\ell}_k^{-1},\,f){\ls}_{\text{I}} \, =\, {P_k }\,  f\left({\vec{P}_k}\right)\ ,
\end{equation}
where $P_k = |{\vec{P}}_k|$ with 
\begin{equation}\label{E:PkXk}
{\vec{P}}_k = \frac{\!{\vec{X}}'_k}{\,|{\vec{X}}'_k|^2}\ \,.
\end{equation}
It also turns out that the same basic relationship between interpolation and source points 
(\ref{E:PkXk}) holds for the interior case.
  Two other relationships exist tying the I norm to both unweighted and weighted Dirichlet integrals.  The relationship to the standard Dirichlet integral is
\begin{equation}\label{E:rel2}
 (f,\,g){\ls}_{\text{I}}\ =\ ({1}/{2\pi})\,\, \text{D}[f,\,g,\,1,\,\widehat{\Omega}_{\sps}]\,  - \,(f,\,g){\ls}_{\sigma}\ ,
\end{equation}
where the second factor on the right hand side (RHS) here contains the surface inner product, which is defined by 
\begin{equation}\label{E:SURFint}
(f,\,g){\ls}_{\sigma}\, \eq\, (1/{4\pi})\iint\limits_{\sigma} f(r, \theta,\,\phi)\,g(r, \theta,\,\phi)\,d\,\sigma\ =\, \int\limits_{\theta=0\ \ }^{\pi}\negthickspace\negthickspace\!\!\int\limits_{\ \phi=0}^{\ \ 2\pi} \negmedspace\left[f\,g\right]{\Big|}_{r=1}\negmedspace\,\, \sin\,\theta\,\,d\,\theta\,d\,\phi\ .
\end{equation}
The relationship to the Dirichlet integral with weight $\mu_{\sps} = 1/r$ is
\begin{equation}\label{E:inorm2d}
(f,\,g){\ls}_{\text{I}}\, = \, ({1}/{2\pi})\,\text{D}[f,\,g,\,\mu_{\sps},\,\widehat{\Omega}_{\sps}] \,
= \, \frac{1}{2\pi}\negthinspace\iiint\limits_{\widehat{\Omega}_{\sps}} r^{-1}\, \vec{\nabla}f\!\cdot\!\vec{\nabla}g\,\,dV\ . 
\end{equation}

  The above relationships for $\widehat{\Omega}_{\sps}$ can be compactly stated in terms of Dirichlet and surface integrals by using (\ref{E:inorm2d}) to eliminate the I norm expressions:
\begin{equation}\label{E:firstExterior} 
 \text{D}[f,\,\ell^{-1}_k,\,\mu_{\sps},\,\widehat{\Omega}_{\sps}]\,\, = \, 2\pi\,{{P}}_k\,f({\vec{P}}_k)
\end{equation}
and
\begin{equation}\label{E:secondExterior}
\text{D}[f,\,g,\,1,\,\widehat{\Omega}_{\sps}]\,\, = \,\text{D}[f,\,g,\,\mu_{\sps},\,\widehat{\Omega}_{\sps}]  \, + \, 2\pi\,(f,\,g){\ls}_{\sigma}
\end{equation}
with, of course, $\mu_{\sps} = 1/r$ and (\ref{E:PkXk}) understood.

  For the interior of a sphere, analogous expressions exist.  First, consider the actual definition of the II norm mentioned earlier in Section~\ref{S:intro}:
\begin{equation}\label{E:IInorm}
(f,\,g){\ls}_{\text{II}} \eq \,\,\frac{1}{4\pi}\negthinspace\iint\limits_{\sigma} \Dr (rf\,g)\,\, d\,\sigma\ .
\end{equation}
While (\ref{E:IInorm}) is formally similar to (\ref{E:Inorm}) they are, in fact, quite different since the domain variables for $f$ and $g$ here are understood to satisfy $|\vec{X}| \leqq 1$ rather than $|\vec{X}| \geqq 1$.  [The difference in sign between the RHSs of (\ref{E:IInorm}) and (\ref{E:Inorm}) obviously arises from the difference in direction of the outward pointing unit normal vectors for the common spherical boundary surface of the two harmonic regions.]  Of course with this understanding, the same definition of the surface integral (\ref{E:SURFint}) and the same evaluation convention specified by (\ref{E:sigeqn}) are employed.

  The same basic form of replication equation holds for the interior setting as held for the exterior, (\ref{E:IntRep}):
\begin{equation}\label{E:IIntRep}
({\ell}_k^{-1},\,f){\ls}_{\text{II}} \, =\, {P_k }\,  f\left({\vec{P}_k}\right)\ ,
\end{equation}
where ${\vec{P}}_k$ and ${\vec{X}}'_k$ are again related by the same equation (\ref{E:PkXk}), even thought there locations have basically been interchanged.

   The relationship for interior regions corresponding to (\ref{E:rel2}) is similar, but the surface integral enters with a positive sign rather than a negative one:
\begin{equation}\label{E:IIrel2}
 (f,\,g){\ls}_{\text{II}}\ =\ ({1}/{2\pi})\,\text{D}[f,\,g,\,1,\,\widehat{\Omega}_{[\sps]}]\,  + \,(f,\,g){\ls}_{\sigma}\ ;
\end{equation}
however, this sign difference is significant since it immediately shows that the II norm is positive definite so that $\|f\|{\ls}_{\text{II}} = 0$ if and only if $f = 0$.  Notice (\ref{E:IIrel2}) is slightly reminiscent of the first Sobolev norm where a volume integral of $fg$ is added to  $\text{D}[f,\,g,\,1,\,\widehat{\Omega}_{[\sps]}]$.

The relationship of the II inner product to the weighted interior Dirichlet integral has the same weight function $\mu_{\sps} = 1/r$ and is also like the corresponding exterior relationship (\ref{E:inorm2d}), but with an added term that makes it transparently positive definite:
\begin{equation}\label{E:IInorm2d}
(f,\,g){\ls}_{\text{II}}\, =\, (1/{2\pi})\,\text{D}[f,\,g,\,\mu_{\sps},\,\widehat{\Omega}_{[\sps]}]\, + \, f_o\,g_o
\, = \, \frac{1}{2\pi}\negthinspace\iiint\limits_{\widehat{\Omega}_{[\sps]}} r^{-1}\, \vec{\nabla}f\!\cdot\!\vec{\nabla}g\,\,dV\ + f_o\,g_o\ ,
\end{equation}
where $f_o$ and $g_o$ denote values of $f$ and $g$, respectively, at the origin: $f_o = f(\vec{0})$ and $g_o = g(\vec{0})$.  For bounded connected regions it has long been recognized that stipulating that admissible functions vanish at some interior point, which is equivalent to adding a point  evaluation term, transforms a Dirichlet integral into a positive definite norm \cite[p. 109]{Schmeidler}.  This clearly holds true for weighted Dirichlet integrals as well.
In passing, notice that the singularity of $\mu_{\sps}$ at the origin causes no problems since  $d\,V/r = r\, dr\, d\sigma$. 

  The above equations also immediately imply that compact analogs of (\ref{E:firstExterior}) and (\ref{E:secondExterior}) exist for the interior case:
\begin{equation}\label{E:IIfirstExterior} 
 \text{D}[f,\,\ell^{-1}_k,\,\mu_{\sps},\,\widehat{\Omega}_{[\sps]}]\,\, = \,\, 2\pi{{P}}_kf({\vec{P}}_k)\, -\, 2\pi{{P}}_kf_o
\end{equation}
and
\begin{equation}\label{E:IIsecondExterior}
\text{D}[f,\,g,\,1,\,\widehat{\Omega}_{[\sps]}]\,\, =\, \text{D}[f,\,g,\,\mu_{\sps},\,\widehat{\Omega}_{[\sps]}] \, - \, 2\pi\,(f,\,g){\ls}_{\sigma}\, + \,\, 2\pi f_o\,g_o
\end{equation}
where, of course, $\mu_{\sps} = 1/r$ and (\ref{E:PkXk}) are again understood.  As an aside, observe that substituting a spherical harmonic expansion of $f$ into $\|f\|{\ls}_{\sigma}^2$ shows that $\|f\|{\ls}_{\sigma}^2 \geq f_o^2$ and thus from (\ref{E:IIsecondExterior}) that
\begin{equation}\label{E:IIgt}
\text{D}[f,\,f,\,\mu_{\sps},\,\widehat{\Omega}_{[\sps]}] \geq \text{D}[f,\,f,\,1,\,\widehat{\Omega}_{[\sps]}]\ .
\end{equation}

  It is a simple matter to perform either an exterior or interior fit by substituting the appropriate inner-product expression [i.e., (\ref{E:IntRep}) or (\ref{E:IIntRep})] into (\ref{E:LLSQ}).  The resulting fit can be regarded as an interpolating function when the ${\vec{P}}_k$'s are given and the ${\vec{X}}'_k$'s are determined by (\ref{E:PkXk}) or as a approximation when the  ${\vec{X}}'_k$'s are given and the ${\vec{P}}_k$'s are determined by (\ref{E:PkXk}).  Additional implementation details for scalar point source fits are given in Section~\ref{S:Tests}.

  The basic formalism can also be immediately extended to interpolate for point values of the partial derivatives of $f$, which entails using other types of point source basis functions.  The point is that since first-order partial derivatives of ${\ell}_k^{-1}$ with respect to source coordinates and field coordinates differ by only a sign and differential operators with respect to source coordinates can be moved outside or inside of an inner-product expression at will, closed-form expressions for all of the required inner products can easily be obtained by taking the appropriate partials of both sides of (\ref{E:IntRep}) or (\ref{E:IIntRep}) with respect to source coordinates.   The corresponding interpolating basis functions are various components of higher order point multipoles, such as point dipoles and point quadrupoles.  
There is, however, one complication: for higher-order multipole fits all the derivatives of the potential for all the lower orders are also required in the spherical case because ${\vec{P}}_k = {\vec{P}}_k({\vec{X}}'_k)$ yields additional terms on the RHS of (\ref{E:IntRep}) or (\ref{E:IIntRep}) when partials with respect to the components of ${\vec{X}}'_k$ are taken.  This means, for example, that a dipole fit requires not only point gradient of potential information, but point potential information as well.  For example, it is thus natural in this case to perform not only a point dipole fit, but a combined point mass/dipole fit. 

 As a concrete example, consider how a combined point mass/dipole fit can be implemented.  First, observe that an electrostatic or point mass dipole term is proportional to $\vec{D}_k\cdot\vec{\nabla}{\ell^{-1}_k}$ with a dipole vector source strength of $\vec{D}_k \in \mathbb{R}^3$.  The approximating (or interpolating) potential thus becomes $\varphi \eq \sum^{N_k}_{k=1}\{q_k\ell^{-1}_k + \vec{D}_k\cdot\vec{\nabla}{\ell^{-1}_k}\}$ and the matrix $\mathbf{T}$ in (\ref{E:LLSQ}) thus becomes a block matrix with four partitions: one corresponding to the point source strengths and three for the various vector components of each dipole strength.  Since $\vec{\nabla}{\ell^{-1}_k} = - \vec{\nabla}'_k{\ell^{-1}_k}$, where $\vec{\nabla}'_k \eq ({\partial\,}/{\partial {x'_k}},\,{\partial\,}/{\partial {y'_k}},\,{\partial\,}/{\partial {z'_k}})^T$ and here components of $\vec{X}'_k$ serve only as parameters when they occur inside the inner products for $\mathbf{T}$ and $\mathbf{A}$, all $\vec{X}$ dependent derivative factors operating on ${\ell^{-1}_k}$ that occur inside inner products can be replaced with $\vec{X}'_k$ derivative factors; consequently, these differential operators can be moved inside or outside the inner products entirely as desired so that all needed inner products for $T_{k, k'}$ and $A_k$ can be easily evaluated in closed form.   The associated point measurement quantities and point sources up through composite quadrupole fits are displayed for both the spherical interior and exterior cases in Table~\ref{Ta:MeasSour}.  Analogous possibilities exist for half-space fits, where  matters are even simpler since in this case for a dipole fit, quadrupole fit or other higher-order fit (or any combination thereof) there is no need to include the lower order point source types.

\vskip .2in
\begin{table} 
\begin{center}
\begin{tabular}{|c|c|}\hline
   Required Point Data Types                     & Composite Source Types             \\ \hline\hline
 $f$                                             & Scalar Point Source                \\ \hline
 $f$ \& $\vec{\nabla}\!f$                        & Scalar Point Source \& Dipole             \\ \hline
 $f$, $\vec{\nabla}\!f$ \& Gradient of $\vec{\nabla}\!f$ & Scalar Point Source, Dipole \& Quadrupole \\ \hline
\end{tabular}
\caption{Point Data/Source Correspondences for Spherical Geometries}\label{Ta:MeasSour}
\end{center}
\end{table}

\section{Spherical Interior ($\widehat{\Omega}_{[\sps]}$)  Relationships}\label{S:InteriorRelationships}

  This section derives the three primary relationships for the II norm: (\ref{E:IIntRep}), (\ref{E:IIrel2}) and (\ref{E:IInorm2d}).  Although they are of secondary interest here, the corresponding relationships for the I norm \{(\ref{E:IntRep}), (\ref{E:rel2}) and (\ref{E:inorm2d})\} can be derived in a similar fashion.  In what follows liberal use will be made of the evaluation convention implied by (\ref{E:sigeqn}) and all functions considered will be assumed to be harmonic in the domain $\widehat{\Omega}_{[\sps]}$ and have bounded partials.

  First consider the derivation of (\ref{E:IIntRep}).  From definition (\ref{E:IInorm}) it follows immediately that
\begin{equation}\label{E:IIa}
(g,\,f){\ls}_{\text{II}}\, = \,\,\frac{1}{4\pi}\negthinspace\iint\limits_{\sigma} f\,g\,\, d\,\sigma\, + \,\frac{1}{4\pi}\negthinspace\iint\limits_{\sigma} rg\,\Dr(f)\,\, d\,\sigma\, + \,\frac{1}{4\pi}\negthinspace\iint\limits_{\sigma}rf\, \Dr(g)\,\, d\,\sigma\ . 
\end{equation}
Since ${\nabla}^2 f = 0$ and ${\nabla}^2 g = 0$, applying Green's second identity\footnote{\ $\iint_{\partial{\Omega}} \left(\phi\frac{\partial\psi}{\partial n} - \psi\frac{\partial\phi}{\partial n}\right) \,dS = \iiint_{{\Omega}}(\phi{\nabla}^2\psi - \psi{\nabla}^2\phi)\,\,dV$} with $dS = d\,\sigma$ and ${\partial\,}/{\partial\, n} \eq \Dr$ shows 
\begin{equation}\label{E:IIb}
 \iint\limits_{\sigma} g\,\Dr(f)\,\, d\,\sigma\, = \,\iint\limits_{\sigma}f\, \Dr(g)\,\, d\,\sigma 
\end{equation}
and thus that (\ref{E:IIa}) can be rewritten as 
\begin{equation}\label{E:IIc}
(g,\,f){\ls}_{\text{II}}\, = \,\,\frac{1}{4\pi}\negthinspace\iint\limits_{\sigma}f\,\{2\,\Dr(rg)\, -\, g\}\,\, d\,\sigma\ . 
\end{equation}

The strategy will be to show that when $g$ is replaced by ${\ell}_k^{-1}$ in (\ref{E:IIc}),
then (\ref{E:IIntRep}) follows from Poisson's integral formula  for the interior of a sphere \cite[p. 526]{Korns}.  Towards that end let $r'_k \eq |{\vec{X}}'_k|$ and $\cos {\psi}_k\, \eq  \,(\vec{X}\!\cdot\!{\vec{X}}'_k)/(r\,r'_k)$.  Then substituting ${\ell}_k^{-1} = 1/\sqrt{r^2 + {r'_k}^2 - 2 rr'_k\cos {\psi}_k}$ for $g$ into (\ref{E:IIc}) and performing the indicated evaluation leads to
\begin{equation}\label{E:IId}
({\ell}_k^{-1},\,f){\ls}_{\text{II}}\, = \,\,\frac{1}{4\pi}\negthinspace\iint\limits_{\sigma}\,\frac{({r'_k}^2 -\, 1)\,f}{{{L}}^3_k}\,\, d\,\sigma\,,\ \ \ \text{where}\ \ \ L_k \eq \sqrt{1 + {r'_k}^2 - 2r'_k\cos {\psi}_k}\ . 
\end{equation}
From (\ref{E:PkXk}) $\,\,r'_k = 1/P_k$, so that (\ref{E:IId}) can be easily reexpressed in terms of ${\vec{P}}_k$:
\begin{equation}\label{E:IIe}
({\ell}_k^{-1},\,f){\ls}_{\text{II}}\, = \,\,P_k\left(\frac{1 - P_k^2}{4\pi}\right)\negthinspace\iint\limits_{\sigma}\,\frac{f}{\sqrt{1 + P_k^2 - 2P_k\cos {\psi}_k}}\,\, d\,\sigma\,\ . 
\end{equation}
From Poisson's integral formula for the interior of a sphere, the RHS of (\ref{E:IIe}) can immediately be recognized as $P_k$ times $f({\vec{P}}_k)$, which proves (\ref{E:IIntRep}). 
Although $r'_k < \infty$ [and thus from (\ref{E:PkXk}) that $P_k > 0$] has always been assumed, for completeness the relationship corresponding to (\ref{E:IIntRep}) for the case $\vec{P}_k = 0$ will be derived at the end of the section.

  Next consider the derivation of (\ref{E:IIrel2}).  From Green's first identity\footnote{\ $\iiint_{{\Omega}}(\phi{\nabla}^2\psi + \vec{\nabla}\psi\cdot\vec{\nabla}\phi)\,\,dV = \iint_{\partial\Omega} \phi\frac{\partial\psi}{\partial n}\,dS$}
\begin{equation}\label{E:E0norm}
 \iiint\limits_{\widehat{\Omega}_{[{\sps}]}} \vec{\nabla} f\cdot\vec{\nabla} g\,\,\, d\,V\, =
\iint\limits_{\sigma} g\,\Dr\!f\,\,d\,\sigma\ . 
\end{equation}
Thus (\ref{E:IIrel2}) follows directly from (\ref{E:IIa}), (\ref{E:IIb}) and (\ref{E:E0norm}). 

  Proving (\ref{E:IInorm2d}) is a bit more involved.  First recall the standard identity
\begin{equation}\label{E:DelIden}
{\nabla}^2\psi\phi = \psi{\nabla}^2\phi\, +\, \phi{\nabla}^2\psi\, +\, 2\,\vec{\nabla}\psi\cdot\!\vec{\nabla}\phi\,\ .
\end{equation}
Setting $\psi = fg$ and $\phi = r^{-1}$ in (\ref{E:DelIden}) gives 
\begin{equation}\label{E:fgr}
{\nabla}^2(fg/r) = fg{\nabla}^2r^{-1}\, +\, r^{-1}{\nabla}^2fg\, +\, 2\,\vec{\nabla} r^{-1}\cdot\vec{\nabla}fg\,\ .
\end{equation}
Using (\ref{E:DelIden}) to reexpress ${\nabla}^2fg$ as  $2\,\vec{\nabla}f\cdot\!\vec{\nabla}g\,$, along with the facts that
$\vec{\nabla} r^{-1}\cdot\vec{\nabla} = -(\vec{X}/r^3)\cdot\vec{\nabla} = -(1/r^2)\partial/\partial r$ and that ${\nabla}^2(r^{-1}) = -4\pi\,\delta(\vec{X})$, where $\delta$ is the Dirac delta function, allows (\ref{E:fgr}) to be rewritten as
\begin{equation}\label{E:rfg}
\frac1r\vec{\nabla}f\cdot\!\vec{\nabla}g\, =\, \frac12{\nabla}^2\bigg(\frac{fg}r\bigg)\, +\, 2\pi fg\,\,\delta(\vec{X})\, +\, \frac{1\,\,}{r^2}\frac{\partial (fg)}{\partial r\,\ }
\end{equation}
Next consider Gauss's divergence theorem\footnote{\ $\iiint_{{\Omega}}\vec{\nabla}\cdot \vec{F}\,\,dV = \iint_{\partial{\Omega}}\vec{F} \cdot \vec{dS}$} for the region of interest with $\vec{F} = \vec{\nabla}(fg/r)$:
\begin{equation}\label{E:GaussDiv}
\iint\limits_{{\widehat{\Omega}}_{[\sps]}} {\nabla}^2\bigg(\frac{fg}r\bigg)\,\, d\,V \, = \, \iint\limits_{\sigma} \frac{\partial\ }{\partial r}\bigg(\frac{fg}r\bigg)\,\, d\,\sigma
\end{equation}
Integrating (\ref{E:rfg}) over $\widehat{\Omega}_{[\sps]}$ and using (\ref{E:GaussDiv}) gives
\begin{equation}\label{E:Dfgmu1}
\iint\limits_{{\widehat{\Omega}}_{[\sps]}} r^{-1}\vec{\nabla}f\cdot\!\vec{\nabla}g\,\, d\,V  = \frac12\negthinspace\iint\limits_{\sigma} \frac{\partial\ }{\partial r}\bigg(\frac{fg}r\bigg)\,\, d\,\sigma\, +\, 2\pi f_o\,g_o\, +\, \iint\limits_{\sigma} \Bigg{\{}\,\,\int\limits_{r=0}^{\ \ 1}\ \frac{\partial (fg)}{\partial r\ }\,\,d\,r\Bigg{\}}\,\, d\,\sigma\ ,
\end{equation}
where $d\,V = r^2\,d\,r\,\,d\,\sigma$ was used on the last term on the RHS here. [With respect to the last term on the RHS of (\ref{E:Dfgmu1}), since the $r$ limits for the integrand occuring inside the integral with respect to $d\,\sigma$ are fixed it is clear that the evaluation convention implied by (\ref{E:sigeqn}) is immaterial for this particular integrand.]

  Since $\Dr(fg/r) = -r^{-2}fg\, +\, r^{-1}\Dr(fg)\,$, (\ref{E:Dfgmu1}) can be rewritten as 
\begin{equation}\label{E:Dfgmu2}
\begin{split}
\text{D}[f,\,g,\,1,\,\widehat{\Omega}_{[\sps]}]\,\, &= -\frac12\iint\limits_\sigma fg\,\,d\,\sigma\, + \,\frac12\!\!\iint\limits_\sigma\frac{\partial (fg)}{\partial r\ }\,\,d\,\sigma\, +\, 2\pi f_o\,g_o\, + \,\iint\limits_\sigma fg\,\,d\,\sigma\, -\,4\pi f_o\,g_o 
\\
 & = \,\frac12\,\iint\limits_\sigma \Dr\,(rfg)\,\,d\,\sigma\, -\,2\pi f_o\,g_o\ ,\\
\end{split}
\end{equation}
which immediately shows (\ref{E:IInorm2d}) from the definition of the II norm, (\ref{E:IInorm}).

  Finally consider the analog of (\ref{E:IIntRep}) for the case $\vec{P}_k = 0$.  First obverse that since $\vec{X}'_k$ and $\vec{P}_k$ are related by (\ref{E:PkXk})
\begin{equation}\label{E:Xinfty}
\lim\limits_{P_k \to 0}\ \frac{1}{P_k{\ell}_k} = \lim\limits_{P_k \to 0}\ \frac{1}{|P_k\vec{X} - P_k\vec{X}_k|} = 1\ ,
\end{equation} 
thus allowing interpolation at $\vec{P}_k = 0$ corresponds to adding a constant basis function to $\varphi$.  Thus using (\ref{E:IInorm}), consider the following inner product
\begin{equation}\label{E:Basis1}
(1,\,f){\ls}_{\text{II}}\, = \,\,\frac{1}{4\pi}\negthinspace\iint\limits_{\sigma}\Dr(rf)\,\, d\,\sigma\, = \,\,\,\frac{1}{4\pi}\negthinspace\iint\limits_{\sigma}f\,\, d\,\sigma\, +  
\,\frac{1}{4\pi}\negthinspace\iint\limits_{\sigma}\Dr(f)\,\, d\,\sigma\ .
\end{equation}
Next consider a spherical harmonic expansion of $f$.  From the spherical harmonic
orthogonality conditions it is obvious that
\begin{equation}\notag
\frac{1}{4\pi}\negthinspace\iint\limits_{\sigma}f\,\, d\,\sigma\ = f_o\ ,
\end{equation}
while from Green's first identity or Gauss's divergence theorem it is clear that
\begin{equation}\notag
\frac{1}{4\pi}\negthinspace\iint\limits_{\sigma}\Dr(f)\,\, d\,\sigma\, = 0\ .
\end{equation}
Hence the inner product analog of (\ref{E:IIntRep}) for the case $\vec{P}_k = 0$ is
\begin{equation}\label{E:Basis2}
(1,\,f){\ls}_{\text{II}}\, = \, f_o\,,
\end{equation}
which also holds for the case $f =1$, as a direct substitution into (\ref{E:IInorm}) shows:
\begin{equation}\notag
(1,\,1){\ls}_{\text{II}}\, = 1\ .
\end{equation}
(If the evaluation of any inner product expression is ever in question, then the definition of the inner product itself should be used for the evaluation.)

\section{Synopsis of the Collocation Minimum Norm Property}\label{S:Analysis}

 This section gives an overview of the collocation minimum norm property that includes an indication of its utility.  A fairly detailed mathematical discussion of the collocation minimum norm property for a general setting with various proofs and counter-examples is given in Appendix~A; however, this appendix is intended for a limited audience.  The discussion here and in Appendix~A has implications for geophysical collocation (GC) (perhaps more commonly called least-squares collocation \cite{Moritz}), general kernel fitting theory (specifically, RKHS collocation fits), interpretation of inverse source theory and  MFS like approaches (including BEM and related approaches for certain relevant geometries).  Special care and attention is needed here since this result must be extended to DIDACKS theory. but the generalizations of it considered here may also prove to be useful in other contexts.
   
  For the II norm, the collocation minimum norm property for scalar point sources over $\Omega_{[\sps]}$ can be stated as the fact that of all admissible functions $g$ such that
\begin{equation}\label{E:Pcoll}
g({\vec{P}}_k) = C_k\ \ \text{for}\ \ k = 1,\,2,\,3,\,\ldots,\,N_k\,\,,
\end{equation}
where the $C_k$ are arbritrary bounded constants, a DIDACKS scalar point source fit with coefficients $q_k$ specified by (\ref{E:LLSQ}) is the function of smallest norm: $\|g\|{\ls}_{\text{II}} \geq \|\varphi\|{\ls}_{\text{II}}$.  Obviously this means that there are two distinct minimum norm conditions simultaneously satisfied by a DIDACKS fit that is based on $1/{\ell}_k$:
\begin{itemize}
\item
  $\|f - \varphi\|$ is minimized directly by (\ref{E:LLSQ}). 
\item
 $\|\varphi\|$ is minimized by (\ref{E:Pcoll}).
\end{itemize}
(Trivially, if $a^2 > b^2$ then $|a| > |b|$ and conversely, so that $\|f - \varphi\|^2$ and $\|\varphi\|^2$ are also concurrently minimized.)
The first of these two norm properties is obviously naturally satisfied by any DIDACKS fit.  As shown in Appendix~A, the second minimum norm property also holds simultaneously for DIDACKS scalar point source fits over exterior geometries.  RKHS fits that meet certain minimum requirements also display both of these two minimum norm properties.  For a general RKHS setting, although they are usually not addressed in the general literature in conjunction with demonstrations of the collocation minimum norm property, the precise theorem requirements necessary are explicitly spelled out and discussed at length in Appendix~A and the topic of counter-examples is also raised there in order to bolster the case for normally including these conditions.  (Unfortunately no references known to the author discuss these conditions.) 

  There are two primary lines of argument associated with the collocation minimum norm property (\ref{E:Pcoll}) that are of interest here:
\begin{description}
\item[\ \ \textbullet \ Type \,I\,]proofs, which are based primarily on various properties of the fit $\varphi$ itself and
\item[\ \  \textbullet \ Type II]proofs, which are based primarily on the properties of the function to be fit, $f$, \\ 
      \text{ \ \ \ \ \ \ \ \ \ \ \ \ \,}as well as its associated function space.
\end{description}
Appendix~A details various aspects of each of these approaches in turn.  In order to clarify the various relevant aspects, it is useful to give a brief overview of each of these overall lines of development.    Type I arguments are generally used to prove RKHS collocation minimum norm property (i.e., GC) analogs of (\ref{E:Pcoll}) \cite{Moritz}.  Type II arguments are generally found in conjunction with proofs of associated results that are often found in mathematics monographs and texts and, in a sense, are otherwise hard to characterize.  Both of these two lines of argument have been slightly abused in much of the literature, but the first one perhaps much more so than the second one.  The main abuse of for both lines of argument in the existing literature has been to state results as though they were perfectly general without warning the reader that, while they may be true in the particular context under consideration, they cannot be applied to other contexts without closer examination and may, in fact, well not hold at all in these other contexts.  For example in the main-stream GC literature, historically in proofs of collocation minimum norm results it has been the convention to implicitly assume that the analog of $\mathbf{T}^{-1}$ exists when, in fact, as it should be apparent from the sequel, this actually deserves some sort of proof and, furthermore, it clearly is not true for some common RKHS settings of interest---see, for example, Moritz's \emph{Advanced Physical Geodesy}, which is a well written and well respected GC reference \cite[pp. 207--220]{Moritz}.  (Helmut Moritz, along with Torbin Krarup, actually founded GC, hence Moritz is rightly considered the leading expert in the field.)  Generally the places where the existence of $\mathbf{T}^{-1}$ enters as an unstated assumption in these sorts of proofs is not obvious since the proofs themselves are usually longer or more abstract than they need to be and, as the line of proof given immediately below shows, the implicit point of entry of this assumption itself is not at all obvious---except after the fact and when one knows to look for it.

  Before proceeding, consider the overall way that the above two types of arguments will be employed.  The overall distinction between the two types of arguments above may be slightly unclear for now, but this distinction should be clearer from the sequel since these arguments are used in a fairly clearcut way.  First, a Type I argument will be used to show that when $\mathbf{T}^{-1}$ exists the collocation minimum norm property exemplified by (\ref{E:Pcoll}) holds for DIDACKS scalar point source fits and that, moreover, from results in the literature $\mathbf{T}^{-1}$ can be shown to always exist for this case.  All this results in Theorem~\ref{T:DIDACKSfit}, which is generalized in Appendix~A. In general, the major problem here is that due to properties of (or restrictions on) the space of admissible functions (\ref{E:Pcoll})  may represent an overdetermined or nearly overdetermined system and the easiest way to prove $\mathbf{T}^{-1}$ exists is by examining the properties of the space of admissible functions itself.  This Type II line of thought culminates in Theorem~\ref{T:FkBk}.  Since, as discussed below, scalar point sources are linearly independent and since DIDACKS point replication conditions exist, in Appendix~A it is shown that \riii\ harmonic functions over $\Omega_{\sps}$, which is the main geometric backdrop for GC, satisfy a certain property (namely pointwise independence), which, in turn, implies that if no factors including differential operators are involved, then the GC analog of $\mathbf{T}^{-1}$ always exists.  Thus, at least for this special case, one can then say that the usual GC collocation minimum norm proofs do indeed hold.

 Thus first consider an outline of how (\ref{E:Pcoll}) might be demonstrated using the Type I approach.  For convenience it will be assumed that $\vec{P}_k \neq 0$, although it should be apparent that the same line of argument holds when one of the points in (\ref{E:Pcoll}) is zero.
Here, of course, it is assumed that $f$ is a fixed specified function and that $\varphi = \sum^{N_k}_{k=1}q_k/\ell_k$ is also a fixed specified function, where the source strengths $q_k$ are determined by (\ref{E:LLSQ}).  Hence (\ref{E:LLSQ}) can be written out explicitly as
\begin{equation}\label{E:PhiBif}
\sum\limits_{k'=1}^{N_k}({\ell}_{k}^{-1},\,{\ell}_{k'}^{-1}){\ls}_{\text{II}}\,q_{k'}
 = ({\ell}_{k}^{-1},\,f){\ls}_{\text{II}}\,\,,
\end{equation}
or as 
\begin{equation}\label{E:PhiBif2}
({\ell}_{k}^{-1},\,\varphi){\ls}_{\text{II}}
 = ({\ell}_{k}^{-1},\,f){\ls}_{\text{II}}\,\,,
\end{equation}
 since the $q_k$'s are assumed to be fixed by (\ref{E:LLSQ}) itself.  Because $P_k > 0$, (\ref{E:IIntRep}) then immediately implies 
\begin{equation}\label{E:PfColl}
\varphi\big({\vec{P}_k}\big)\, = \,f\big({\vec{P}_k}\big)\ .
\end{equation}
Let $C_k \eq f\big({\vec{P}_k}\big)$, then (\ref{E:PfColl}) directly shows that $\varphi$ satisfies the point matching requirement of (\ref{E:Pcoll}). 

 It remains to be shown that no other function that satisfies the same matching condition has a larger norm than $\varphi$.  First, for convenience rewrite (\ref{E:PhiDef}) as
\begin{equation}\label{E:PhiDefinit}
\Phi(f) \eq \|f - \varphi\|{\ls}_{\text{II}}^2 =  \|f\|{\ls}_{\text{II}}^2 - 2(f,\,\varphi){\ls}_{\text{II}} + \|\varphi\|{\ls}_{\text{II}}^2 \geq 0\ .
\end{equation}
Multiplying (\ref{E:PhiBif2}) by $q_k$ and summing over $k$ yields
\begin{equation}\label{E:PhiBif3}
\|\varphi\|{\ls}_{\text{II}}^2 = (\varphi,\,f){\ls}_{\text{II}}\ .
\end{equation}
Using $2(\varphi,\,f){\ls}_{\text{II}} = 2\|\varphi\|{\ls}_{\text{II}}^2$ in (\ref{E:PhiDefinit}) produces
\begin{equation}\label{E:fGEQphi}
\Phi(f) \eq \|f - \varphi\|{\ls}_{\text{II}}^2 =  \|f\|{\ls}_{\text{II}}^2 - \|\varphi\|{\ls}_{\text{II}}^2 \geq 0\ \ {\text{or}}\  \ \|f\|{\ls}_{\text{II}} \geq \|\varphi\|{\ls}_{\text{II}}\ .
\end{equation}

  Next let $g$ be some arbritrary admissible function that satisfies the point matching equation 
(\ref{E:Pcoll}) at the specified locations.  Consider a DIDACKS fit to $g$ with basis functions ${\ell}_{k}^{-1}$ and coefficients $q'_k$: ${\varphi}' \eq  \sum^{N_k}_{k=1}q'_k/\ell_k$.  Since the fit of ${\varphi}'$ to $g$ is completely analogous to that of  ${\varphi}$ to $f$, it follows that a relationship corresponding to (\ref{E:fGEQphi}) also obviously holds for $g$ and ${\varphi}'$:
\begin{equation}\label{E:fGEQphiP}
\Phi(g) \eq \|g - {\varphi}'\|{\ls}_{\text{II}}^2 =  \|g\|{\ls}_{\text{II}}^2 - \|{\varphi}'\|{\ls}_{\text{II}}^2 \geq 0\ \ {\text{or}}\  \ \|g\|{\ls}_{\text{II}} \geq \|{\varphi}'\|{\ls}_{\text{II}}\ .
\end{equation}
Next consider the actual form that the fit ${\varphi}'$ takes.  For $k = 1,\,2,\,3,\,\cdots\,N_k$, since 
\begin{equation}\notag
g\big({\vec{P}_k}\big) = C_k = f\big({\vec{P}_k}\big)
\end{equation}
it follows from (\ref{E:IIntRep}) that
\begin{equation}\notag
({\ell}_{k}^{-1},\,g){\ls}_{\text{II}} = ({\ell}_{k}^{-1},\,f){\ls}_{\text{II}}
\end{equation}
and thus it follows from (\ref{E:PhiBif}) that $q'_k = q_k$ so that ${\varphi}' = {\varphi}$.
(\ref{E:fGEQphiP}) then shows $\|g\|{\ls}_{\text{II}} \geq \|{\varphi}\|{\ls}_{\text{II}}$, as desired.  This proof is a streamlined and much more transparent version of a standard type of proof of the collocation minimum norm property that is found, for example, in discussions of GC.

  Consider a general RKHS setting.
  On first examination, it might seem that this line of proof just given is perfectly general and can be directly adapted to any RKHS setting without the need for additional stipulations or conditions.  Upon closer examination and upon considering possible counter-examples---as in Appendix~A---it is apparent that at several places $\mathbf{T}^{-1}$ was assumed to exist and that this is needed as a separate assumption.  As discussed in Appendix~A an independent proof exists in the literature that scalar point source basis functions are independent, which implies here that $\mathbf{T}^{-1}$ always exists, so this is not needed as a separate assumption for the special case dealt with above.  (Here linear independence of a set of functions $\{f_j(\vec{X})\}_{j=1}^N$ means that $\sum_{j=1}^N c_jf_j(\vec{X}) = 0$ implies $c_j = 0$ for all $j$ and all $\vec{X} \in \Omega$.  The relevant points pertaining to linear independence of functions can be gathered, for example, from Korn \& Korn at various places \cite[p. 488, p. 253]{Korns}.)
Since it is obvious that the above line of argument also holds for any appropriate DIDACKS setting, this can all be summed up in the following theorem:
\begin{theorem}\label{T:DIDACKSfit}
A finite \R{3} DIDACKS scalar point source (i.e., point charge or point mass) fit $\varphi$ exactly replicates the data values that result from the point evaluation of some specified function to be fit $f$; moreover, the resulting fit minimizes not only $\|f - \varphi\|$, but also $\|\varphi\|$ is less than or equal to the norm of that of any other admissible function that also matches the specified data point values, where $\|\,\cdot\,\|$ is an appropriate DIDACKS norm.
\end{theorem}

 In general
it is obvious that if the dimension of the family of admissible functions is $N_k$ or less, then the point matching condition (\ref{E:Pcoll}) determines (at most) one unique function so that at best $\varphi = g = f$ and thus $\Phi(f) = 0$.   When there is only one unique function specified by (\ref{E:Pcoll}) the two minimum norm conditions become the trivial statements that $\|f - \varphi\| = \|0\| = 0$ and $\|g\| = \|\varphi\|$.  It is useful for having a way of preventing this, but the concept of linear independence of functions is not strong enough.  Thus a new condition is needed here to quantify the relevant linear-independence like property of the class of admissible functions.  Thus, consider the following definition that is stated for convenience in terms of \riii\ points, but holds for scalar valued functions over \R{n}:

\vskip 7pt

\noindent
{\bf{Definition \ref{S:Analysis}.1}}\ \ 
A class of scalar valued functions $\mathscr{F}$ is said to be \emph{uniformly $N$ pointwise independent} over a region $\Omega$ if for $N + 1$ arbritrary points ${\vec{X}}_j \in \Omega$ and arbritrary bounded constants $C_j$, there always exists a function $f \in \mathscr{F}$ such that $f({\vec{X}}_j) = C_j$ for $j = 1,\,2,\,3,\,\ldots,\,N+1$.  For any finite $N$, if $\mathscr{F}$ is uniformly $N$ pointwise independent, then $\mathscr{F}$ will be called simply \emph{uniformly pointwise independent}.
\vskip 7pt

\noindent
The relationship of this definition to linear independence of a set of functions will be briefly considered in the next paragraph after a few other points are made here.  As a slight abuse of nomenclature, if no confusion can arise a function $f$ may be labeled uniformly ($N$) pointwise independent rather than the class of functions that $f$ belongs to.  Clearly, if the class of admissible functions is uniformly $N$ pointwise independent, then it has a dimension of at least $N$ and there is more than one function in the class of admissible functions that satisfies (\ref{E:Pcoll}) when $N_k = N$.  Finally as an aside, although it might seem obvious that most classes of functions satisfy Definition \ref{S:Analysis}.1 in one sense or another so that it is a trivial property, it is actually a fairly strong property.  Suppose, for example, that a family of functions is uniformly pointwise independent over $\Omega$ and that a complete orthonormal set of basis functions exists for this family over the same region, then it is obvious that at any arbritrary point of $\Omega$ all of these basis functions cannot simultaneously be zero (i.e., at any point at least one of the basis functions must be non-zero).  It is also easy to show that a restricted class of functions that necessarily is either even or odd with respect to any of its variables cannot be uniformly pointwise independent.   Also the demonstration that a class of functions is uniformly pointwise independent usually requires a constructive proof.

     The concept of a uniformly $N$ pointwise independent span of functions is stronger than that of set of span of $N + 1$ linearly independent functions.  One type of counter-example will have to suffice here.  The functions under consideration need not be harmonic, thus let $f_j \in \{f_j(\vec{X})\}_{j=1}^{N+1}$ be a function that has the value $1$ at $\vec{Q}_j$ and that rapidly falls off to zero away from $\vec{Q}_j$ so that $f_{j}(\vec{Q}_{j'}) = 0$ for $j' \neq j$.  This set of functions is obviously linearly independent.  However, for most of these types of functions one would generally expect to find some points in $\Omega$ such that $f_j(\vec{X}) = 0$ for all $j$ and when this happens the set $\mathscr{F} \eq$ linear span of $\{f_j(\vec{X})\}_{j=1}^{N+1}$ is obviously not uniformly $N$ pointwise independent.

  Next briefly consider Type (II) approaches to collocation minimum norm related results.  One common feature of these approaches is that while mathematicians generally use them correctly, the results obtained are, in the end, only loosely related to the least-squares collocation minimum norm principle.
The Type (I) approach used above to obtain the collocation minimum norm result given in Theorem~\ref{T:DIDACKSfit} will be the main one used here and in Appendix~A, so the Type (II) approach used below will be primarily employed as a tool for obtaining a different perspective.  This idea  is that if the class of admissible functions is uniformly pointwise independent (or uniformly $N$ pointwise independent) then the existence of a point replication or reproducing kernel has strong consequences, that include, among other things, the linear independence of associated basis functions.  Given, then, that the basis functions are necessarily independent the above proof suffices to show $\|g\| \geq \|\varphi\|$; hence, the issue of linear independence of basis functions is the only one that will be followed up on.

  Since little additional effort is required it is useful to consider the case of both point replicating and reproducing kernels simultaneously over some domain $\Omega$; hence, consider a linear expansion in terms of basis functions of two arguments: $\psi \eq \sum_{k=1}^{N_k}\,q_k\,{\mathcal{B}}(\vec{X},\, {\vec{P}}_k)$, where $\vec{X} \in \Omega$ is regarded a an independent vector variable and $\vec{P}_k \in \Omega$ is regarded merely as a vector parameter.  Here (point) replication conditions are generalized so as to handle both point replicating and reproducing kernels by assuming that for $\mathcal{B} = {\mathcal{B}}(\vec{X},\, {\vec{P}})$
\begin{equation}\label{E:Bhf}
({\mathcal{B}},\,f) = h({\vec{P}})f({\vec{P}})
\end{equation}
where  $h({\vec{P}}) \neq 0\,$---\,or more specifically for cases of interest that $h({\vec{P}})$ has the form $h({\vec{P}})= \lambda$ or $h({\vec{P}})= {P}$ for some nonzero $\lambda$.  Thus, for example, (\ref{E:IIntRep}) can be reexpressed in this form by first using (\ref{E:PkXk}) to eliminate the source location $\vec{X}'$ in ${\ell}^{-1}$, which results in the required ${\mathcal{B}}$ form with $h({\vec{P}})\eq {P}$. [Here the case ${\vec{P}} = 0$ can be handled by using (\ref{E:Basis2}).]
 Likewise, for a reproducing kernel $K(\vec{X},\, {\vec{P}})$, since $(K,\,f) = f({\vec{P}})$ the identifications ${\mathcal{B}}(\vec{X},\, {\vec{P}}) \eq K(\vec{X},\, {\vec{P}})$ and $h({\vec{P}}) = 1$ suffice.

 In Appendix~A the basis functions are defined in a slightly more general way where the vector parameter(s) need not be explicitly introduced as an argument.  Thus, in this sense, let ${\mathcal{B}}_{k}(\vec{X}) \eq {\mathcal{B}}(\vec{X},\, {\vec{P}}_k)$ where the points ${\vec{P}}_k$ are assumed to be distinct, then the following relevant theorem holds:
\begin{theorem}\label{T:FkBk}
If the family of admissible functions is uniformly pointwise independent and if there is a point replicating or reproducing kernel characterized by (\ref{E:Bhf}) then the set of basis functions $\{{\mathcal{B}}_{k}\}_{k=1}^{N_k}$ are linearly independent.
\end{theorem}
\begin{proof}
 Suppose to the contrary that the ${\mathcal{B}}_{k}$'s are linearly dependent and thus that some set of $q_k$'s exist such that $\sum_{k=1}^{N_k}{\mathcal{B}}_{k}(\vec{X})\,q_k = 0$, where at least one of the $q_k$'s are non-zero.  Specifically, let $q_{k'} \neq 0$ and then reorder the basis functions so that ${\mathcal{B}}_{k'}$ is the first one and thus $q_{1} \neq 0$.  (\ref{E:Bhf}) implies that $\sum_{k=1}^{N_k}q_k({\mathcal{B}}_{k},\,f) = \sum_{k=1}^{N_k}q_kh({\vec{P}}_k)f({\vec{P}}_k) = 0$.  Since $f$ is uniformly pointwise independent, choose it such that $f({\vec{P}}_k) = 1$ if $k=1$ and $f({\vec{P}}_k) = 0$, if $k > 1$, then $\sum_{k=1}^{N_k}q_kh({\vec{P}}_k)f({\vec{P}}_k) = q_1h({\vec{P}}_1) \neq 0$, which is a contradiction. 
\end{proof}

  Finally briefly consider the significance of the collocation minimum norm property.  As noted above, the collocation minimum norm property used in conjunction with various covariance norms over $\Omega_{\sps}$ is a cornerstone of GC \cite{Moritz}.  It is also well known result from general RKHS theory within various other contexts, such as analytic function theory and functional analysis: given certain conditions, of all those functions that match some specified function ($f$) at $N_k$ points, a linear combination of $N_k$ (symmetric) reproducing kernels is the one that has the smallest norm.  There are also two broad areas where this minimum norm property strongly suggests that DIDACKS theory has results of direct interest for other mathematical specialties:
\begin{itemize}
 \item
  As previously mentioned, for MFS and BEM-like methods DIDACKS theory implies that for many Laplacian problems there is a not only an implied minimization of (\ref{E:PhiDef}), but also that the MFS or BEM solution, $\varphi$, is the one of smallest possible norm; i.e., any other harmonic function $g$ that matches the prescribed boundary  conditions (which are generally discretized or otherwise incomplete) must satisfy $\|g\| \geq \|\varphi\|$.  Observe that this condition provides a very strong interpretational basis to MFS and BEM methods for appropriate geometries that, so far, has gone unnoticed.  Moreover, it probably implies that there is a certain inherent numerical robustness in these approaches.  Furthermore, one could argue that even when these approaches are based on some other criteria than the direct matching of $f$ at prescribed points, the possibility of such a match is always implicitly there and thus the associated theoretical implications are also, in some sense or other, present. 
 \item
It seems likely that various (weighted) energy norm inequalities can be derived and that they have both theoretical and practical importance (c.f., Raleigh-Ritz approaches). 
\end{itemize}
Connections of these minimum norm conditions to MFS approaches will be dealt with briefly in Section~\ref{S:MFS} and those of BEM-like approaches are presumably analogous, which leaves only the energy inequalities to discuss.  

It is clear that DIDACKS theory is useful for deriving various energy (norm) based inequalities from an examination of the interactions of (\ref{E:rep1}), (\ref{E:firstExterior}), (\ref{E:secondExterior}), (\ref{E:IIfirstExterior}), (\ref{E:IIsecondExterior}) and (\ref{E:IIgt})
 in concert with these two minimum norm results.  A single example here will have to suffice.  Thus consider the exterior of a sphere (i.e., the I norm over $\widehat{\Omega}_{{\sps}}$) and a harmonic function $f$ that falls off sufficiently fast as $r \rightarrow \infty$.  First, from the collocation minimum norm property applied to the exterior I-norm setting
 \begin{equation}\label{E:inequal1}
 \|f\|{\ls}_{\text{I}}^2 \geq \|{\varphi}{\ls}_{\text{I}}\|{\ls}_{I}^2 = {\mathbf{q}}^T\,\mathbf{T}\,\mathbf{q} = {\mathbf{q}}^T\,\mathbf{A}
 \end{equation}
where $\mathbf{q}$ and ${\varphi}{\ls}_{\text{I}}$ are assumed to be fixed by (\ref{E:LLSQ}) for the I norm and thus, for example, from (\ref{E:IntRep}) $\mathbf{A}$ is the vector whose components are $A_k = ({\ell}_k^{-1},\,f){\ls}_{\text{I}} = P_kf({\vec{P}}_k)$.  Since here $\text{D}[f,f] = \text{D}[f,f,1,\widehat{\Omega}_{\sps}] > \text{D}[f,f,{\mu}_{\sps},\widehat{\Omega}_{\sps}]$\,, from inequality (\ref{E:inequal1}) it follows immediately that
 \begin{equation}\label{E:inequal2}
\iiint\limits_{\widehat{\Omega}_{\sps}} \,\vec{\nabla} f\cdot\vec{\nabla} f \,\,dV \, > \, 
 {\mathbf{q}}^T\,\mathbf{A}\ .
 \end{equation}
For $N_k$ less than or equal to three $\mathbf{T}^{-1}$ can be evaluated in closed form and thus a closed form expression for inequality (\ref{E:inequal2}) can easily be found, while for larger $N_k$ a numerical lower bound can be easily computed.  (The explicit determination of these $1$-point, $2$-point and $3$-point forms is left to the reader.)

\section{Implementation Issues and Numerical Tests}\label{S:Tests}

\newcommand{\cn}{$C_{{\#}}$} 
\newcommand{\cns}{{\cn}$\vphantom{s}'\!$\,s}
 
  This section presents and draws conclusions from numerical interpolation examples based on (\ref{E:LLSQ}).  While the interpolation region of interest is always taken here to be the entire interior of a unit sphere, it is also obviously possible with the present approach to  interpolate over only some portion of this interior.  As mentioned in Section~\ref{S:intro}, only interpolation using scalar point sources will be considered here since the author has had extensive experience with dipoles and quadrupoles for the exterior of \riii\ spheres and for \riii\ half-space \cite{Ruf1} and this experience strongly indicates that with a little care others should be able to apply DIDACKS theory to standard LLSQ dipole interior problems without untoward difficulty; moreover, the difficulties that might normally be expected are mostly the same ones encountered in performing LLSQ scalar point source DIDACKS fits. 

 Before detailing the actual specifications of the test functions to be fit ($f$) and of the interpolating functions ($\varphi$) used to match them, several computational and conceptual details need to be briefly addressed since the matrix $\mathbf{T}$ in (\ref{E:LLSQ}) often has large condition numbers.  After this discussion various other side issues will also be addressed before the actual numerical examples are considered. 

\begin{center}
\ \\
 \begin{large}\underline{\textbf{Side Issues}}\end{large}\\
\ \\
\end{center}

  Large condition numbers are not unexpected since point source problems are known for their ill-conditioned nature; moreover, the confluence of large condition numbers and good results has been long noted in conjunction with the somewhat analogous MFS approach \cite{MFS1}.  The magnitude of this confluence and the necessity of pairing a suitable $\varphi$ to $f$, based on the properties of $f$, are two of the major results that follow from the numerical examples.  A simple explanation for this confluence exists and is given next.  Here the condition number ({\cn}) is taken to be the largest eigenvalue divided by the smallest eigenvalue and absolute values are not needed since all of the eigenvalues encountered here are positive. 

  First, the same basic confluence of large condition numbers and good fits occurs for point source fits over any region and since physical geodesy and MFS approaches commonly use exterior spherical settings and the language is a little less cumbersome for these settings, for simplicity $\widehat{\Omega}_{\sps}$ will be assumed in the present paragraph.  (Recall that in this setting $\vec{X} \geq 1 > \vec{X}'_k > 0$.)  Second, it is easy to see that it is \underline{possible} for $\varphi$ to match $f$ even in the presence of large \cns\ since (\ref{E:LLSQ}) directly minimizes $\|f - \varphi\|{\ls}_{\text{I}}^2$ and there are no constraints put on the source parameters themselves.  Hence, the values of $|q_k|$ may be large and yet the $q_k$ may still produce a $\varphi$ that matches $f$ closely.  Third, when normalized basis functions are used, the resulting $\mathbf{T}$ matrix will be denoted $\widetilde{\mathbf{T}}$ and the mathematical form of its elements are suggestive:
\begin{equation}\label{E:Ttilde}
\widetilde{T}_{k\,k'} \eq ({\ell}_k^{-1},\,{\ell}_{k'}^{-1}){\ls}_{\text{I}}/\{\|{\ell}_k^{-1}\|{\ls}_{\text{I}}\,\|{\ell}_{k'}^{-1}\|{\ls}_{\text{I}}\}\ .
\end{equation}
From an  examination of the explicit closed-form expression for (\ref{E:Ttilde}) [which can be easily determined from (\ref{E:inorm2d}) and (\ref{E:firstExterior})]  it is obvious that deep source placements correspond to large inner products when $k' \neq k$, which translates directly into large \cns.  Fourth and finally, if $f$ is smooth, then the basis functions $\ell_k^{-1}$ that are used to match $f$ must also be smooth, which means that the sources must be deep.  One important point here is that if $f$ is not smooth then shallow point source placements may be called for and, in this case, the \cns\ may be much smaller.

  Consider this last point.  In the MFS literature it has often been observed that the deeper the sources are the better the results and it has occasionally been speculated that this is a general characteristic.  To see that this is not so and that it is dependent on the smoothness of $f$, only one simple counter-example needs to be considered.  Thus consider a function $f$ to be fit with the form $f = q_f/|\vec{X} - \vec{X}'_f|$, where $q_f$ is some appropriate constant and $\vec{X}'_f$ is fixed in what follows.  Suppose, that $\varphi$ has the usual form: 
$\varphi \eq \sum^{N_k}_{k=1}q_k/\ell_k$ and that the ``depth'' of the sources is adjusted simply by simultaneously rescaling the $\vec{X}'_k$'s by $\vec{X}'_k \rightarrow \lambda\vec{X}'_k$, so that the various source locations determine fixed directions even as $\lambda$ is varied.  If for some fixed constant $r'_f$ and some $k'$, $\vec{X}'_f/r'_f = \vec{X}'_{k'}/r'_{k'}$ then as $\lambda$ is adjusted the miss-match in the fit will improve or worsen.  In fact, $\|f - \varphi\|{\ls}_{\text{I}} \rightarrow 0$ as $r'_k \rightarrow r'_f$ either from above or bellow and, to within allowed numerical accuracy, a perfect fit results when $r'_k = r'_f$.  A similar example obviously results when $f$ is defined by an appropriate surface distribution of points.  Observe that when expanded in terms of a spherical harmonic series $q_f/|\vec{X} - \vec{X}'_f|$ has spherical harmonics of all degrees and orders and that as $|\vec{X}'_f| \rightarrow 1$ the relative size of the higher degree and order terms increases.  In short, to properly fit a choppy $f$ requires many sources on the surface, but to fit a smooth $f$ requires fewer, but deeper sources.  In order to underscore and explore the phenomena of deep source placement here smooth test functions for $f$ are used.  While the above discussion may seem obvious after a little reflection, what is unexpected is the actual magnitude of the \cns\ that are encountered here in the presence of good fits. 

  Several questions naturally arise here:  What is a precise quantitative statement of the underlying phenomena, which has been noted in the MFS literature, that clearly links deep fundamental solution source placements, large \cns\ and smooth functions to be fit?  Moreover what are the actual limits, in both a practical and theoretical sense, of actual depth of point source placements here?  For the particular examples considered, the issue of deep source placement is explored to the numerical limits allowed by the hardware, software and operating system used in this study; however, it would seem quite likely that other operating environments allowing for more significant digits and thus deeper source placements might yield even better results for the test cases studied here.  [Here, it seems reasonable to speculate from the source statistics given in Table~\ref{Ta:CaseCondit}, that some sort of aggregate multipoles are being formed in the usual sense of the standard limiting process employed by physicist to define specific multipole types (for an electrostatic dipole with a specific orientation, this process consists of letting two point charges of equal and opposite sign approach each other along a specific direction, while simultaneously letting the magnitude of the charge approach infinity in such a way as to give a well defined dipole limit).  Following up on these possibilities is beyond the scope of the present article and it probably will not be fully addressed by the author in the future either.]

  One more surprising aspect associated with large \cns\ uncovered by this study is worth explicitly noting:  After the onset of numerical roundoff due to large \cns, the actual modeling error of the resulting fits degrades gracefully.  [This is undoubtedly due not only to fitting properties of the II norm, but also to properties of the Householder triangulation algorithm used in solving (\ref{E:LLSQ}) and it strongly hints at underlying numerical robustness.] 

 Obviously, to obtain reasonable results in the presence of large \cns\ attention must be paid to equation solving techniques employed and to computational word length(s) used.  Toward this end all computations were performed using a 128-bit word length (33-34 significant digits). (On many systems 128-bit words have fewer significant digits than this.)  The linear equations themselves were then solved using Householder triangulation.   The eigenvalues, and thus the \cns\ themselves were computed using singular value decomposition (SVD) software implemented with the same word length.  

  Next consider the overall strategy employed for placing the interpolation points.  Observe from the collocation minimum norm result for scalar point sources---Theorem~\ref{T:DIDACKSfit}---it is obvious for $\widehat{\Omega}_{\sps}$ that when the interpolation points $\vec{P}_k$ are placed on a 
sphere of radius $R_p < 1$ that, for well-behaved functions $f$, $\varphi$ will match $f$ on and over the interior of this sphere of radius $R_p$ if the interpolation points are sufficiently dense on this sphere.  As an aside, the properties of $f$ that are necessary to insure that it can be considered well behaved can be made fairly precise.  From subsequent considerations based on the Nyquist frequency, it is apparent that if $f$ is to be matched over some grid with a specified angular spacing, then it must have no frequency content above a certain limit (i.e., its spherical harmonic expansion must not have terms larger than some specified degree and order).  Conversely, if $f$ doesn't have such frequency limits, then from Theorem~\ref{T:HarmonicSurprise} in Appendix~A and DIDACKS theory examples displaying arbitrarily large modeling errors can be constructed.  Also by harmonic outward continuation it seems reasonable to expect that $\varphi$ and well behaved $f$'s will match to a certain extent over the region characterized by  $1 \geq r > R_p$.  In fact, contingent on a good match for the region $1 \geq r > R_p$, this interior spherical surface interpolation point placement strategy is perfectly reasonable in practice.  (At least for the functions tested here, this is indeed found to be the case.)  While this strategy will clearly be generally suboptimal in the sense that for any specific $f$ under consideration if, as discussed immediately below, a nonlinear least squares (NLLSQ) fit of $\varphi$ to $f$ is performed a better fit can almost always be found; there are many circumstances where a generalized LLSQ fit based on (\ref{E:LLSQ}) is not only adequate, but preferable.  Not only are these NLLSQ fits much more difficult to handle, but they introduce a number of unique pitfalls of their own due to the real possibility of large condition numbers and their ramifications, which include the implications of extreme nonlinearity.  One major point is that, when the spectral properties of $f$ are well known, an appropriate fixed grid can be selected and the resulting fitting properties of (\ref{E:LLSQ}) can be exhaustively analyzed so that a repeatable fitting algorithm can be produced, which is only possible for NLLSQ fits if the very greatest of care is taken.  (Observe that the counter examples discussed above based on $f = q_f/|\vec{X} - \vec{X}'_f|$ do not apply in cases where spectral limits are imposed since, for finite nonzero $|\vec{X}'_f|$ the form $1/|\vec{X} - \vec{X}'_f|$ has an infinite spherical harmonic expansion.)  A generalized LLSQ fit based on (\ref{E:LLSQ}) also has the following advantages:
\begin{itemize}
 \item
  A natural correspondence to spherical BVPs is set up.
 \item
  Interpolation depth ($R_p$) and angular spacing parameters can be easily understood and implemented.
 \item
 Regular samplings or griddings of point masses have a long history with a rich associated (geophysical) literature, which aside from elementary considerations is not required here, but that can be called on as needed. (A reference to this literature can be found in \cite{Ruf1}.)  An underlying assumption of this literature seems to be a completely regular gridding and, as such, the theory is unambitious for a plane, but the full implications for a spherical gridding are unclear.  The development here is completely independent of the ideas in this literature and, as such, it would be interesting to consider the of ideas of this literature within a DIDACKS context, but this can be said of many other research areas as well. 
 \item
 The types of functions $f$ that can be fit for a given angular spacing can be easily characterized and understood, so that an appropriate point source spacing can be chosen for any specified $f$, or if $f$ cannot be fit, then the reasons for this can be easily understood. 
 \end{itemize}

  A NLLSQ fit of $\varphi$ to $f$ entails iteratively adjusting the source locations to minimize $\Phi$, so that the resulting $\vec{P}_k$'s will end up on a spherical surface for only very specialized types of $f$'s.  A further consideration of NLLSQ fits, however, makes it obvious that there are no universally valid interpolation point placements, which are optimal for all $f$'s, since for any initial $\vec{P}_k$ values an $f$ can be found that will produce a lower value of $\Phi$ when the $\vec{P}_k$'s are changed.  To see that this is so, suppose that the $N_k$ fixed locations $\vec{P}_k$ have been chosen as a hypothesized good universal source placement candidate, so that they also serve as an initial NLLSQ guess.  Further suppose that $f$ itself is specified by $N_k$ sources, which are all at distinct locations form the chosen values for the $\vec{P}_k$.  In this case an adequate NLLSQ fit will obviously produce a considerably better value of $\varphi$ than a simple linear fit will. This might seem to argue for the necessity of performing NLLSQ fits and, in fact the spherical exterior formalism was developed by the author over 25 years ago expressly for the purpose of performing NLLSQ point mass fits to spherical harmonic representations of the Earth's gravity field \cite{Ruf1}.  Although these NLLSQ fits were, and are, of very high quality, the resulting NLLSQ problem itself is a very difficult one and it is perhaps best skipped whenever possible, unless the primary intent is to seek out a challenging problem (which is a worthy goal in-and-of itself).

One recurring problem encountered in practice is that of implementing the analog of a regular gridding on the surface of a sphere.  For example, severe angular distortion enters when a straight-forward evenly spaced angular grid is used, since the resulting sample points near the poles or $z$-axis are obviously closer than might normally be considered acceptable.  While these unwelcome variations obviously do not occur for gridding on a plane, they are an integral part of the geometry of the sphere and cannot generally easily be overcome.  The only way to completely overcome this problem is to place points at the vertices of some specified regular geometric solid figure (for example, at the vertexes of a cube), which fixes the number of interpolation points; however, this is not only very restrictive but more time consuming to implement for realistic general cases than it might at first appear. \{Other strategies exist.  One to consider might be that of placing a number of equal (repulsive) charges on a sphere and ascertaining the resulting configurations on the assumption that, in some sense or other, the resulting final equilibrium configuration will be that of a set of points that have the largest spacing possible.  The theory of minimum energy point configurations on a sphere has its own extensive literature (see \cite{Cohn} and references therein), but even given this pre-existing literature there are greater implementation complications than one might want to deal with and, moreover, it is unclear whether this is really a general suitable point selection strategy to adopt for those not already familiar with the relevant literature in this area.  A second strategy to consider might be one based on spherical $t$-designs---see for example, the first paragraph of \cite{Tdesign} and the citation given---but again it is unclear whether this is really a good strategy to adopt for someone who is not already familiar with the relevant literature.\}  For deep point source placements the issue of obtaining a more-or-less regular angular separation (i.e., one where the points are distributed somewhat uniformly over the entire surface of a sphere) is an important one.  Specifically, from an examination of the values of the elements of the matrix $\widetilde{\mathbf{T}}$ given by (\ref{E:Ttilde}), it is clear that as $\vec{P}_{k'}$ approaches $\vec{P}_k$,\ $\widetilde{T}_{k\,k'} \rightarrow 1$ (of course, $\widetilde{T}_{k\,k'} < 1$ must hold for all $\vec{P}_{k'} \neq \vec{P}_k$).  This means that if two of the sample points on the sphere are close together, then the resulting \cn\ will be large.  By direct experimentation using irregular spacing on a sphere of radius $R_p$ and comparing these results with results presented below, the reader can see for him or herself that somewhat regular spacing is very desirable.  The question, then, is how to do this.  An algorithm is introduced below in conjunction with Table~\ref{Ta:Pks} that makes a very good trade-off between ease of implementation and spacing regularity for the resulting grid.  Obviously the issue of point source spacing in the horizontal or tangential direction is the same for both the spherical exterior and interior problems, hence only the spherical interior setting will considered in the rest of this section.

 Having temporarily settled the gridding requirements for $\varphi$, next consider the issue of how to conveniently mathematically characterize the suitability of $f$, with regards to the overall angular separation of the spherical grid used.  First, as alluded to above, consider the elementary well-known aspects of sampling along a straight line or of a 2-dimensional gridding in the plane.  From the 1-dimensional theory of Fourier series and digital signal processing, the relevant concept for determining the appropriate spacing of an even grid of data is the concept of Nyquist frequency.  Also the corresponding 1-dimensional criteria for smoothness of $f$ is simply the highest allowed non-zero frequency term in a Fourier series expansion of $f$, which can be characterized by some integer $n$. The 2-dimensional Fourier series generalizations of this idea for a plane are obvious.   For the surface of a sphere, the relevant concept might                seem to be the maximum degree and order of a spherical harmonic expansion of $f$ since a spherical harmonic expansion is the direct spherical analog of a 2-dimensional Fourier series expansion for the plane.  It is clear, however, that $\phi$ and $\theta$ are treated very          differently in spherical harmonics since the undulations of say $P_n(\cos \theta)\cos m\phi$ correspond to very short distances at the poles for large $n$ and $m$ values; however, for small values of $n$ (and thus $m$) this angular distortion is not quite so marked (i.e, for $n$ and $m$ less than about $7$),.  Thus for cases where $n$ is small (or for cases where there is no $\phi$ dependence), one can attempt to directly carry over the concept of a Nyquist frequency from lines in $\mathbb{R}^1$ or planes in     $\mathbb{R}^2$ to spheres in $\mathbb{R}^3$ and compute a Nyquist frequency based on $\cos m\phi$, for the maximum $m$, (or on $\cos n\theta$ for the maximum $n$) that may serve as a reliable guide for the sampling of $f$ on a sphere [i.e., sampling of $f(R_p,\,\theta,\,\phi)$].  It is clear that this spherical Nyquist criteria, which is characterized by the minimum great circle over which $f$ varies appreciably, is the more meaningful one, rather than a direct characterization in terms of maximum allowed degree and order.  For low degrees and orders the reader can easily test out the utility of this criteria for him or herself as follows.  The idea behind this testing procedure is that it is a simple matter to distribute points evenly on an $\mathbb{R}^2$ unit disk and simple $\mathbb{R}^2$ harmonic test functions can be used to first see and then explicitly check out the role of the Nyquist frequency in this case; moreover, this simple testing algorithm need not be limited to $\mathbb{R}^2$ cases since any $f$ that is a function of only $x$ and $y$ and is harmonic in $\mathbb{R}^2$ is also harmonic in $\mathbb{R}^3$.  Looking slightly ahead, a more-or-less regularly spaced grid can be computed from the algorithm given in conjunction with Table~\ref{Ta:Pks}.  Suppose that such a semi-regular grid has been implemented, then consider the following simple harmonic functions in the interior of an $\mathbb{R}^2$ unit disk:
\begin{equation}\label{E:H1h2}
 H_{1\,M} \eq {\rho}^M \sin M\vartheta \ \ {\text{and}}\ \ 
 H_{2\,M} \eq {\rho}^M \cos M\vartheta 
\end{equation}
where $\rho = \sqrt{x^2 + y^2} < 1$, $\vartheta = \tan^{-1}y/x$ and $M$ is an integer.
Consider the natural extension of these functions to the interior of a unit sphere, where the  $\mathbb{R}^2$ $x$ and $y$-axes are mapped into the \riii\ $x/y$ plane.  In this case $\partial H_{1\,M}/\partial z = 0$ and $\partial H_{2\,M}/\partial z = 0$ so ${\nabla}^2H_{1\,M} = {\nabla}^2H_{2\,M} = 0$ for the interior of an  \riii\ unit sphere.  Since an assumed grid of $\vec{P}_k$ values specifies the form of $\varphi(\vec{X})$, it is a simple matter to fit to either $H_{1\,M}$ or $H_{2\,M}$ for various integer values of $M$.  Then, as the reader can then directly verify, the quality of such fits rapidly decline as the ``spherical Nyquist sampling criteria'' is violated.  It might be argued that this is due to the special orientation that $H_{1\,M}$ and $H_{2\,M}$ have with respect to the $z$-axis, but it is a simple matter to rotate the coordinate axes of either the sources or the  $H_{j\,M}$'s to show that this is not true.  To further test the validly of this ``spherical Nyquist sampling criteria'' the reader can use a full spherical harmonic expansion to some appropriate degree and order in place of the $H_{j\,M}$'s if he or she desires, but the results should be obvious.

\pagebreak
\begin{center}
\ \\
 \begin{large}\underline{\textbf{Specification of $f$ and $\varphi$}}\end{large}\\
\ \\
\end{center}
   
  Since a discussion of most of the side issues is complete, consider the specification, in turn, of the actual forms used for $f$ and $\varphi$ in testing. 
 Fits to two different basic reference functions were performed: $f = F_j$ for $j = 1, 2$, where 
\begin{itemize}
\item
$F_1 \eq \frac12(x^2 + y^2) - z^2$
\item
$F_2 \eq \frac32{r} P_1(\cos \theta) - \frac78{r}^3 P_3(\cos \theta) + \frac{11}{16}{r}^5 P_5(\cos \theta)$.
\end{itemize}
Here the $P_n$ are the usual Legendre polynomials [i.e., $P_1(x) = x$, $P_3 = (5x^3 - 3x)/2$ and $P_5(x) = (63x^5 - 70x^3 + 15x)/8$].  [For consistency with earlier notation, where $f_o$ was used to denote an evaluation at the origin, and for the reader's convenience, the symbol $F_j$ is used instead of $f_j$, since $f_j$ is reserved to mean $f_j \eq f(\vec{P}_j)$.]

 Both $F_j$'s were chosen because they are smooth functions and thus the assumed form for $\varphi$ can reasonably be expected to work well with a small number of point sources.  While the form of $F_1$ was chosen mainly for its simplicity, the motivation for the choice of the form of $F_2$ can also be easily understood.  Thus let $U_a(\vec{X})$ denote the potential function that is harmonic for $r < a$ and that is specified by the boundary condition $U_a(a,\,\theta,\,\phi) = 1$ for $\pi/2 >\theta \geq 0$ and by $U_a(a,\,\theta,\,\phi) = -1$ for $\pi > \theta \geq \pi/2$.  $F_2$ then corresponds to the first three terms of the spherical harmonic expansion of $U_a(\vec{X})$, with $a = 1$ \cite{Jackson}.   While $F_2$ is smooth at $r = 1$, it is does have a somewhat abrupt change in transitioning from $\theta < \pi/2$ to $\theta > \pi/2$.  Notice that when $a = 1$ the full expansion for $U_a$ is not harmonic on the surface of the unit sphere due to the discontinuity at $\theta = \pi/2$ and thus it is not an allowed function (since $|{\vec{X}}'_k| > 1$, only functions that are somewhat smooth on the unit sphere itself can be accurately modeled).  Furthermore, partials with respect to $z$ are not continuous at $z = 0$ and $x^2 + y^2 = 1$. 

  Next, consider the actual specification of $\varphi$.  The scalar point source basis function sets can be fixed by specifying the location of interpolation points (${\vec{P}}_k$) and whether a constant basis function term is present or not in the fitting form.  First consider only implementations without a constant term.  Further, as discussed above, suppose that the interpolation points form a grid of points on some sphere of radius $R_P < 1$.  Then, as previously noted, from the collocation property one might reasonably expect that $f$ itself will be accurately matched over the entire sphere of radius $R_P < 1$ from the uniqueness of Dirichlet boundary conditions; furthermore, one might expect that $f$ and $\varphi$ will match reasonably well over the part of the interior region specified by $R_P \geq r$, due to the matching properties of the II norm itself.  Thus, it is a reasonable hypothesis, pending numerical testing, to assume that $\varphi$ and $f$ will also match over all of the interior region, with the largest excursions occuring near the surface of the unit sphere.  On this theory the basic configuration tested will thus be that of a set of interpolation points positioned on a sphere of radius $R_P < 1$.  This is done not only for convenience, but also for conceptual clarity since there is otherwise no obvious strategy, at present, for efficiently placing point sources so as to produce matches to $f$'s that share common characteristics.  Also as noted above, while it is clearly desirable that this set of interpolation points be uniformly spaced over this sphere, this cannot be done in general exactly for an arbritrary number of points; moreover, it is a nontrivial matter to implement the particular cases where such a regular spacing is possible. To address this problem a very simple algorithm is given in the next paragraph that produces a sufficiently regular spacing of points.

\pagebreak
\begin{center}
\ \\
 \begin{large}\underline{\textbf{Specification of Interpolation Point Gridding}}\end{large}\\
\ \\
\end{center}

  The interpolation (and source) placement algorithm places points on various rings and is based on maintaining a minimum angular separation of points while simply adding as many points on each ring as can be added without violating this spacing constraint. Since these points on the sphere of radius $R_P$ are determined by a grid of $\theta$ and $\phi$ values, which are fixed, in turn, by an array of delta spacing values $\Delta \theta$ and $\Delta \phi$, it is convenient to define this grid of points on a unit sphere and then simply rescale them to the appropriate value of $R_P$, as needed. 

\vskip .2in
\begin{table} 
\begin{center}
\begin{tabular}{|c|c|c|c|c|}\hline
    Symbol  &  Input $N_{\theta}$ & $N_k$ &  Min. Ang. Sep.  &  Max. Min. Ang. Sep. \\ \hline\hline
   $C_{P_1}$&   8                 & 58    & 25.71 Deg.           & 27.96 Deg         \\ \hline
   $C_{P_2}$&   36                & 1542  & 5.14 Deg.           & 5.30 Deg           \\ \hline
\end{tabular}
\caption{Point Configurations, $C_{P_j}(R_P)$, for Sphere of Radius $R_P$}\label{Ta:Pks}
\end{center}
\end{table}

  First, the angular separation, ${\beta}_{k\,k'}$, of two points ${\vec{P}}_k$ and ${\vec{P}}_{k'}$ will prove to be a very useful concept and it is defined by 
\begin{equation}\label{E:AngSep}
 {\beta}_{k\,k'} \eq {\cos}^{-1} \left(\tfrac{{\vec{P}}_k \cdot {\vec{P}}_{k'}}{|{\vec{P}}_k|\,|{\vec{P}}_{k'}|}\right)\ .
\end{equation}
 Second, the algorithm to fix this grid of $\theta$'s and $\phi$'s is uniquely determined by the input number $N_{\theta}$, which specifies the number of rings parallel to the $x/y$-plane and  whose $z$ coordinate locations are fixed by the different values of $\theta$ (i.e., $z = \cos \theta$).  Let the indexes $n$ and $m$ be associated with the various $\theta$ and $\phi$ values respectively.  Then $\Delta \theta \eq \pi/(N_{\theta} - 1)$. The first ``ring'' ($n = 1$) consists of the single point $(0,\,0,\,1)^T$ along the positive $z$-axis and the last ring ($n = N_{\theta}$) consists of the single point $(0,\,0,\,-1)^T$ along the negative $z$-axis.  For $1 < n < N_{\theta}$ the ring locations along the $z$-axis are given by ${\theta}_n = (n - 1) \Delta \theta$.  For each value of $n$ such that  $1 < n < N_{\theta}$, there are associated $\phi$ values that fix the points on ring $n$.  For a given $n$, let $\Delta {\phi}_n$ denote the angular spacing for ring $n$.  For a given $\Delta {\phi}_n$ the values of $\phi$ along the ring will then be given by ${\phi}_{n\,m} \eq (m-1)\Delta {\phi}_n$, where $m = 1,\,2,\,3,\,\ldots,\,N_{\phi}$ and $\Delta {\phi}_n \eq 2\pi/N_{\phi}$.  This leaves only $N_{\phi}$ to be determined, which is the maximum integer that will ensure that all points on the ring obey the minim angular separation requirements.  To compute $N_{\phi}$ simply test out successive values until the inequality ${\beta}_{k\,k'} < \Delta \theta$ is violated for two typical adjacent points on the ring and then set $N_{\phi}(n)$ to the last value for which this inequality held.  Processing the various rings in sequence and then rescaling to the appropriate $R_P$, this algorithm then completely specifies the ${\vec{P}}_k$ by $(r,\,\theta,\,\phi) = (R_P,\,{\theta}_n,\,{\phi}_{n\,m})$. 

   For $N_{\theta} = 8$, this algorithm gives $N_k = 58$.  This configuration of points is labeled $C_{P_1}(R_P)$ and is completely specified by an input value of $R_P$.  For simplicity this configuration is the one that is used to define $\varphi$. For $N_{\theta} = 36$, this algorithm gives $1,542$ points.  This configuration of points is the one used for evaluating the quality of the resulting fits and is labeled $C_{P_2}(R_E)$.  It is completely specified by an input for the evaluation radius, $R_E$.  For any of these point configurations there is a minimum angular separation of distinct points and a maximum minimum angular separation of distinct points, which together give some indications of the uniformity of the distribution of points on the sphere.  The minimum angular separation is defined by the minium of ${\beta}_{k\,k'}$ for all $k$ and $k' \neq k$.  For each $k$, let ${\gamma}_k \eq$ minimum of ${\beta}_{k\,k'}$ for all $k' \neq k$. Then the maximum minimum angular separation is given by the maximum of ${\gamma}_k$ for all $k$.   These angular separations for $C_{P_1}(R_P)$ and $C_{P_2}(R_P)$ are given in Table \ref{Ta:Pks}.

\begin{center}
\ \\
 \begin{large}\underline{\textbf{Numerical Results and Discussion}}\end{large}\\
\ \\
\end{center}

  Interpolation results for fits to $F_1$ and $F_2$ using $C_{P_1}$ at different $R_P$'s for the configuration of $\varphi$ are presented in Table \ref{Ta:CaseResults}.  The first three columns of  
Table \ref{Ta:CaseResults} are self-explanatory.  The fourth column gives the value of $\varphi(\vec{0}) = \varphi(\vec{0}) - F_1(\vec{0}) = \varphi(\vec{0}) - F_2(\vec{0})$, since  $F_1(\vec{0}) = 0$ and  
$F_2(\vec{0}) = 0$.  The fifth and sixth columns then give statistical parameters for the fit  based on $C_{P_2}$ evaluated at $R_E = 1$.  Like results for $R_E = 1/2$ are then given in columns seven and eight.

  Additional information about these fits is given in Table \ref{Ta:CaseCondit}, where for convenience the first three columns are repeated.  In Table \ref{Ta:CaseCondit} the fourth column is the collocation check number, which is the maximum of $|f({\vec{P}}_k) - \varphi({\vec{P}}_k)|$ for all $k$, and gives some measure of numerical roundoff.  The fifth column is the condition number as previously defined.  The sixth and seventh columns gives the statistical parameters for the source strength ($q_k$), while the eighth and ninth columns give the fit error parameters evaluated at $R_E = 1/4$.

\vskip .02in
\begin{table} 
\begin{center}
\begin{tabular}{|c||c|c|c|c|c|c|c|}\hline
 &   &    &  &\multicolumn{2}{c|}{Fit Errors at $R_E = 1$} & \multicolumn{2}{c|}{Fit Errors at $R_E = 1/2$}\\\cline{5-8} 
 \raisebox{1.6ex}[0cm][0cm]{Case} & \raisebox{1.6ex}[0cm][0cm]{$f$} & \raisebox{1.6ex}[0cm][0cm]{$R_{P}$} & \raisebox{1.6ex}[0cm][0cm]{$\varphi$ at}  & Standard & Maximum & Standard & Maximum \\ 
 \raisebox{2.6ex}[0cm][0cm]{Number}& \raisebox{2.6ex}[0cm][0cm]{Choice}& \raisebox{2.6ex}[0cm][0cm]{(${P}_k$)} &\raisebox{2.6ex}[0cm][0cm]{Origin} & \raisebox{1.4ex}[0cm][0cm]{Deviation} & \raisebox{1.4ex}[0cm][0cm]{Magnitude} & \raisebox{1.4ex}[0cm][0cm]{Deviation} & \raisebox{1.4ex}[0cm][0cm]{Magnitude} \\ \hline\hline
 1      & $F_1$ & 3/4    & .755$\times 10^{-4}$  & .100$\times 10^{-1}$ & .314$\times 10^{-1}$   &.270$\times 10^{-3}$    & .496$\times 10^{-3}$ \\ \hline
 2      & $F_1$ & 1/2    & .100$\times 10^{-6}$  & .362$\times 10^{-3}$ & .103$\times 10^{-2}$ &.132$\times 10^{-5}$    & .293$\times 10^{-5}$ \\ \hline
 3      & $F_1$ & 1/10   & .244$\times 10^{-17}$ & .983$\times 10^{-8}$ & .407$\times 10^{-7}$ &.374$\times 10^{-10}$   & .145$\times 10^{-9}$ \\ \hline
 4      & $F_1$ & $\frac1{100}$ & 0.0$\times 10^{0}$  & .947$\times 10^{-14}$ & .387$\times 10^{-13}$ 
& .370$\times 10^{-16}$  & .151$\times 10^{-15}$  \\ \hline
 5      & $F_1$ & $\frac9{1000}$ & .197$\times 10^{-29}$  & .511$\times 10^{-14}$ & .208$\times 10^{-13}$ & .200$\times 10^{-16}$ &  .826$\times 10^{-16}$  \\ \hline\hline
 6      & $F_2$ & 3/4  & -.389$\times 10^{-30}$ & .519$\times 10^{-1}$ & .131$\times 10^{0}$ &.827$\times 10^{-3}$   & .127$\times 10^{-2}$  \\ \hline
 7      & $F_2$ & 1/2  & -.277$\times 10^{-31}$   & .931$\times 10^{-2}$ & .371$\times 10^{-1}$ & .304$\times 10^{-4}$ & .587$\times 10^{-4}$ \\ \hline
 8      & $F_2$ & 1/10 & .473$\times 10^{-29}$ & .135$\times 10^{-4}$ & .593$\times 10^{-4}$ &.265$\times 10^{-7}$  & .114$\times 10^{-6}$  \\ \hline
 9      & $F_2$ & $\frac1{100}$ &  0.0$\times 10^{0}$ & .137$\times 10^{-8}$ & .589$\times 10^{-8}$ & .277$\times 10^{-11}$  & .114$\times 10^{-10}$  \\ \hline
 10     & $F_2$ & $\frac9{1000}$ & .290$\times 10^{-23}$  & .109$\times 10^{-8}$ & .427$\times 10^{-8}$ & .311$\times 10^{-11}$  & .107$\times 10^{-10}$  \\ \hline
\end{tabular}
\caption{Fit Results to $f = F_1$ or $F_2$ with 58 Interpolation Points}\label{Ta:CaseResults}
\end{center}
\end{table}

\vskip .01in

\begin{table} 
\begin{center}
\begin{tabular}{|c||c|c|c|c|c|c|c|c|}\hline
 &   &  &   &  &\multicolumn{2}{c|}{Source ($q_k$) Statistics} & \multicolumn{2}{c|}{Fit Errors at $R_E = 1/4$}\\\cline{6-9} 
 \raisebox{1.6ex}[0cm][0cm]{Case} & \raisebox{1.6ex}[0cm][0cm]{$f$} & $R_{P}$ & \raisebox{1.6ex}[0cm][0cm]{Coll.} & \cn  & &  Standard & Standard & Maximum \\ 
 \raisebox{2.6ex}[0cm][0cm]{\#}& \raisebox{2.6ex}[0cm][0cm]{Fit}& &\raisebox{2.6ex}[0cm][0cm]{Check}  &  & \raisebox{2.4ex}[0cm][0cm]{Mean} & \raisebox{1.4ex}[0cm][0cm]{Deviation} & \raisebox{1.4ex}[0cm][0cm]{Deviation} & \raisebox{1.4ex}[0cm][0cm]{Magnitude} \\ \hline\hline
 1      & $F_1$ & 3/4  & .16$\times 10^{-32}$ & .66$\times 10^{3}$  & .17$\times 10^{-5}$ & .911$\times 10^{-1}$ & .983$\times 10^{-4}$  & .209$\times 10^{-3}$  \\ \hline
 2      & $F_1$ & 1/2  & .13$\times 10^{-32}$ & .49$\times 10^{6}$  & .35$\times 10^{-8}$ & .310$\times 10^{0}$ & .125$\times 10^{-6}$  & .235$\times 10^{-6}$  \\ \hline
 3      & $F_1$ & 1/10 & .71$\times 10^{-31}$ & .80$\times 10^{17}$& .42$\times 10^{-18}$ & .388$\times 10^{2}$ & .149$\times 10^{-12}$   & .469$\times 10^{-12}$  \\ \hline
 4      & $F_1$ & $\frac1{100}$ &.68$\times 10^{-29}$ & .80$\times 10^{33}$ & .98$\times 10^{-30}$ & .388$\times 10^{5}$ & .145$\times 10^{-18}$   &.592$\times 10^{-18}$  \\ \hline
 5      & $F_1$ & $\frac9{1000}$ & .14$\times 10^{-28}$ & .43$\times 10^{34}$  & .28$\times 10^{-29}$ & .532$\times 10^{5}$ & .781$\times 10^{-19}$ & .333$\times 10^{-18}$  \\ \hline\hline
 6      & $F_2$ & 3/4  & .18$\times 10^{-32}$ & .66$\times 10^{3}$ & -.90$\times 10^{-32}$ & .253$\times 10^0$ & .107$\times 10^{-3}$  & .270$\times 10^{-3}$ \\ \hline
 7      & $F_2$ & 1/2  & .43$\times 10^{-31}$  & .49$\times 10^{6}$ & -.10$\times 10^{-32}$ & .267$\times 10^{1}$ & .170$\times 10^{-5}$  & .330$\times 10^{-5}$ \\ \hline
 8      & $F_2$ & 1/10 & .28$\times 10^{-28}$  & .80$\times 10^{17}$ & .87$\times 10^{-30}$ & .402$\times 10^{5}$ & .519$\times 10^{-10}$ & .204$\times 10^{-9}$ \\ \hline
 9      & $F_2$ & $\frac1{100}$ &  .15$\times 10^{-23}$& .80$\times 10^{33}$ & .45$\times 10^{-24}$ & .402$\times 10^{11}$ & .614$\times 10^{-14}$  & .219$\times 10^{-13}$  \\ \hline
 10     & $F_2$ & $\frac9{1000}$ & .10$\times 10^{-22}$& .43$\times 10^{34}$ & .37$\times 10^{-23}$ & .756$\times 10^{11}$ & .120$\times 10^{-13}$ & .404$\times 10^{-13}$ \\ \hline
\end{tabular}
\caption{Ill-conditioning Indicators for Table \ref{Ta:CaseResults} Fits}\label{Ta:CaseCondit}
\end{center}
\end{table}

   There are several significant observations that can be immediately made from Tables \ref{Ta:CaseResults} and \ref{Ta:CaseCondit}, which tie into previous discussions:
\renewcommand{\theenumi}{\alph{enumi}}  
\renewcommand{\labelenumi}{(\theenumi)} 
\begin{enumerate}
  \item
    Placing the interpolation points on a sphere of radius ($R_P < 1$) rather than distributing them throughout the whole interior of the unit sphere seems to work well for the tested functions. 
  \item
Both $F_1$ and $F_2$ are very smooth functions.  Small choices of $R_P$, large $C_{\#}$'s and accurate fits occur together for these test functions.  As previously discussed, while small values of $R_P$ and large $C_{\#}$'s are unavoidably linked for smooth functions, for irregular functions better results generally occur for larger values of $R_P$ and more interpolation points.
  \item
    From the fourth column of Table \ref{Ta:CaseCondit} it is apparent that all of the fits satisfy the collocation condition, but as one might expect a small degradation occurs for large condition numbers due to the onset of numerical roundoff.
  \item
   $C_{\#}$'s are sufficiently large in Table \ref{Ta:CaseCondit} that the issue of finding ways to naturally reduce them arises.  One way will be considered below, but other ways will not be addressed until subsequent articles.
  \item
 As noted earlier, and as indicated in Table \ref{Ta:CaseCondit}, large condition numbers are associated with large excursion in the source parameters.
  \item
   Since the results of Table \ref{Ta:CaseResults} indicate that the approach works well for smooth functions, this suggests the strategy of residual fitting: i.e., first fitting the smooth part of a harmonic function and then removing this part and then performing a fit the remaining part, etc.  The final result can be reconstructed as a sum of these two (or more) parts. (As just noted, these and other similar strategies will be addressed in another venue.)
\end{enumerate}
Points related to the above items dealing with large \cns, as well as to the ideas discussed earlier, can be found in, for example, the pair of articles \cite{MFS4} and \cite{MFS7}.

  To bring other points to light, additional testing was done based on the results of Tables \ref{Ta:CaseResults} and \ref{Ta:CaseCondit} (most of these tests are modifications to the basic configuration of test Case 3).  First, to see how an additional sphere of point sources would effect the results, the configuration $C_{P_1}(1/2) \cup C_{P_1}(1/10)$ was run and, as one might expect, resulted in a slight improvement over the results of Case 3:  A standard deviation of .202$\times 10^{-9}$ resulted, but with a larger associated $C_{\#}$ of .43$\times 10^{22}$.

  Second, there are three ways to account for the occurrence of constant terms in $f$:
\renewcommand{\theenumi}{\Alph{enumi}}  
\renewcommand{\labelenumi}{(\theenumi)} 
    \begin{enumerate}
    \item
      Use an unmodified basis function configuration to see if constant terms are naturally handled properly. 
    \item
      Add a constant basis function directly to $\varphi$.
     \item
      Just subtract off the constant term from $f$ and then add it back on.
    \item
      Add a point source term near the origin. 
    \end{enumerate}
   To test out the how fits work when a constant bias is added to $f$, the following fits were done using approaches (A) and (B), which both seemed to work well:
 \begin{itemize}
    \item
     Case 3 results were replicated with $F_1$ replaced by $F_3 \eq F_1 + 1$. The quality of the fit was unchanged, which indicates that constant terms can be naturally handled if the point sources are reasonably positioned.
    \item
   The Case 3 fit to $F_3$ was rerun with a constant basis function added to $\varphi$.  An improvement in the quality of the fit resulted (with a standard deviation of .304$\times 10^{-8}$ at $R_E = 1$) since the added basis function implies more fitting degrees of freedom; however, the $C_{\#}$ was somewhat larger (.11$\times 10^{22}$).
 \end{itemize}
Approaches (C) and (D) were not tested.

  Third, to test out the effects of normalization of basis functions, the following experiments were run:
  \begin{itemize}
    \item
    The results of Tables \ref{Ta:CaseResults} and \ref{Ta:CaseCondit} were rerun using normalized basis functions and the condition numbers were unchanged.  Furthermore, aside from expected numerical changes (in, for example, columns labeled ``$\varphi$ at the origin'', ``Coll. Check'' and the Mean value of $q_k$) the results were almost identical.
    \item
    The configuration $C_{P_1}(1/2)\,\cup C_{P_1}(1/10)$ mentioned above was rerun with normalized basis functions and had an improved $C_{\#}$ (of .29$\times 10^{21}$).
    \item
    A fit using the $\varphi$ specified  by $C_{P_1}(1/10)$, with a constant term added yielded a larger $C_{\#}$ (of .15$\times 10^{23}$).  This indicates that normalization does not necessarily always improve the $C_{\#}$; however, as previously noted it is definitely called for when mixed dipole and point source fits are performed due to the disparate nature of the physical scaling involved.
 \end{itemize}

   A final word of caution is in order here.  Although the quality of the match to the sampled values of a smooth function may be very good in the presence of large \cns, small deviations due to either a rougher function than expected or to the existence of measurement sampling errors can be of special concern.  In particular, it is obvious that the propagation of errors always deserves special attention when large \cns\ are present.  As the reader can verify for him or herself directly by perturbing individual components of the $\mathbf{A}$ vector, any miss-match here, even a very small one, can result in an extremely poor quality fit and, in fact, a small error at some interpolation point(s) may well result in a fit error at certain field points $\vec{X}$ that is much larger than the value of $|f(\vec{X})|$ itself, even though the fit still matches at all of the interpolation points:  That is, in the presence of large \cns, if measurement errors are present then $|\varphi(\vec{X})| >> |f(\vec{X})|$ for some values $\vec{X} \neq \vec{P}_k$ may well occur.   One obvious strategy is to increase the number of interpolation points and place them closer to the surface of the sphere.  There are various other strategies and implementation procedures that can be considered, but as mentioned in Section~\ref{S:intro}, errors and ways to treat them are not considered in the present article. Finally, as discussed above, if $f$ is unsuitable or if the sampling points are inappropriately located $|\varphi(\vec{X})| >> |f(\vec{X})|$ for some values $\vec{X} \neq \vec{P}_k$ may also happen due to other causes and this is one of the main reasons that application of the text approach have been limited here to modeling problems where these possibilities can be explicitly checked for by means of direct comparisons.

\section{DIDACKS MFS/BVP Connections}\label{S:MFS}

At the end of Section~\ref{S:intro} it was pointed out that the technique presented here can be profitably used in conjunction with MFS BVP approaches for certain problems, thus the use of the current method in conjuction with BVPs, and especially the MFS approach, is singled out for further discussion in this section.  Only a preliminary analysis of DIDACKS connections to MFS approaches will be undertaken here.

  Prior to considering MFS approaches themselves, consider how most of the numerical test cases given in Section~\ref{S:Tests} can be reinterpreted as solutions to spherical BVPs for some interior region of $\widehat{\Omega}_{[\sps]}$.  First recall that the general source basis function placement strategy used in Section~\ref{S:Tests} was to place sources on an exterior sphere of radius $1/R_P > 1$, which corresponds by (\ref{E:PkXk}) to placing interpolation points on a sphere of radius $R_P < 1$.  From Theorem~\ref{T:DIDACKSfit},
 since the fitting function $\varphi$ matches the function to be fit $f$ at the interpolation points, when $f$ satisfies an appropriate spherical Nyquist criteria, as discussed in  Section~\ref{S:Tests}, $\varphi$ and $f$ match over the sphere of radius $R_P$ and  $\varphi$ is a solution to the Dirichlet BVP for $r < R_P$.  Case~2 and 7 of Table~\ref{Ta:CaseResults} correspond to solutions of Dirichlet BVPs for the interior of a sphere of radius $R_P = 1/2$, where $\varphi$ matches $F_1$ over $r = R_P$ for Case~2 and $\varphi$ matches $F_2$ over $r = R_P$ for Case~7.  The last two columns of Table~\ref{Ta:CaseResults} give the corresponding mismatch over the boundary of this sphere of radius $r = R_P$, while the last two columns of Table~\ref{Ta:CaseCondit} give the corresponding mismatch over an interior sphere of radius $r = R_P/2$.  From the discussions in Section~\ref{S:Analysis} and Appendix~A these BVP solutions also satisfy the following two norm minimization criteria:
\begin{itemize}
\item[(A)]
  \ $\|f - \varphi\|{\ls}_{\text{II}}$ is minimized. 
\item[(B)]
 \ $\|\varphi\|{\ls}_{\text{II}}$ is minimized.
\end{itemize}
While these norm minimization conditions hold for the whole of $\widehat{\Omega}_{[\sps]}$ and not just that part inside the sphere of radius $R_P$, they still are still clearly very significant properties of the BVP solution $\varphi$.  While it is possible to analyze the various mathematical implications of (A) and (B) and then derive related inequalities for the subregion $r < R_P$; for our purposes here a direct analysis of (A) and (B) will have to suffice.  Thus notice, for example, that statement (B) makes a fairly strong assertion about the harmonic extension of $\varphi$ from the region $r < R_P$ to the whole of region $\widehat{\Omega}_{[\sps]}$.  (For a further discussion of the properties of harmonic extrapolation see Appendix~B.)  Finally, it is obvious that if $\varphi$ matches $f$ over the whole of $\widehat{\Omega}_{[\sps]}$ then it must match over the subregion $r < R_P$ also:  In fact, from the properties of harmonic functions, since the largest excursion in $f - \varphi$ must occur at $r = 1$ one would naturally expect a much better fit over the subregion $r < R_P$ than over the subregion $R_P < r < 1$, as seen in Tables~\ref{Ta:CaseResults} and \ref{Ta:CaseCondit} for the two cases noted above.  Third and finally, as indicated in Table~\ref{Ta:CaseCondit}
 the \cns\ for most of the cases dealt with here are much larger than those normally encountered in MFS applications where, as noted below, many more sample points are commonly used.

 Several generalizations of the basic BVP approach just outlined are obviously possible.  First, as mentioned  before, it is a trivial matter to rescale the results in order to handle interior problems for spheres with an arbitrarily specified radius.  Second, since the various partials of $f$ correspond to higher multipole fits it is easy to consider BVPs for interior spheres that have other specified boundary information than potential.  Third, as previously mentioned, corresponding approaches to BVPs over unbounded domains can be based on the I norm.  Fourth, the above ideas, including the discussion of norm criteria, apply to other DIDACKS settings, such as to the interior and exterior of units disks in the real and complex plane \cite{DIDACKSI}.  Fifth, since DIDACKS theory does not otherwise constrain interpolation point locations, it is quite possible to consider a like approach to BVP problems with more general shaped boundaries.  This last topic will be taken up below.

  Next specifically consider how the MFS approach relates to the basic BVP technique just outlined.  In what follows, it is assumed that the reader has at least some familiarity with MFS approaches---the reader who is not familiar with MFS approaches can consult any of the outlines that are normally contained in the front of current articles dealing with MFS techniques, such as \cite{MFS5}, \cite{MFS2}, \cite{MFS3}, \cite{MFS4} or \cite{MFS6} (all these references also give additional references).  Bogomolny \cite{MFS1} is a frequently referenced article that contains probably the first mention of the confluence of good solutions and large \cns\ for MFS-like approaches.  One normal way that MFS is applied to the interior or exterior of a sphere where Laplace's equation holds and Dirichlet boundary conditions are specified at some radius, say $r = R_P$, is by simply postulating a linear combination of the form $\varphi \eq \sum^{N_k}_{k=1}q_k/\ell_k$ and then determining the associated source strengths $q_k \in \mRR$ from a forced collocation point matching condition analogous to (\ref{E:Pcoll}), which is a strategy called \emph{direct collocation} by Saavedra and Power \cite{MFS3} and that is discussed at length in Appendix~A in related contexts:
\begin{equation}\label{E:MFScoll}
\varphi({\vec{P}}_k) = f({\vec{P}}_k)\ \ \text{for}\ \ k = 1,\,2,\,3,\,\ldots,\,N_k\,\,.
\end{equation}
Here the interpolation points are on the surface $r = R_P$ so $P_k = R_P$ and the source points are located on some other spherical surface, say $r = R_{X'}$, so that  $r'_k \eq |\vec{X}'_k| = R_{X'}$.  For $\mathbb{R}^3$ MFS problems with this geometry, it is usually further assumed that ${\vec{P}}_k/P_k = {\vec{X}}'_k/r'_k$, which means that for any $R_P$ and $R_{X'}$ when the problem is rescaled in an appropriate way (\ref{E:PkXk}) can be assumed to hold.  When this geometry holds then the fit corresponding to (\ref{E:MFScoll}) is a DIDACKS scalar point source fit that, as discussed above, is a solution to the associated Laplacian BVP.  For this geometry, given this direct equivalence of MFS and DIDACKS approaches it follows immediately that the minimum norm conditions (A) and (B) outlined above apply and that they have the indicated interpretation.  Furthermore, other conclusions can be drawn from discussions of DIDACKS theory given previously, such as the explanation for the confluence of large \cns\ and well behaved solutions.  Finally, even after taking the implied correspondence between DIDACKS fits and MFS fits into account, observe that in the MFS literature there are several differences between the way fundamental solution basis function fits are commonly implemented and the way they were implemented in Section~\ref{S:Tests}.  First, many more fundamental solution basis function are generally used than indicated by the configuration $C_{P_1}(R_P)$ of Table~\ref{Ta:Pks}.  Second, the source locations indicated in Table~\ref{Ta:CaseResults} are generally much deeper than is normal in the MFS literature.  For example, for spherical interior problems \cite{MFS3} cites a general rule-of-thumb that the magnitude of the source point vectors should be five times the magnitude of the interpolation point vectors or $r'_k = 5P_k$.  From (\ref{E:PkXk}) this means 
that $P^2_k = 5$ or that $R_P = \sqrt{5} \approx .45$, which corresponds closest to Cases~2 and 7 in Table~\ref{Ta:CaseResults}.  

  In the literature MFS approaches to $\mathbb{C}$ or $\mathbb{R}^2$ unit disks are also common.  For  complex or real $\mathbb{R}^2$ unit disk geometries the same conclusions can be drawn, but complications may enter since sometimes the interpolation points are offset by a certain angular amount from the direction implied by the corresponding source vector so that the relationship for the plane corresponding to (\ref{E:PkXk}) fails to hold.  Besides the simple direct collocation approach outlined by (\ref{E:MFScoll}) and the complication just mentioned, there are two other common ways that MFS techniques are applied.  In thee cases, it is still often possible to draw significant implications from DIDACKS theory about the associated MFS solutions; however, these discussions are outside the scope of the present article.  There is one area that is of more immediate concern---namely the question of how general the boundary surface can be with regards to DIDACKS approaches.  Notice that even when the interpolation points are placed on a different shaped interior surface in $\widehat{\Omega}_{[\sps]}$, norm conditions (A) and (B) still hold and while they still have considerable general significance their physical meaning becomes less clear.

The possibility of applying DIDACKS theory to general bounded domains is easy to understand. 
  Consider a (possibly dense) set of interpolation points specified on some closed star-shaped surface $\Sigma$ (c.f, Definition B.1), which lies solely within some spherical domain.  For a suitable given spherical domain, DIDACKS theory then prescribes the locations and strengths of a set of point sources that will replicate the specified information on $\Sigma$.  The sources are located on some corresponding surface ${\Sigma}'$, which is assumed to be wholly in the exterior of the given sphere.  Clearly if the interpolation points are properly chosen then the DIDACKS fit yields a function that is harmonic inside the region bounded by $\Sigma$ and that matches the boundary value information specified on it.  Since the underlying fitting process reconstructs the BVP as a Poissonian inverse source problem and these particular inverse source problems are notoriously ill-conditioned, some restrictions on the shape of $\Sigma$ and on the locations of the interpolations points, which are on $\Sigma$, are clearly called for.  For example, for a given choice of sphere due to the fact that the source locations themselves are directly specified by this BVP information, the resulting source locations may be entirely inappropriate.  This clearly opens up a number of issues related to both the applications and theory of (\ref{E:LLSQ}).  

  Additional issues also merit further examination here.  For purposes of discussion, consider only the exterior problem for unbounded regions, which have a single bounding interior surface.  First, even for BVPs with fairly irregular geometries there may be no real problem if the underlying field function to be matched displays very benign behavior and the source locations are appropriately chosen---that is, if various interpolation points on the boundary surface are naturally related to the given (benign) attenuation properties of the harmonic function that is to be reconstructed.  (It is worth emphasizing that even in this situation the choice of implicit sphere used for positioning the sources must be made fairly carefully.)  For optimal solutions, some problems require both fairly irregular shallow point source placement as well as deep source placement.  For such problems, while it may be that the natural interpolation points fall on an appropriate prescribed BVP surface that has suitable boundary conditions, this clearly would be very much the exception rather than the rule.  As indicated in Appendix~A, when the properties of $f$ are unsuitable and the sources are badly placed the resulting error excursions may even be many times larger than the maximum of $|f|$ itself in much of $\Omega$.

  Second, for nearly spherical geometries the problems encountered may be minimal, so fairly direct schemes can be used to overcome any residual problems.  In geophysics, for example, there are fairly standard techniques for addressing gravitational problems that can be adapted to some other scenarios \cite{PGII} when the geometries are not to irregular.  The techniques used there include always subtracting off a standard ellipsoidal reference model from the total field to obtain what is called gravity anomaly (or a related quantity called gravity disturbance).  Often a further subtraction, which is called remove-and-restore, is carried out that results in a localized field.  At several other places in this article the advantages of such residual fitting techniques for DIDACKS applications have been pointed out.  General issues associated with sources near or slightly outside the harmonic region of interest are also common to physical geodesy; for example, over land where mountains protrude above the reference ellipsoid, reduction techniques are employed to obtain equivalent sources that do not extend above the ellipsoid---for additional points see \cite{PGII,Moritz}.

 Third, and finally, there are certain geometric consequences of the DIDACKS relationship (\ref{E:PkXk}) between field and source points that must be considered.  For the sake of argument, consider an exterior \riii\ harmonic BVP problem for some surface $\Sigma$ that has been positioned and scaled so that a unit sphere that is constrained to be inside $\Sigma$ is a reasonable candidate choice for the source region.  Suppose that the sample point $\vec{P} \in \Sigma \subset \mriii$ is located at some peak of $\Sigma$, then from (\ref{E:PkXk}) the corresponding source placement will not be closer to the surface as one might expect, but further away and, in fact, deeper than the other source points.  This geometric effect is labeled the reverse-dimple effect.  For nominal DIDACKS fits this reverse-dimple effect is quite natural since sample points that are further away will be markedly attenuated and thus deep sources are much more appropriate; however, for BVPs, or more specifically BEM or other source coating techniques, one would nominally expect to find sources at and around excursion points like $\vec{P}$.  Hence, the reverse-dimple effect poses a conceptual problem in adapting DIDACKS theory to these BVP approaches for non-spherical geometries.   While there may be various ways of overcoming these serious conceptual issues for non-spherical problems, any real progress in this area is outside the scope of the present work and must await the efforts of others.  (Keep in mind, however, in the end that the underlying DIDACKS formalism makes no restrictions on the locations of sample or source points, aside from the obvious requirement that the be in the correct region.  Also, in passing, notice that in the appendices $\vec{P}$ will not take on the special meaning that has been assigned to it in this section.)

\newpage

\appendix
\begin{center}
 \begin{Large}{\textbf{Appendix A}}\end{Large}
\end{center}

\renewcommand{\theequation}{A-{\arabic{equation}}}
\setcounter{equation}{0}

\section*{\hfil The Collocation Minimum Norm Property\hfil}\label{S:A}
\ \hfill  \\
\vskip -.2in 

\noindent
  In Section~\ref{S:Analysis}, the collocation minimum norm property for scalar point sources over $\widehat{\Omega}_{[\sps]}$ equipped with the II norm was stated by (\ref{E:Pcoll}) and in a slightly more general way by Theorem~\ref{T:DIDACKSfit}.  This appendix entertains the possibilities of various other generalizations.  It also discusses the specific ramifications for RKHS based fits.  Much of the exposition here has a work-in-progress feel to it and the bulk of this appendix is not aimed at the general reader, but rather to those either engaged in or contemplating research in the areas of RKHS theory, Bergman kernel theory, general kernel theory or in the areas dealing with the interaction of general collocation kernel theory and operator theory.

  For convenience this appendix is largely independent of Section~\ref{S:Analysis}, so a restatement of (\ref{E:Pcoll}) for a slightly generalized setting is given first.  Thus, here the minimum norm property means that of all admissible functions, $g$ such that
\begin{equation}\label{E:APcoll}
g({\vec{P}}_k) = C_k\ \ \text{for}\ \ k = 1,\,2,\,3,\,\ldots,\,N_k\,\,,
\end{equation}
where the $C_k$ are arbritrary constants, a DIDACKS scalar point source fit with coefficients $q_k$ specified by (\ref{E:LLSQ}) or a proper RKHS fit is the function of smallest norm: $\|g\| \geq \|\varphi\|$.  The general plan of approach will be to explore the limits of what can be reasonably obtained in the way of possible generalizations, before giving formal results.  Thus (\ref{E:APcoll}) will be analyzed and restated in slightly different ways as the section progresses.  Although the resulting exposition may be somewhat circuitous, various subtleties will emerge as the section unfolds and the reader should finally end up with a better overall understanding of the relevant issues involved than might have resulted from a more direct exposition.

    The corresponding minimum covariance norm result for various norms over $\widehat{\Omega}_{\sps}$ is a cornerstone of GC \cite{Moritz}, which as previously noted, is often called linear least squares collocation.  Related results are known to hold for other reproducing kernel contexts, such as analytic function theory and functional analysis.  For a general RKHS setting this minimum norm condition can be stated as: Of all those functions that match some specified function ($f$) at $N_k$ points, a linear combination of $N_k$ (symmetric) reproducing kernels is the one that has the smallest norm.  This will be labeled the RKHS minimum norm property (this section presupposes a rudimentary understanding of RKHS theory---see, for example, \cite{Aronszajn,Mate,SandW}.)  There are several relevant issues that need to be considered in association with the conditions required for this RKHS minimum norm property to hold, which are not explicitly addressed in standard references.  Furthermore, not only is there no existing reference known to the author that discusses these subtleties, but many references seem to be, at best, slightly misleading.   Some additional attention and special care is needed here since this result must be extended to DIDACKS theory.  

 This appendix is fairly long since a somewhat thorough, but informal, analysis is given.  Prior to Theorem~\ref{T:CollRep} most of the discussion is an effort to establish a proper framework and to determine the appropriate limits of its generality.  In the end, the easily codified results are directly linked to the definitions of properties of the class of admissible functions and of the class of operators that are allowed to act on them.  With regards to relevant additional trait that any allowed family of admissible functions is assumed to satisfy, while other characterizations of it are considered prior to introducing the idea of pointwise independence, this is the trait that is deemed most appropriate in the end due to the various considerations brought up.   With regards to the relevant characterization of linear differential operators, due to other types of considerations, a slight generalization of the one considered in Section~\ref{S:Analysis} is finally settled on and introduced by Definition A.4.  These two concepts immediately yield Theorem~\ref{T:Fk}, although it is, in reality, only a slight generalization of the results of Section~\ref{S:Analysis}.   On first reading the reader may wish to skip directly to the \textbf{\underline{Formal Result}} given at the end of the appendix and in so doing the exposition gains a certain measure of precision, cohesion and closure, but the reader will do this at the expense of a deeper overall understanding and knowledge base.  The intervening material here before this last part of the appendix is not especially well polished or organized and includes some topics, such as geographic factors, that are entertained primarily to simply gain a different perspective on existing material.  Finally, what is labeled the collocation minimum norm property or simply the minimum norm property might just as well have been labeled the RKHS minimum norm property except for the fact that a more generalized setting is addressed here.

\begin{center}
\ \\
 \begin{large}\underline{\textbf{Preliminary Comments}}\end{large}\\
\ \\
\end{center}

   As previously noted in Section~\ref{S:Analysis}, there are essentially two ways to approach issues associated with minimum norm properties:  Type I proofs, which are based primarily on various properties of the fit and its norm $\|\varphi\|$.  Type II proofs, which are based primarily on the properties of the function to be fit, $f$, as well as its associated function space.  The former is the one taken by proponents of GC, while the second is the one frequently taken by mathematicians.  The informal analysis undertaken here starts by examining Type I approaches in detail before finally considering Type II approaches, which in turn leads back to considerations associated with Type I approaches.  As noted below, there are issues with both types of approaches.  Although Type II approaches seem to be safer, one historical problem with  Type II approaches is that results are generally associated with a functional analysis setting, where it is obvious that each point of $f$ can be varied in dependently and this attribute is taken for granted.  This may not be true even when very natural restrictions are implied by the nature of the family of admissible functions.  For both types of approaches there seems to be a need for a commonly unstated requirement that $N_k < N_f$, where $N_f$ is the dimension of the span of the family of admissible functions.  Otherwise, for example, if $N_k = N_f$ a trivial and uninteresting situation results where there is at most only one function that satisfies (\ref{E:APcoll}), so that $\varphi = f$ and thus $\|\varphi - f\| = 0$ and $\|\varphi\| = \|f\|$.
 
  For Type I approaches various other issues are involved.
  Hille seems to be fully aware of all the relevant issues involved and gives a clear and rigorous proof of this RKHS minimum norm property for a somewhat limited case (i.e., for $N_k = 1$ only in the one dimensional complex setting) \cite[p. 333]{Hille}; however, as noted above,
there appears to be no reference that discusses the relevant subtleties involved and simultaneously gives a clear, concise and rigorous discussion for general cases.  The authors that are mindful of the issues involved generally seem to give either somewhat limited or specialized piecemeal discussions.  (Furthermore, even for the restricted RKHS setting, a discussion of the collocation minimum norm property often does not appear in the mainstream literature at places where one might expect to find mention of it, such as \cite{Aronszajn} and \cite{SandW}.)  Most, if not all, of the issues involved concern either the class of admissible functions or the invertibility of the associated Gram matrix [$\mathbf{T}$ in (\ref{E:LLSQ})] or both.  For RKHS theory the usual form of the underlying Gram matrix can be inferred from (\ref{E:LLSQ5}).  Let $\mathbf{G}$ denote this matrix in general.  Since a Gram matrix is the Hermitian matrix of inner products of basis functions its, eigenvalues are greater than or equal to zero and either $|\mathbf{G}| > 0$ or $|\mathbf{G}| = 0$.  In keeping with current conventions, when  $|\mathbf{G}| > 0$ holds $\mathbf{G}$ will said to be positive definite and if $|\mathbf{G}| = 0$ cannot be ruled out, then it will be called positive semi-definite.  It is thus the case $|\mathbf{G}| = 0$ that causes problems, since it implies that the associated basis functions are not linearly independent.   Historically, many authors, such as Aronszajn \cite{Aronszajn}, used the term positive definite for what is labeled positive semi-definite here and presumably this variation in nomenclature has given rise to some of the confusion that has occasionally arisen in the past.  As is often the case with matters pertaining to linear independence, there are greater subtleties involved here than one might at first expect.  After the main results of this section are derived, a discussion of these matters is included in an attempt to provide the reader with some of the tools necessary to sort out the main issues involved on a case-by-case basis---especially with regards to the RKHS context.  Hopefully, it should be apparent from the sequel that simply adding the explicit assumption of invertibility of the underlying linear systems to existing RKHS minimum norm demonstrations is all that is needed in order to take care of all, or at least most, of the issues involved.  With regards to RKHS collocation, although the discussion here may seem to be overly elaborate, the relevant points involved are considered well worth making.

\begin{center}
\ \\
 \begin{large}\underline{\textbf{Point Collocation Without Differential Operators}}\end{large}\\
\ \\
\end{center}

  Before considering the complications that result from allowing point collocation or replication conditions that have associated operators, it is useful to consider the implications of a direct point matching condition like (\ref{E:APcoll}).  At the end of this discussion a simple way of describing most of the relevant points will emerge from an examination of standard RKHS fits.

  In what follows a general \R{n} setting is assumed and in order to set this setting apart from that of the rest of the paper a new notation is used for it, but only where it is appropriate.  Thus, at the risk of being repetitious, let $\Omega$ now denote some appropriate region in \R{n}, $z$ a point in $\Omega$ and $f(z)$ a typical member of some admissible class of functions defined over $\Omega$.  It is assumed that $\Omega$ is connected and does not consist of the entire space \R{n} (so that if the class of admissible functions is harmonic, then $f(z)$ will not necessarily be zero).  In what follows all the basis functions and kernels will be assumed to be bounded members of the associated function space.  Further for some relevant inner product and positive definite norm structure ($\|f\| > 0$ for all $f\neq 0$), let the cost function to be minimized be
\begin{equation}\label{E:PHI}
 \Phi \eq \|f - \varphi\|^2 = \|f\|^2 - 2(f,\,\varphi) + \|\varphi\|^2\ \!,
\end{equation}
where, as usual, $f$ is the function to be approximated and $\varphi$ is the approximating function.
Here $\varphi$ is assumed to be a linear combination of fixed basis functions of the form
\begin{equation}\label{E:LLSQbasis}
\varphi(z) = \sum\limits_{k=1}^{N_k} {\mu}_k{\mathcal{B}}_k(z)
\end{equation}
and these basis functions themselves are assumed to also be in the class of admissible functions.  Since the class of admissible functions is assumed to admit linear superposition, a finite linear span of admissible functions yields an admissible function and so $\varphi$ is an admissible function.  For convenience, it is further assumed that $f(z)$, ${\mathcal{B}}_k(z)$ and ${\mu}_k$ are real valued.  

   As before, the values of ${\mu}_k$ that minimize $\Phi = \Phi({\mu}_k)$ are: 
\begin{equation}\label{E:LLSQ4}
\sum\limits_{k=1}^{N_k}\, ({\mathcal{B}}_{k'},\,{\mathcal{B}}_k)\,{\mu}_k = ({\mathcal{B}}_{k'},\,f) \ ,
\end{equation}
which can be written more compactly as
\begin{equation}\label{E:TmuA}   
\boxed{\,\mathbf{T}\,\mathbf{\mu} = \mathbf{A}\,} 
\end{equation}
 where $\mathbf{T}$ again denotes the matrix whose elements are $T_{k',\,k} \equiv ({\mathcal{B}}_{k'},\,{\mathcal{B}}_k)$, $\mathbf{A}$ the vector whose element are  $A_{k} \equiv ({\mathcal{B}}_{k},\,f)$ and $\mathbf{\mu}$ the vector whose elements are ${\mu}_k$.  Functional minimization processes with an underpinning inner product structure will labeled generalized least squares (GLLSQ) processes, so as to distinguish them from the more common processes where a sampled sum of squared residual errors is minimized.  This standard sampling least-squares process will be labeled linear least squares (LLSQ).  For a LLSQ fit using a combination of $N_k$ basis function, if there is a sum of $N_R$ squared residuals then in general $N_R >> N_k$.

  Before proceeding to a consideration of the main theorem dealing with minimum norm properties, the following definition is useful:

\vskip 7pt

\noindent
{\bf{Definition A.1}}\ \ 
The \emph{closed form generalized linear least squares} (CFGLLSQ) condition is said to hold for an inner product structure and approximating form specified by (\ref{E:LLSQbasis}) when it is possible to evaluate $({\mathcal{B}}_{k'},\,f)$ in closed form and Equation~(\ref{E:TmuA}) is solvable (i.e., $\mathbf{T}^{-1}$ exists). Further, the vector $\mathbf{\mu}$ is assumed to be fixed by (\ref{E:TmuA}).  When the kernel form ${\mathcal{B}}_{k'}$ is a DACK and produces closed form expressions that are a function of a single vector argument for ${(\mathcal{B}}_{k'},\,f)$ [such as (\ref{E:IIntRep})], then it is called a \emph{point replication kernel}. Also as noted above, $f$ and ${\mathcal{B}}_{k}$ are assumed to be members of some class of bounded admissible functions so that $\|f\| < \infty$ and $\|{\mathcal{B}}_{k}\| < \infty$.

\vskip 7pt

\noindent
  Observe that Definition A.1  only requires that each ${\mathcal{B}}_{k}$ be associated with some argument in $\Omega$ .  For convenience let $t_{k}  \in \Omega$ denote this argument.  In what follows it is assumed that there is a one-to-one correspondence between the 
$\{{\mathcal{B}}_{k}\}$ and $\{t_{k}\}$: $t_{k} = t_{k'}$ if and only if $k = k'$.  Further, in the remainder of this section it is assumed that the values of ${\mu}_k$ are determined by (\ref{E:TmuA}) so that $\varphi$ is a fixed function (c.f., {Definition A.1}) and that $|{\mu}_k| < \infty$ so that $\|\varphi \| < \infty$.

Definition A.1 is quite general and it is worth considering what types of kernels are allowed.  First, all of the DIDACKS examples considered in Section~\ref{S:Recap} satisfy the CFGLLSQ condition, including dipoles and higher order multipoles, but this fact will not be shown here.  Mixed types of fits are also allowed since each ${\mathcal{B}}_{k}$ can be a different type of basis function.
 Second, reproducing kernels, $K(x,\,t_{k'})$ for $t_{k'} \in \Omega$, with an associated RKHS and norm are allowed if the associated $\mathbf{T}^{-1}$ exists, since by definition $(K(t_{k'},z),f) = f(t_{k'})$; moreover, in the real setting when a linear combination of reproducing kernels of the form 
\begin{equation}\label{E:Kbasis}
 \varphi(z) = \sum\limits_{k=1}^{N_k}\, {\mu}_k\,K(t_k,\,z)
\end{equation}
can be fit to $f$, then (\ref{E:LLSQ4}) can be rewritten as
\begin{equation}\label{E:LLSQ5}
\sum\limits_{k=1}^{N_k} K(t_{k'},\,t_k){\mu}_k = f(t_{k'}) 
\end{equation}
since $T_{k'\,k} = (K(t_{k'},z),\,K(t_k,z)\,) = K(t_{k'},\,t_k)$ and $K(t_{k'},\,f) = f(t_{k'})$.
Third, Dirac delta function forms, such as $\delta(x,\,t_{k'}) = \delta(x - t_{k'})$, are not allowed since they are unbounded, even though they obviously satisfy $(\delta(t_{k'},z),f) = f(t_{k'})$ (they also cannot be considered reproducing kernels for the same reason).

   Notice that a GLLSQ setting and its solution epitomized by (\ref{E:LLSQ4}) entails little more than a positive definite norm and inner product structure.  Both  DIDACKS and RKHS approaches simultaneously display both GLLSQ and standard collocation aspects; hence, there is an underlying Gram matrix in both cases.  This dual aspect of both DIDACKS and RKHS approaches comes together in Definition A.1, but there a number of side issues to be dealt with.  For example, while a choice of orthonormal basis functions usually makes sense for many GLLSQ processes and the trivial situation $\mathbf{T} = \mathbf{I}$ occurs in this case, for RKHS and DIDACKS approaches matters are somewhat different.  Although a set of orthonormal basis functions may be known for the family of admissible functions and it may be used to reexpress the replicating or reproducing kernel, as discussed below, ${\mathbf{T}}^{-1}$ need not necessarily exist in such cases and the actual cases where $\mathbf{T} = \mathbf{I}$ occurs are very much the exception rather than the rule. One case where $\mathbf{T} = \mathbf{I}$ could occur in a reproducing-kernel like environment is if $K(z,\,t)$ were a Dirac delta function, but as just noted a Dirac delta functions are not an allowed kernel choice.  Likewise, in the DIDACKS environment a set of point sources located on $\partial \Omega$ would produce a diagonal $\mathbf{T}$ matrix, but in DIDACKS theory it is assumed that the kernel functions are harmonic in not only the interior of $\Omega$, but also on $\partial \Omega$ so this is not an allowed case either.

  Before proceeding, notice that there is an entirely different standard strategy for forcing a candidate linear combination of basis functions $\varphi$ to match some given function $f$, which is characterized by directly imposing the function matching condition implied by (\ref{E:APcoll}), (\ref{E:Pcoll}) or some analogous relationships and then simply solving the resulting linear equation set for the coefficients of $\varphi$.  In general, no inner product structure is required to perform this standard type of collocation fit, so there is usually no associated Gram matrix. In the literature such approaches are referred to by the broad rubric collocation, but here is it useful to label such approaches direct collocation, when the need to distinguish them arises at all.  Other approaches also use this same strategy, but do not use the label collocation.  One such approach is MFS, as commonly understood, which was addressed in Section~\ref{S:MFS} and where connections to material in other sections of this appendix were pointed out.

   To put all of this in proper context, first an analysis and extension of (\ref{E:APcoll}) is called for, in order to better correlate it with Definition A.1.  Since a general context is still being assumed, it is useful to first break out the collocation function matching condition associated with (\ref{E:APcoll}) and introduce a separate name for it.  Thus the criteria that some arbritrary, but admissible, function $g(t)$ with $t \in \Omega$ matches some prescribed constants at the $N_k$ specified points $t_k$
\begin{equation}\label{E:pointmatch}
 f(t_k) = C_k\ ,
\end{equation}
 is labeled the point collocation condition.  For purposes of discussion only one condition of the form (\ref{E:pointmatch}) will be entertained at a time so that the $C_k$ are to be regarded as fixed real constants throughout the discussion.

  For $\eta \in \Omega$, let the values of $\eta$ that satisfy $f(\eta) = 0$ be called node points.  Obviously, when two separate functions $f$ and $g$ both satisfy the point collocation condition $g(t_k) = f(t_k)$ and one of these functions will be said to satisfy the point collocation condition with respect to the other.  In this case $f$ and $g$ can differ by an arbritrary admissible function $\chi(t)$, $g = f + \chi$, provided that the $t_k$ are node points of $\chi$: $\chi(t_k) = 0$.

Next, in order to amplify on the foregoing and to consider it from a slightly different perspective, first notice that when $U = U(z)$ is expanded as a linear combination of $N_k$ basis functions the coefficients are uniquely determined by the point collocation condition provided that the resulting system of equations is invertible.  Here this collocation fitting process is often considered without assuming explicit norm and inner product structures.  Specifically, let ${\{h_k\}}_{k=1}^{N_k}$ be a set of suitable basis functions that are not necessarily orthogonal.  The term suitable here generally means that this set of basis functions are linearly independent and are appropriate to the situation being modeled.  Then, as in (\ref{E:LLSQbasis}), let 
\begin{equation}\label{E:Uh}
U(z) = \sum\limits_{k=1}^{N_k} {\mu}_kh_k(z) \ .
\end{equation}
(\ref{E:pointmatch}) then becomes 
\begin{equation}\label{E:HmuC}   
{\,\mathbf{H}\,\mathbf{\mu} = \mathbf{C}\,} 
\end{equation}
where $\mathbf{H}$ is the matrix whose components are $H_{k\,k'} = h_{k}(t_{k'})$ and $\mathbf{C}$ is the vector whose components are $C_k$.  For the sake of discussion assume $\mathbf{H}^{-1}$ always exists.   Here the number of basis functions used has been constrained to be the same as the number of sample points on the RHS.  In general, if the number of sample points is larger than the number of basis points then a LLSQ fitting process is implied rather than the process (\ref{E:HmuC}), while if the number of basis functions is larger than the number of sample points then the associated $\mathbf{H}$ matrix does not have full rank.  Both of these cases are of no particular interest here and are not considered in the sequel.  If the $h_k$ are expressed as is, and not in kernel form, (\ref{E:Uh}) will be called a \emph{direct collocation} process.  It also represents what might also be labeled a forced least squares LLSQ fit, where the number of sampled residuals is $N_k$.  Next suppose that a kernel, ${\mathcal{H}}(\rho,\,z)$ can be associated with the $h_k$'s by the ansatz  $h_k(z) = {\mathcal{H}}({\rho}_k,\,z)$.  The idea here is that when a proper kernel function exists the index $k$ can be naturally transferred to a set of indexed points ${\{{\rho}_k\}}_{k=1}^{N_k}$, where it is assumed that ${\rho}_k \neq {\rho}_{k'}$ whenever $k \neq k'$.  Here ${\rho}_k \in \Omega$ need not be assumed, but ${\rho}_k \in \mR{n}$ generally is assumed.  In this case and ignoring ill-conditioning issues, since $H_{k\,k'} = {\mathcal{H}}({\rho}_{k},\,t_{k'})$ the main criteria for a successful kernalization of $h_k$ is that $\mathbf{H}^{-1}$ exist.  The matrices $\mathbf{H}$ and $\mathbf{T}$ obviously play the same role, so only the matrix $\mathbf{T}$ will be referred to in the sequel.  As noted above, when neither an associated reproducing or replicating property nor an associated norm is assumed, then (\ref{E:Uh}) exemplifies a direct collocation kernel based process and direct collocation, in one form or another, is pervasive in the applied scientific literature where it is usually simply called collocation. 

  Reproducing kernel and replicating kernels fits can be considered kernelized examples of (\ref{E:HmuC}).  First, as an aside, evaluate Equation~(\ref{E:Kbasis}) at the points $t_{k'}$ to obtain
\begin{equation}\label{E:Kbasis2}
 {\varphi}(t_{k'}) = \sum\limits_{k=1}^{N_k}\, {\mu}_k\,K(t_k,\,t_{k'})\ .
\end{equation}
This equation, in conjunction with Equation~(\ref{E:LLSQ5}), then implies that $\varphi$ satisfies the point collocation condition with respect to $f$, since $K(t_k,\,t_{k'}) = K(t_{k'},\,t_k)$ holds by assumption.  

   There is one additional distinction to be considered for replicating and reproducing kernels with regards to the constraint condition (\ref{E:pointmatch}).  As noted, for these cases, while it is a Gram matrix, the $\mathbf{T}$ matrix  is only indirectly related to the set of orthogonal basis functions that span the family of functions and, moreover, is always a square matrix of size $N_k$ for any specified set of $N_k$ evaluation points.  If only the general form of the kernel function is known, then unless the explicit dimension, $N_f$, of the span of the family of basis functions is known then $N_k \geq N_f$ may well occur.  In this case (\ref{E:pointmatch}) either completely determines or overdetermines the coefficients of $\varphi$, which is a misapplication of the underlying technique.  In this situation (\ref{E:pointmatch}) will be said to overwhelm the family of admissible functions.  A separate condition must be imposed in this case since it is a property of the class of admissible functions and not necessarily the kernel basis functions themselves, but it is dependent on the number, $N_k$, of evaluation points.  Crudely put, as discussed in Section~\ref{S:Analysis} and below, the requirement is that $N_f \geq N_k$. (It will soon be painfully apparent that the real relevant concepts are that of uniformly $N$ pointwise independent family of admissible functions over $\Omega$, which was introduced by Definition \ref{S:Analysis}.1, and it is difficult to avoid it.)

  Much of the foregoing discussion can be understood by examining a reproducing kernel collocation fit and in the ensuing discussion it should be plain why certain conditions (such as $|\mathbf{T}| \neq 0$) must be assumed.  Suppose, as usual that $\Omega$ is the region of interest. First, consider the kernel form:
\begin{equation}\label{E:regK}
K(z,\,t) = \sum\limits_{j=1}^{N_u} {\hat{u}}_j(z){\hat{u}}_j(t)\ ,
\end{equation}
where ${\{{\hat{u}}_k\}}_{j=1}^{N_u}$ are a complete set of orthonormal basis functions over $\Omega$:
\begin{equation}\label{E:Uhat}
({\hat{u}}_j,\,{\hat{u}}_{j'})_z = {\delta}_{j\,j'}\ ,
\end{equation}
where as in \cite{Aronszajn}, the bound variable(s) for the inner product in question are indicated by a subscript.  Notice that $K$ obviously has the reproducing property:
\begin{equation}\label{E:RegRep}
(K(z,\,t),\,f(z))_z = \sum\limits_{k=1}^{N_u} {\hat{u}}_j(t)\ ({\hat{u}}_j(z),\,f(z))_z = f(t)\ .
\end{equation}
A reproducing kernel expression of this type will be called regular.

 An RKHS fit corresponding to (\ref{E:Kbasis}) and (\ref{E:LLSQ5}) can be rewritten in a revealing way for regular kernel expressions.  First rewrite (\ref{E:Kbasis}) as
\begin{equation}\label{E:Kregbasis}
 \varphi(z) = \sum\limits_{k=1}^{N_k}\, {\mu}_k\,K(t_k,\,z) =  \sum\limits_{k=1}^{N_k}\, {\mu}_k\, v_k(z) 
\end{equation}
where
\begin{equation}\notag
 v_k(z) = K(z,\,t_k) = \sum\limits_{j=1}^{N_u} {\hat{u}}_j(t_k){\hat{u}}_j(z)
\end{equation}
or in matrix form as
\begin{equation}\label{E:U2V}
 \mathbf{V} = \mathbf{W}\mathbf{U}
\end{equation}
where $\mathbf{V}(z)$ is the $N_k$ vector whose $k$'th component is $v_k(z)$,  $\mathbf{U}(z)$ is the $N_u$ vector whose $j$'th component is $u_j(z)$ and $\mathbf{W}$ is the $N_k$ by $N_u$ 
matrix whose components are  ${\hat{u}}_j(t_k)$.  Equation (\ref{E:U2V}) clearly represents a transformation to a new set of basis functions.  Unless $\mathbf{W}$, which does not depend on $z$, has rank $N_k$ the new set of basis functions will not be linearly independent.  Observe that (\ref{E:LLSQ5}) can be written in the form
\begin{equation}\label{E:LLSQ9}
\sum\limits_{k=1}^{N_k} G_{k'\,k}{\mu}_k = f(t_{k'}) 
\end{equation}
where $G_{k'\,k} \eq ({v}_{k'},\,{v}_{k})_z$ corresponds to a Gram matrix $\mathbf{G}$ and this matrix will be non-singular only if $\mathbf{W}$ has rank $N_k$.  Further notice that the rank of $\mathbf{W}$ may depend on the actual interpolation point locations $t_k$, which corresponds to the ``geographic factors'' discussed below.

\begin{center}
\ \\
 \begin{large}\underline{\textbf{Point Collocation Conditions With Operators}}\end{large}\\
\ \\
\end{center}

  Next consider the analog of (\ref{E:pointmatch}) for operators that is implied by the point replication condition Definition A.1.  Thus suppose that this condition can represented by the action of a global linear differential operator $\mathscr{L}_{k}$ that may differ from one  $k$ value to another and that it can be expressed as a sum of partials of various orders with respect to the components of $t$ at each location.  Within the context of Section~\ref{S:Recap}, this means 
that if $\mathscr{L}_{k}$ is the operator associated with $k$ then it can be represented as a sum of partials of various orders with respect to the components of $\vec{P}_k$.  For concreteness, consider the case where a single differential operator exists for all $k$ values so that $\mathscr{L}_{t} =\mathscr{L}_{k}$ and 
\begin{equation}\label{E:Lrep}
\mathscr{L}_{k}f \eq  \mathscr{L}_{t}f(t){\Big|}_{t=t_k} = \mathscr{L}_{t}f(t_k)\ .
\end{equation}
By extension, the notation for the situation where $\mathscr{L}_{t}$ varies from one set of $k$'s to the next will be denoted, for convenience, simply as
\begin{equation}\label{E:Lrep2}
 \mathscr{L}_{k}f(t){\Big|}_{t=t_k} = \mathscr{L}_{k}f\ .
\end{equation}
The point replication analog of (\ref{E:pointmatch}) then takes the form
\begin{equation}\label{E:Lrep3}
 \mathscr{L}_{k}f = {\alpha}_k\\,\,,
\end{equation}
for fixed real constants ${\alpha}_k$.  Just as before, when $g$ satisfies the condition (\ref{E:Lrep3}), it can be written as $g = f + \chi$ where the $t_k$'s are now node points of $\mathscr{L}_{k}(\chi)$ rather than $\chi$.  Expression (\ref{E:Lrep3}) will be called the generalized point collocation condition.
 
  At the risk of being overly abstract, the exact nature of the linear differential operator $\mathscr{L}_{k}$ and its relationship to the condition of Definition A.1 has been temporarily left open to allow for the greatest generality.  Presently any residual confusion that enters on this account will be dealt with.  For now $\mathscr{L}_{k}$ can be taken to be an undefined primitive or it can be considered to be a multiplicative constant, whichever  the readers wishes. 

  Since the conditions implied by Definition A.1 hold, it follows that
\begin{equation}\label{E:Bif}
({\mathcal{B}}_{k},\,f) = \mathscr{L}_{k}f = {\alpha}_k\ .
\end{equation}
Also (\ref{E:LLSQbasis}) and (\ref{E:TmuA}) hold.  Here $\varphi$ obviously satisfies the 
 condition implied by (\ref{E:Lrep3}) since (\ref{E:TmuA}), as expressed in the form (\ref{E:LLSQ4}), can be immediately be rewritten as $({\mathcal{B}}_{k'},\,\varphi) = ({\mathcal{B}}_{k'},\,f)$ so that condition (\ref{E:Lrep3}) expressed for $f$, along with (\ref{E:Bif}), also implies it holds for $\varphi$:
\begin{equation}\label{E:Lrep4}
 \mathscr{L}_{k}\varphi = {\alpha}_k\ .
\end{equation}
 Likewise multiplying (\ref{E:TmuA}), as expressed in the form (\ref{E:LLSQ4}), by ${\mu}_{k'}$ and summing over $k'$ gives
\begin{equation}\label{E:Bof}
(\varphi,\,\varphi) = (\varphi,\,f)\ ,
\end{equation}
which can be used to eliminate $(\varphi,\,f)$ on the RHS of (\ref{E:PHI}):
\begin{equation}\label{E:PHI2}
 \|f - \varphi\|^2 = \|f\|^2 - \|\varphi\|^2\ \!.
\end{equation}
Since the norm has been assumed to positive definite obviously
\begin{equation}\label{E:fGTphi}
  \|f\| \geq \|\varphi\|\ \!.
\end{equation}

It has been assumed that the $\alpha_k$ and $f$ and thus $\varphi$ are fixed.  Consider some other arbritrary admissible function that also satisfies condition (\ref{E:Lrep3}):
\begin{equation}\label{E:Lrep5}
 \mathscr{L}_{k}g = {\alpha}_k\ ,
\end{equation}
then if $\|g\| < \|f\|$ is it possible for $g$ to be less than $\varphi$?  No, for the following reason.  From assumptions about the nature of the action of the  operator and inner product replication condition $({\mathcal{B}}_{k},\,g) = \mathscr{L}_{k}g$, so that (\ref{E:Lrep5}) implies $g$ satisfies all of the conditions assumed for $f$ starting with (\ref{E:Lrep3}) and thus that $f$ can be replaced by $g$ up through and including (\ref{E:fGTphi}):
\begin{equation}\label{E:fGTphi2}
  \|g\| \geq \|\varphi\|\ \!.
\end{equation}
There are several side issues that need to be considered; however, most of them will be addressed after the main theorems are obtained.

   An attempt will now be made to summarize the progress so far in several definitions and theorems.  While the requirement that the underlying vector space of admissible functions have at least dimension $N_k$ might seem to be sufficient, as previously indicated, subsequent discussions and examples show that it is not and the stronger condition that the family of functions is uniformly $N_k$ pointwise independent will have to be assumed.

  Before proceeding it is useful to clarity how the linear differential operator $\mathscr{L}_{k}$ enters here.  Towards that end some additional precision in terminology is in order:

\vskip 7pt

\noindent
{\bf{Definition A.2}}\ \ 
A linear differential operator $\mathscr{L}_{k}(t)$ that is associated with Definition A.1 as discussed above, will be called \emph{$k$-universal} when it is the same basic operator for all $k$, which is to say it has the same form for all values of $k$ [where the evaluation convention implied by (\ref{E:Lrep2}) obviously is still assumed to hold].

\vskip 7pt

\noindent
Next, a definition is introduced to limit the scope of operators under current consideration:

\vskip 7pt

\noindent
{\bf{Definition A.3}}\ \ 
A linear differential operator $\mathscr{L}_{k}(t)$ will be called \emph{primitive} when it is $k$-universal and it has either the form $\mathscr{L}_{t} ={\lambda}$ or $\mathscr{L}_{t} ={\lambda}|t|$, for some non-zero real constant ${\lambda}$ and $0 < |t| < \infty$.  From the evaluation convention mentioned in Definition A.2, this means that for $\lambda \neq 0$ either $\mathscr{L}_{k} = {\lambda}$ for all $k$ or $\mathscr{L}_{k} = {\lambda}|t_k|$ for all $k$ and $0 < |t_k| < \infty$.

\vskip 7pt

In Definition A.1, while an explicit side mention is made of the point replication property satisfied by all DIDACKs, it clearly also applies to reproducing kernels.  When Definition A.1 is understood in the sense of a primitive operator, as outlined in  Definitions A.2 and A.3, the various cases of immediate interest for standard RKHS and scalar point source fits can be dealt with.  For example, a RKHS collocation fit corresponds to $\mathscr{L}_{k} = 1$ for all $k$ and (\ref{E:IIntRep}) implies that a scalar point source fit corresponds to $\mathscr{L}_{k} = |t_k|$ in the current notation.  It is worth noting that if a linear differential operator is not suitable, then there is a real chance of obtaining meaningless and trivial results.  For example, if $\mathscr{L}_{k}f \eq 0$, then (\ref{E:fGTphi2}) becomes $\|g\| \geq 0$.  The index $k$ dependence of $\mathscr{L}_{k}$ also leaves open ``geographic implications'' that must be considered.  (``Geographic implications,'' are discussed below in a related context.)  These geographic implications can be overcome by generalizing  the idea of uniformly $N$ pointwise independent families of functions to include the action of operators.  For example, if it can be shown that for some $k$-universal operator $\mathscr{L}$, $\mathscr{L}f$ maps a given family of uniformly $N_k$ pointwise independent functions into a corresponding family of $N_k$ pointwise independent functions, then the resulting $\mathbf{T}$ matrix will be invertible and the replication or reproducing property (\ref{E:Lrep4}) will be non-trivial. 

\newcommand{\Di}[1]{\mathscr{D}_{{#1}}}

  For general linear differential operators that are non $k$-universal issues that must also be considered.  Observe, for example, that $\mathscr{L}_{k}$ itself may be a linear combination of other differential operators and thus it may have internal parameters.  If these other parameters are to be fit, then it is a simple matter to separate out the resulting relevant types of operator basis functions, but in this case having multiple types of operators associated with a single location produces non-trivial complications because it is generally harder to prove that these various types of operators all act independently of each other.  An example of one such situation arises is for a combined point mass/dipole associated with location $\vec{P}_k = \vec{X}_k/{r^2_k}$.  It is obviously much harder to prove the linear independence all of the component parts of this sort of combination than it is when only simple operators are present; moreover, it unclear how to even generalize existing point mass linear independence proofs when even only simple dipole terms appear. (Current plans are to present a more general line of proof that includes dipole linear independence as a special case in a future article.)  Other complexities also enter when considering compound linear differential operators, since they may well not actually yield linearly independent basis functions and, in fact, the issue of linear independence may depend on the underlying properties of the family of admissible functions itself.  For example, first let $(\Di{1},\,\Di{2},\,\Di{3})^T \eq ({\partial\,}/{\partial {x}},\,{\partial\,}/{\partial {y}},\,{\partial\,}/{\partial {z}})^T = \vec{\nabla}$, then consider the operators $L_{i\,j} \eq \Di{i}\Di{j}$.  There are at most six independent components since (for any well behaved family of functions) $L_{j\,i} = L_{i\,j}$; however, if $f$ is harmonic then there will be at most five independent operators of this type at any given point due to Laplace's equation.  For this example, the interaction of the operator form $L_{i\,j}$ with the class of admissible functions is fairly obvious, but for certain operators and classes of admissible functions this interaction may not be so transparent. (For example, as pointed out later $\Dr f$ is not harmonic when $f$ is harmonic and thus it is not in the family of admissible harmonic functions defined over, say, $\Omega_{\sps} \subset \mathbb{R}^3$; hence, it cannot rigorously be represented as a harmonic expansion.) 

For the above reasons, as well as ones discussed below, it is expedient to restrict ourselves to  
 relatively easily obtained results.  When this is done, all of the above can then be summarized in four main theorems.  Recapping the line of argument leading up to (\ref{E:Lrep4}) then yields:
\begin{theorem}\label{T:CollRep}
When the generalized point collocation condition is satisfied by a primitive linear operator and the conditions of {Definition A.1} are also satisfied for a uniformly $N_k$ pointwise independent kernel environment, then the resulting fit, $\varphi$, also satisfies the same generalized point collocation condition; moreover, a point collocation condition of the form (\ref{E:Lrep3}) is also implied. 
\end{theorem}
\noindent
This theorem guarantees that the given data set is replicated by the resulting fit.
Here, notice that if $t_k \neq 0$ then for primitive operators ${\alpha}_k = \lambda f(t_k)$  or ${\alpha}_k = \lambda |t_k|f(t_k)$ in (\ref{E:Lrep3}) and thus the condition (\ref{E:Bif}) implies the point collocation matching condition (\ref{E:pointmatch}) and conversely.

  The line of argument leading up to (\ref{E:fGTphi2}) yields:
\begin{theorem}\label{T:gGTphi}
As in Theorem~\ref{T:CollRep}, given that the generalized point collocation condition is satisfied by a primitive linear operator and the conditions of {Definition A.1} are also satisfied for a uniformly $N_k$ pointwise independent kernel environment, then the resulting fit $\varphi$ also satisfies the $\|g\| \geq \|\varphi\|$ for any other admissible function $g$ that satisfies either  the same generalized point collocation condition or point collocation condition.  (Equivalence of the conditions (\ref{E:pointmatch}) and (\ref{E:Lrep3}) is also implied in this case). 
\end{theorem}
\noindent

  Next it is useful to consider Theorems~\ref{T:CollRep} and \ref{T:gGTphi} specifically in the context of DIDACKS point mass fits:
\begin{theorem}\label{T:DIDACKSfits}
 \R{3} DIDACKS point mass fits replicate the specified data points of $f$ used; moreover, the resulting fits minimize not only $\|f - \varphi\|$ but $\|\varphi\|$ is less than or equal to that of any other admissible function that also matches the specified data points.
\end{theorem}
\begin{proof}
As previously noted in Section~\ref{S:Analysis}, there is a proof in the literature that a finite set of point mass basis functions are linearly independent for the situation where the sources are contained inside some sphere and the harmonic region consists of the exterior of this sphere.  If linear independence of point mass basis function can be shown in the complimentary case where the sources are in $\widehat{\Omega}_{\sps}$ and  $\widehat{\Omega}_{[{\sps}]}$ is the region of interest then all of the conditions for Theorems~\ref{T:CollRep} and \ref{T:gGTphi} to hold will obviously be satisfied and thus the desired result will follow.  Towards this end consider the method of images for the \R{3} sphere \cite{Jackson}.  This technique states that to each charge in the interior of the sphere there corresponds a charge of a prescribed strength of opposite sign and at a determined location such that the net effect of these two changes together gives a zero potential on the surface of the sphere in question.  [The relative positions of these two charges is prescribed by (\ref{E:PkXk}).]  Suppose to the contrary that a counter example did exist where a collection of non-zero changes in  $\widehat{\Omega}_{[{\sps}]}$ produced a zero potential over $\partial{\widehat{\Omega}}_{[{\sps}]}$.  Then from the method of images each individual charge in this combination would have a image charge that also produces a zero field over $\partial{\widehat{\Omega}}_{[{\sps}]}$  and, furthermore, the sum of all of these charges would produce a linearly dependent counter example to the existing theorem for point mass basis functions.  From this contradiction the desired linear independence result follows.
\end{proof}

\begin{center}
\ \\
 \begin{large}\underline{\textbf{Geometric Interpretation of Collocation Fits}}\end{large}\\
\ \\
\end{center}

 Here, temporarily leaving aside the issue of general operators, it is worth observing that it is possible to give an explicit geometric interpretation for collocation fits, but the correct one appears to be slightly more involved than the generally well-known one.  Before proceeding, as a side issue, first consider the geometry that occurs in $n$-dimensional Euclidian space (${\mathbb{R}}^n$) when the values of $N < n + 1$ vectors are specified as a constraint condition and one wants to find the vector of minimum length that satisfies this constraint.  (As in other parts of the article, at several places here the reader may wish to draw a diagram for her or himself.)   Since $N$ points determine an $N - 1$ dimensional hyperplane, this condition of constraint determines an $N - 1$ dimensional hyperplane.  The vector of minimum length from the origin to this hyperplane is then the one that is orthogonal to it. There is, however, a notable difference between this situation and the present one exemplified by (\ref{E:pointmatch}) or (\ref{E:HmuC})---equations (\ref{E:pointmatch}) and (\ref{E:HmuC}) are conditions of constraint that directly determine the various basis function coefficients themselves.  Thus the closest ordinary vector analog of (\ref{E:pointmatch}) is the one where the \emph{components} of some arbritrary vector are fixed with respect to a subspace spanned by a set of basis vectors (corresponding to the various chosen basis vectors, $h_k$, themselves), but not fixed with respect to the whole space.  To see this, for example, consider the following hypothetical situation that uses basis functions that have a coordinate representation.  Thus suppose that, contrary to the assumption of boundedness, Dirac delta functions are allowed basis functions and that an $n$-dimensional volume integral of $f\,g$ specifies the inner product for $f$ and $g$, then the minimum norm solution to (\ref{E:pointmatch}) that results is obviously $f(z) = \sum_{k=1}^{N_k}\ C_k\delta(z - t_k)$.  In this degenerate case the $\delta(z - t_k)$'s are the basis functions and the $C_k$'s are the projections of $f$ onto them.  Thus (\ref{E:pointmatch}) is not a condition that specifies a hyper-plane of dimension $N_k -1$ in the function space, but rather it is a condition that implies the projection of $f$ onto a set of basis functions, which may be as yet undetermined.  For this whole geometric situation not to collapse, it is necessary that $N_k < n$ hold.  Here $N_k = n$ is not of interest since it completely fixes $f$ and $g = f = \varphi$.  To see that the proper dimension is indeed only $N_k$ and not $N_k -1$, in (\ref{E:Uh}) and (\ref{E:HmuC}) consider a function of the form $U(z) = \sum_{k=1}^{N_k}\ {\mu}_k\delta(z - t_k)$.  In this case $\mathbf{H} = I$ and each individual constraint condition in (\ref{E:HmuC}) fixes the magnitude of the associated $\mu_k$ for the basis function $\delta(z - t_k)$.  Hence (\ref{E:HmuC}) is not a condition specifying the points themselves (i.e., vectors) in the function or Hilbert space, but it is a condition specifying the \emph{components} of the basis vectors.
 
  Finally, observe that if additional Dirac delta basis functions of the form $\delta(z - z_{k'})$ for $N_{k'} \geq k' > N_k$ are added, where  $z_{k'} \neq t_k$ for all $k'$ and $k$, then the associated coefficients ${\mu}_{k'}$ are not in any way constrained.  Specifically, let $U'(z)$ denote this new functional form with the added basis functions, then at any of the evaluations points of (\ref{E:HmuC}), say $p$:
\begin{align}\notag
 U'(p) &= \sum_{k=1}^{N_k}\ {\mu}_k\delta(p - t_k) + \sum_{k'=N_k+1}^{N_k'}\ {\mu}_{k'}\delta(p - z_{k'})\\
&= \sum_{k=1}^{N_k}\ {\mu}_k\delta(p - t_k) + \sum_{k'=N_k+1}^{N_k'}\ {\mu}_{k'}\cdot 0
 = \sum_{k=1}^{N_k}\ {\mu}_k\delta(p - t_k) = U(p)\ .
\end{align}
Only the components of $U(z)$ that fall within the constraint subspace are fixed and the other projections of $U(z)$ into the compliment of this space is not fixed at all---some sort of least squares or minimum norm solution is needed to fix that part of $U(z)$ that is in this complimentary space.  Clearly, in so far as possible, these minimum norm solutions will tend to minimize the part of $U(z)$ that is in this complimentary subspace so that the chosen solution will be the one where $\mu_{k'} = 0$ for all $N_{k'} \geq k' > N_k$.  Within the context of equation~(\ref{E:fGTphi2}),  $U(z) \rightarrow \varphi$ and  $U'(z) \rightarrow g(z)$. 
 Observe that when overlapping basis functions---such as fundamental solutions---are used this subspace decomposition is obviously not nearly as clean.

  With the foregoing as backdrop, consider now the situation at hand.  First, take two different arbritrary admissible functions $f$ and $g$ that satisfy (\ref{E:pointmatch}).  Here $\chi$, the difference of these two functions, was introduced above.  When either $f$ or $g$ is fixed and the other is varied, $\chi = f - g$ spans a linear vector space since $a{\chi}_1(t_k) + b{\chi}_2(t_k) = 0$ for $a$ and $b \in \mRR$, provided of corse that ${\chi}_1(t_k) = {\chi}_1(t_k) =  0$.  Let $\{\chi\}$ denote the span of this full vector subspace.  Temporarily leaving aside the question of uniqueness, of all possible functions that satisfy this condition one will have minimum norm; let $\varphi$ denote this function.  The requirement that $\varphi$ yields the minimum $\|\varphi\|$ of all those functions that satisfy condition (\ref{E:pointmatch}) is equivalent to the requirement that it must be orthogonal to the any vector in $\{\chi\}$: $(\varphi,\,\chi) = 0$ or $(\varphi,\,f) = (\varphi,\,g)$.  Suppose $f$ is the fixed element in the representation of $\{\chi\}$ and $g$ is the variable one.  Clearly $g = \varphi$ is one possible choice so $(\varphi,\,\varphi) = (\varphi,\,f)$ must hold, which is (\ref{E:Bof}).  It is then obvious that $\|\varphi\|$ is a minimum from the steps leading up to (\ref{E:fGTphi2}).

   When the CFGLLSQ condition holds for primitive operators, solutions that satisfy the collocation condition also are GLLSQ fits.  Suppose that $f$ is to be fit with a linear combination of the form (\ref{E:LLSQbasis}).  Thus $\|f - \varphi\|^2$ is a minimized when the projection of $f$ onto the subspace spanned by the set of basis functions $\{{\mathcal{B}}_{k}\}_{k=1}^{N_k}$ is matched exactly in the sense of (\ref{E:pointmatch}).  Geometrically this occurs when $f - \varphi$ is orthogonal to the plane spanned by the set of basis functions ${\{{\mathcal{B}}_{k}\}}_{k=1}^{N_k}$: $(\varphi,\,\varphi) = (\varphi,\,f)$.  Furthermore, since $f - \varphi$ is orthogonal to this subspace it must also be orthogonal to each of the component ``basis vectors'' of this subspace:  $({\mathcal{B}}_{k},\,f - \varphi)$, which implies (\ref{E:LLSQ4}).  

\begin{center}
\ \\
 \begin{large}\underline{\textbf{A Simple Example and Related Issues}}\end{large}\\
\ \\
\end{center}

  A brief discussion of counter-examples and associated issues is in order so as to make it plain why certain conditions (such as $|\mathbf{T}| \neq 0$) must be assumed (if it is not already obvious enough to the reader).  Setting the stage just as before, first consider the kernel form:
\begin{equation}\label{E:regK2}
K(z,\,t) = \sum\limits_{k=1}^{N_u} {\hat{u}}_k(z){\hat{u}}_k(t)\ ,
\end{equation}
where $N_u < N_f + 1$ and ${\{{\hat{u}}_k\}}_{k=1}^{N_f}$ are a set of orthonormal basis functions:
\begin{equation}\label{E:Uhat2}
({\hat{u}}_k,\,{\hat{u}}_{k'}) = {\delta}_{k\,k'}\ .
\end{equation}
When $N_u = N_f$\,, $K$ is called regular and obviously it has the reproducing property:
\begin{equation}\label{E:RegRep2}
(K(z,\,t),\,f(z))_z = \sum\limits_{k=1}^{N_f} {\hat{u}}_k(t)\ ({\hat{u}}_k(z),\,f(z))_z = f(t)
\end{equation}
where as in \cite{Aronszajn}, the bound variable(s) for the inner product in question are indicated by a subscript.  If $N_u < N_f$ a projection onto the subspace of $\{f\}$ may occur where $\{f\}$
denotes the complete linear span of the family of admissible functions.  In this case $K$ will be called a pseudo-reproducing kernel and such a kernel may be mistaken for a full reproducing kernel unless careful attention is paid to make sure that the kernel basis functions and $f$ are fully representative members of the family of admissible functions.  This may happen, for example, when $K$ is composed of basis functions obtained through the action of a linear differential operator.  Such an operator may also map $\{f\}$ into some other span of functions: $\{\mathscr{L} f\} \neq \{f\}$.  An associated undesirable situation occurs when $f$ is restricted to be in some sub-span of the family of admissible functions, say ${\{{\hat{u}}_k\}}_{k=N_1}^{N_2}$ where $N_1 > 1$ or $N_2 < N_f$.  Again, while the inner product of a function of this type and a regular reproducing kernel yields a function of this type, it may also be the case that a pseudo-reproducing kernel gives the same result and that simply showing that the reproducing property holds in this situation may not be sufficient.  For these reasons it is clear that the reproducing kernels and $f$ must be general members of the family of admissible functions.

One concrete elementary example should suffice here.  For simplicity only one-dimensional examples need be considered, so let $x,\,y \in \mRR$ represent the two independent associated variables over $\Omega$, rather than $z$ and $t$.  Further, let ${\mathscr{P}}_{n_p}$ denote the family of polynomials in one variable of degree less than or equal to ${n_p}$ for some fixed interval, say $[-1,\,1]$.   Since the set of Legendre polynomials, $P_n(x)$, are a well-known set of orthogonal polynomials over this same interval with inner product specified by 
\begin{equation}\label{E:LegIP}
(f,\,g) \eq \int\limits_{-1}^{1}\ f(x)\,g(x)\ d\,x\ ,
\end{equation}
they can be used to construct the required orthonormal basis vectors and associated regular reproducing kernels:
\begin{equation}\label{E:orthP}
 {\hat{u}}_k(x) \eq P_{k-1}(x)\sqrt{\frac{2k - 1}{2}}\ .
\end{equation}
First consider a pseudo-replicating kernel constructed with this set of basis functions via
(\ref{E:regK2}) with $N_u = 2$ and let $N_k = 3$.  Since $P_0 = 1$ and $P_1 = x$, ${\hat{u}}_1 = 1/\sqrt{2}$ and ${\hat{u}}_2 = x\sqrt{2/3}$, $T_{k\,k'} = 1/2 + (2/3)x_kx_{k'}$.  Consider the sample points $x_1 = 0$, $x_2 = -\sqrt{3}/2$ and $x_3 = \sqrt{3}/2$. It is easy to show that   
$|\mathbf{T}| = 0$ for this case.  This example leads to other questions about pseudo-reproducing kernels and over-represented cases in general, so these cases are best simply avoided at the outset.  Clearly the linear differential operator $d\ /d\,x$ reduces the dimension of the linear span of ${\mathscr{P}}_{n_p}$ by one and, when applied repeatedly, will lead to over-represented cases (i.e., cases where $N_k > N_f$) for any fixed set of polynomials. Observe that this same concern occurs with regards to a harmonic family spanned by spherical harmonics of finite degree and order over $\widehat{\Omega}_{[{\sps}]}$ since such spherical harmonics can be reexpressed as polynomial expressions involving finite powers of $x$, $y$ and $z$.  For the infinite dimensional case the action of a linear differential operator may also reduce or alter the span of the family of admissible functions. Any proof of the collocation and minimum norm property that omits the stipulation that the underlying linear system is invertible and  not over represented is immediately suspect---especially, if it purports to show the minimum norm property for linear operators. 

   It is clear that questions about the nature of the $\mathbf{T}$ matrix arise when several of the $t_k$ are node points or especially when all, or many, of the basis functions share some common node point, since all of the associated basis function terms drop out of the 
$\mathbf{T}$ matrix.  (The converse here is also interesting since $|\mathbf{T}| \neq 0$ implies limits on possible shared node points of orthogonal basis function sets.)  Given that only the form of the kernel function may be known and that connections to the form (\ref{E:regK2}) may be unclear, especially for DACKs, this whole set of issues is best avoided.  For all of these reasons, simply requiring that $\mathbf{T}^{-1}$ exists and that $N_k < N_f$ is much easier than explaining away all the special cases, even when it can be done.

\begin{center}
\ \\
 \begin{large}\underline{\textbf{Geographic and Other Factors}}\end{large}\\
\ \\
\end{center}

 There are also other factors here to consider that have not, as yet, been properly addressed.  Since the primary setting of this article involves \R{3} harmonic functions, some of these additional points will be made specifically with regards to harmonic functions.  First it might be thought that simply requiring harmonicity is, of itself, sufficient, but at the very least, it should be stipulated that the region of interest is contained in \R{n} for $n > 1$, since harmonic solutions in \R{1} satisfy $d^2\,f/d\,x^2 = 0$, which has a general solution that spans only a two dimensional space and so $n_f = 2$ (i.e., $f(x) = ax + b$ for the case $x \in [c,\,d]$).  For the rest of this discussion only the \R{3} harmonic case will be the primary focus. Second, overall constant factors are often of particular concern, especially if linear differential operators are involved.  For example, suppose that as for the I norm, harmonic functions over $\widehat{\Omega}_{\sps}$ that fall off sufficiently fast as $r \rightarrow \infty$ comprises the class of admissible functions, then in general $\mathscr{L}f$ may not be in this class of admissible functions itself---especially if it is an operator that involves integration.  Third, ${\nabla}^2\,f = 0$ does not imply ${\nabla}^2 \partial\,f/\partial\,r = 0$ since $\partial\,f/\partial\,r = r^{-1}\vec{X}\cdot{\nabla}f$ (which follows form basic vector analysis or an application of the chain rule).\footnote{Here an interesting example arises from the spherical approximation of gravity anomaly $\Delta g$, which frequently occurs in physical geodesy. In this approximation $\Delta g \eq \partial V/\partial r - (2/r)V$ where ${\nabla}^2V = 0$.  Although, ${\nabla}^2 \partial V/\partial r \neq 0$, using the product rule for the Laplacian it is easy to show that ${\nabla}^2 \partial V/\partial r =  {\nabla}^2 [(2/r)V]$ so that ${\nabla}^2 \Delta g = 0$; however, some care is required here since, for example,  ${\nabla}^2 [\partial V/\partial r - (2/R)V] = {\nabla}^2 [\partial V/\partial r \neq 0$.}  Fourth, saying that ${\nabla}^2\,f = 0$ over $\Omega$ implies certain geographic or coordinate dependent considerations.  These ``geographic issues'' are worth considering separately in some detail. 

  Two entirely different ``geographic factors'' will be touched on here.
 The first concern is a general one that is somewhat independent of harmonic function considerations.  The concern here is that reproducing kernel counter-examples can be cobbled together for admissible families of functions that are separately defined over individual partitions or subregions of $\Omega$, such that each subregion has a different $n_f$; consequently, simply saying that $N_k < n_f$ is not sufficient unless $n_f$ is taken to be the smaller of these two values.  As an example, let the family of admissible functions be defined as polynomials of fixed degree and order in $x$, $y$ and $z$ for the interior of a unit cube or sphere, such that $f(x,\,y,\,-z) = -f(x,\,y,\,z)$.  Suppose that the highest allowed order is different for $z > 0$ than it is for $z < 0$.  If a reproducing kernel can be defined for $z > 0$ that is different than it is for $z < 0$ and if $n_f$ depends on whether the sample points $\vec{P}_k$ are located in the upper or lower half-plane then a counter-example can be constructed that displays geographic dependence.  It is unclear whether some sort of family of admissible functions and kernels exists that displays a similar geographic variation that is continuous and gradual.

  The second geographic factors center on issues related to (N) pointwise independence, but it is useful to first consider the issues involved independently form this concept in oder to see that in the end for expediency that this concept (or something like it) is probably required.
  From the uniqueness of Dirichlet boundary conditions, if $f$ is specified over some closed surface that is contained inside $\Omega$, then it is fixed in the interior region bounded by this surface; moreover, to the extent that harmonic extrapolation is well defined, its values are also fixed throughout the whole of $\Omega$.  (This issue of harmonic extrapolation is taken up in greater detail in Appendix B.)  This clearly also implies that $\varphi$ is, to some degree or another, fixed by one form of DIDACKS replication result or another and it brings up the possibility of whether there is an associated geographic constraint.  The question as to whether harmonic functions are geographically constrained or not may, at first, seem superficial, but it is not immediately apparent that there is a sufficient amount of implied freedom in the concept of an $\mathbb{R}^3$ harmonic function to guarantee that a harmonic function $f$ in $\Omega$ always exists such that each value of $f({\vec{P}}_k)$ can be assigned an independently for the set of  arbitrarily specified $N_k$ distinct points ${\vec{P}}_k \in \Omega$.  It is assumed that $0 < |{\vec{P}}_k| < \infty$ regardless of $\Omega$.  If this property holds for harmonic functions defined over $\widehat{\Omega}_{[{\sps}]}$, then by the Kelvin inversion theorem it must hold for harmonic functions defined over $\widehat{\Omega}_{{\sps}}$ as well.  The same can be said for ${\Omega}_1$, since the source regions are assumed to be bounded in this case as well.  

   This property does, in fact, have some surprising implications so it is not trivial.  In particular, consider what this implies for harmonic functions defined over $\widehat{\Omega}_{{\sps}}$.  First, in analogy with Table~\ref{Ta:Pks}, let a $\delta/\epsilon$ gridding for a sphere of radius $R_j$ be a set of points on this sphere that are packed as closely as possible with the requirement that the closest two distinct points have an angular separation of $\delta$.  Then $\epsilon$ is the resulting maximum angular separation of any two adjacent points and by closest packing it is meant that $\epsilon$ is as small as possible, in some sense or another.  Practical examples of $\delta/\epsilon$-like griddings are given in Table~\ref{Ta:Pks}. Clearly there is a finite number of points in a $\delta/\epsilon$ griddings for each specified $\delta$.
If admissible families of harmonic functions are uniformly pointwise independent then the following theorem obviously holds:
\begin{theorem}\label{T:HarmonicSurprise}
 Let $\delta$ denote an arbritrary, but fixed, positive parameter and let $\{R_j\}_{j=1}^{N_j}$ be a set of $N_j$ real numbers such that $\infty > R_{j+1} > R_j > 1$.  In \R{3}, given a $\delta/\epsilon$ gridding over each sphere of radius $R_j$,  a harmonic function, $f$, in $\widehat{\Omega}_{{\sps}}$ exists that will match each of these $\delta/\epsilon$ gridding points when arbritrary values are assigned to $f$ at each of these points. 
\end{theorem}
This theorem is somewhat surprising, since one might reasonably expect from the uniqueness of the Dirichlet boundary value problem that specifying values on a dense grid at a sphere of radius $R_1 > 1$ is sufficient to fix the values on all other surfaces; however, Theorem~\ref{T:HarmonicSurprise} implies that when some fixed $f(R_1,\,\theta,\,\phi)$ is given and it is sampled at the points of a $\delta/\epsilon$ grid at a radius $R_1 > 1$ specified by an associated $\delta$, which can be any small fixed number, then other harmonic functions necessarily exist that match $f$ at the given points $R_1$ and take on arbritrary values at specified points on a larger sphere (or even on an array of spheres).   Moreover, from the results discussed above, (\ref{E:LLSQ}) provides a constructive way of obtaining a function of minimum weighted energy norm that will match all of these prescribed points.  As a specific example, consider a fit over various concentric spheres to the nonharmonic function $f = r^m$ with $m > 0$.  Here as $\delta \rightarrow 0$ it is easy to show that although the collocation (or interpolation) values match at collocation points to machine accuracy, away from the collocation points $\varphi$ yields a worse and worse approximation as $\delta \rightarrow 0$, as one would expect, since the natural attenuation of a harmonic function is obviously not being modeled.  (As elsewhere, it is instructive and fairly easy to implement this sort of example and readers are encouraged to test this sort of thing for themselves.)

  From all of the foregoing analysis it is apparent that the best way to make progress is to first settle on some specific allowed attributes that the operators and family of admissible functions must posses and then state theorems that can be easily proven for operators and admissible families of functions with these specific traits.  From all of the issues brought up above, it would seem to be apparent that it necessary to require that the family of admissible functions be either \emph{uniformly pointwise independent} or \emph{uniformly $N$ pointwise independent}, in accord with Definition \ref{S:Analysis}.1.  Future progress in this area beyond the results presented here can then be undertaken by introducing new definitions for linear operators, as the need arises, that are generalizations or distinctions, in some sense or other, of those operators characterized by Definition~A.4, given below.  The above discussion and counter-examples should then serve as a backdrop for these possible future considerations.

\begin{center}
\ \\
 \begin{large}\underline{\textbf{Formal Result}}\end{large}\\
\ \\
\end{center}

  As just noted, in order to introduce a certain measure of clarity and closure it is expedient to specify concrete traits of not only the family of admissible functions to be considered, but of the allowable operators for point reproducing and replication conditions.  Since relevant results can easily be adapted to the case of \emph{uniformly $N$ pointwise independent} families of admissible functions, if they hold for \emph{uniformly pointwise independent} functions, only uniformly pointwise independent functions will be considered in the rest of this appendix.

  With regards to the inner product action of point reproducing or replication basis functions, a slight generalization of primitive operators will be considered, which is labeled semi-primitive.
Thus for an appropriate inner product environment, consider basis functions ${\mathcal{B}}_{k}(\vec{X})$ that have a point reproducing or replication property with an associated semi-primitive operator, so that the point reproducing or replicating property can be written, in analogy with (\ref{E:Bif}), as
\begin{equation}\label{E:Bh}
({\mathcal{B}}_{k},\,f) = h({\vec{P}}_k)f({\vec{P}}_k)\,\,,
\end{equation}
where  $h({\vec{P}}_k) \neq 0$.  (Here for a primitive operator, $h({\vec{P}})$ has the form $h({\vec{P}})= \lambda$ or $h({\vec{P}})= \lambda|\vec{P}|$ for some nonzero $\lambda$.)  Formally:

\vskip 7pt

\noindent
{\bf{Definition A.4}}\ \ 
A linear operon $\mathscr{L}$ that is associated with some family of admissible functions  $\mathscr{F}$ defined over some subregion $\Omega$ of $\mathbb{R}^3$ is said to be semi-primitive if for all $\vec{X} \in \Omega$,\, $\mathscr{L} =  h(\vec{X})$ or $\mathscr{L}f(\vec{X}) =  h(\vec{X})f(\vec{X})$, where $h(\vec{X})\neq 0$.  Likewise, a set of basis functions $\{{\mathcal{B}}_{k}\}_{k=1}^{N_k}$ in $\mathscr{F}$ are said to be semi-primitive if they can be characterized by the action of a semi-primitive operator, as in (\ref{E:Bh}).

\vskip 7pt

\noindent
Here, for concreteness, $\mathscr{F}$ was defined over some subregion of $\mathbb{R}^3$, but the same definition also holds for $\mathscr{F}$ defined over any suitable $\mathbb{R}^n$ domain.

  Since $\mathscr{F}$ is assumed to be uniformly pointwise independent, the ${\mathcal{B}}_{k}$ specified by Definition~A.4 must be linearly independent.  Suppose to the contrary that the ${\mathcal{B}}_{k}$'s are linearly dependent and thus that some set of $q_k$'s exist such that $\sum_{k=1}^{N_k}{\mathcal{B}}_{k}(\vec{X})\,q_k = 0$, where at least one of the $q_k$'s are non-zero.  Specifically, let $q_{k'} \neq 0$ and then reorder the basis functions so that ${\mathcal{B}}_{k'}$ is the first one and thus $q_{1} \neq 0$.  Expression (\ref{E:Bh}) implies then that $\sum_{k=1}^{N_k}q_k({\mathcal{B}}_{k},\,f) = \sum_{k=1}^{N_k}q_kh({\vec{P}}_k)f({\vec{P}}_k) = 0$.  Here choose $f$ such that $f({\vec{P}}_k) = 1$ if $k=1$ and $f({\vec{P}}_k) = 0$, if $k > 1$, then $\sum_{k=1}^{N_k}q_kh({\vec{P}}_k)f({\vec{P}}_k) = q_1h({\vec{P}}_1) \neq 0$, which is a contradiction.  Using a more general notation, in summary:
\begin{theorem}\label{T:Fk}
If the family of admissible functions $\mathscr{F}$ is uniformly pointwise independent and the set of basis functions $\{{\mathcal{B}}_{k}\}_{k=1}^{N_k}$ are semi-primitive reproducing or replicating kernels, which is to say they have an inner-product action characterized by (\ref{E:Bh}), then this set of basis functions $\{{\mathcal{B}}_{k}\}_{k=1}^{N_k}$ is linearly independent and the resulting ${\mathbf{T}}$ matrix is invertible.  Conversely, if for any finite ${N_k}$ a semi-primitive basis function inner-product expression of the form (\ref{E:Bh}) holds for $\{{\mathcal{B}}_{k}\}_{k=1}^{N_k}$ and if the resulting ${\mathbf{T}}$ matrix is invertible for any set of distinct points  $\{{\vec{P}}_{k}\}_{k=1}^{N_k}$ located in the domain of interest, then $\mathscr{F}$ is uniformly pointwise independent.
\end{theorem}
\noindent
The proof of the first part of this theorem was just outlined above and its converse, which is the second part, is obvious.

Thus if it can be shown that harmonic functions are uniformly pointwise independent over some DIDACKS region of interest, say $\widehat{\Omega}_{[{\sps}]}$, then it follows that an associated set of scalar point source basis functions must be linearly independent.  As previously indicated, a proof exists in the literature that scalar point source basis functions are linearly independent; however, another proof of this fact would be enlightening so an attempt is made in Appendix B to show directly that harmonic functions are pointwise independent.  While not entirely successful, this attempt reveals certain hidden dangers in the process of harmonic extrapolation (i.e., downward continuation for unbounded exterior cases or outward continuation for bounded interior cases).

\newpage

\appendix
\begin{center}
 \begin{Large}{\textbf{Appendix B}}\end{Large}
\end{center}

\renewcommand{\theequation}{B-{\arabic{equation}}}
\setcounter{equation}{0}

\section*{\hfil Attempt to Prove Harmonic Functions are Pointwise Independent\hfil}\label{S:B}
\ \hfill  \\
\vskip -.2in 

\noindent
  This appendix makes an attempt to directly prove that harmonic functions defined over an appropriate region $\Omega$ are pointwise independent.  The DIDACKS point replication conditions for a given region $\Omega$, in conjunction with the fact that point source basis functions are linearly independent (which, as previously indicated, has been proven in the literature) imply, by Theorem~\ref{T:Fk} at the end of Appendix~A, that harmonic functions are pointwise independent.  Although the pointwise independence of harmonic functions is not in question, as noted elsewhere in the article, it is a very significant attribute of harmonic functions and  thus one might wish for, and expect to find, a direct proof of it; consequently, this appendix gives an attempt to supply such a proof.  Also, observe that the DIDACKS point replication conditions for a given region $\Omega$, in conjuction with Theorem~\ref{T:Fk} at the end of Appendix~A, imply that if the pointwise independence of harmonic function over $\Omega$ can be shown, then this constitutes an independent proof that scalar point source basis functions are linearly independent, which would also clearly be of interest.  Various side issues and auxiliary theorems of note will emerge as the attempt to prove that $\mathbb{R}^3$ harmonic functions are uniformly pointwise independent unfolds. 
 
This attempt will be approached by way of several lemmas and side theorems specifically for the $\widehat{\Omega}_{[{\sps}]}$ domain.  As previously noted, the results obtained for this special case can be extended to other domains of interest.  For convenience, let  ${\mathcal{H}}_{[{\sps}]}$ denote the family of associated harmonic functions.  Recall that these functions are bounded and harmonic on the boundary $\partial{\widehat{\Omega}}_{[{\sps}]}$ as well as in the interior; moreover, it is assumed that they have bounded partials [they are also smooth (i.e., $C^{\infty}$)].  Since a harmonic function is specified by Dirichlet boundary conditions over some specified surface and there are many more points on such a surface than there are finite interpolation points of interest, it might seem that the result under consideration would be easy to show, but this is not the case and, moreover, there are various consequences of this result that are nontrivial (such as Theorem~\ref{T:HarmonicSurprise} in Appendix~A and the fact that scalar point source basis functions are independent, which is otherwise hard to prove).

  Looking ahead, problems are encountered in the final stages of the proof when formal results dealing with outward continuation are required.  Here outward continuation inside the interior of a sphere is the analog of downward continuation for the exterior of a sphere.  Outward (or downward) continuation is well-known to be an ill-conditioned problem, at best.  Consider, for example, the problem of outward continuation of $f$ to $r = 1$, given that $f$ is specified on some sphere of radius $b$, where $f(b,\,\theta,\,\phi) = U_a(b,\,\theta,\,\phi)$ for $0 < b < a <1$ and where $U_a(\vec{X})$ is specified in terms of a harmonic series for the same function introduced in the discussion after (\ref{E:H1h2}).  Although this leads to a solution that is formally harmonic since it is a linear combination of harmonic functions, one would not expect the partials of $f$ for $ r \geq a$ to be bounded.  
The boundedness conditions on  the class of admissible functions, ${\mathcal{H}}_{[{\sps}]}$, rules out such examples.

   Starting with some $f \in {\mathcal{H}}_{[{\sps}]}$ specified over an appropriate interior boundary surface, if a formal solution to the outward continuation problem can be shown to exist, then provided that no solutions outside of ${\mathcal{H}}_{[{\sps}]}$ are allowed to slip in there is a solution to the outward continuation problem, which is to say that the reconstructed harmonic extension of $f$ is bounded and has bounded partials.  Thus issues concerning ill-posedness versus ill-conditioning have implications that must be dealt with since an ill-conditioned solution processes may produce large but formally bounded results, while an ill-posed one, by definition, produces uncontrolled results.  In this appendix an idealized theoretical perspective is taken so that it is considered to be sufficient to prove that something is possible, regardless of the associated implementation consequences (i.e., only the consequences of ill-posedness are relevant). (In the Section~\ref{S:Tests} actual implementation aspects are dealt with so the consequences of ill-conditioning must be, and thus are, considered.)  The dividing line here between ill-posedness and ill-conditioning is made here according to whether a constructive algorithm does or does not stay within the allowed family of admissible functions. 

  In a general functional analysis setting, the problem of staying within some well behaved family of functions is well-known.  The most common conceptual example encountered occurs when a sequence of smooth functions is considered that has as a step function as a limit, which is clearly outside the family of smooth functions itself.  Here matters are a little more complicated since, besides ${\mathcal{H}}_{[{\sps}]}$, there are various relevant families of functions to consider (such as the family of harmonic functions, the family of smooth functions or the family of functions with bounded norm).

 It will turn out that, in the end, it is very hard to set up an independent constructive interpolation scheme that stays within  ${\mathcal{H}}_{[{\sps}]}$ when outward continuation is entertained.  In spite of any difficulties encountered, observe that the stated theorems and interpolation results are obviously true since DIDACKS theory, in conjunction with the existing proof of point mass independence in the literature, immediately supplies a constructive harmonic interpolation algorithm that guarantees that end results are always in  ${\mathcal{H}}_{[{\sps}]}$; however, the associated proofs arrived at by this route would not then be independent, which is the goal here.  From the conclusions at the end of this appendix, in terms of characterizing outward (or downward continuation) as either an ill-posed or ill-conditioned phenomena, it would seem that it falls somewhere in between these two categories, being worse than what might normally be called ill-conditioned, but in some sense not being fully ill-posed either since a formal unique harmonic solution generally exists, even if it has unbounded partials.  (It is this distinction that underlies the added boundedness requirements stated elsewhere here and in this sequence of articles.)  

   To fix our goals, first consider an actual statement of the theorem of interest and for which an independent proof is sought:
\begin{theorem}\label{T:f-independence}
${\mathcal{H}}_{[{\sps}]}$ is uniformly pointwise independent: An $f \in {\mathcal{H}}_{[{\sps}]}$ always exists that assumes arbritrary (bounded) values at $N_k < \infty$ specified points, $\{{\vec{P}}_k\}_{k=1}^{N_k} \in \widehat{\Omega}_{[{\sps}]}$; i.e., $f({\vec{P}}_k) = {{C}}_k$ for ${k=1}$ to ${N_k}$ with arbritrary $C_k$'s.
\end{theorem}
\noindent
Observe that this theorem is about functions that are harmonic over  $\widehat{\Omega}_{[{\sps}]} =  \widehat{\Omega}_{({\sps})} \cup \partial{\widehat{\Omega}}_{[{\sps}]}$ and that a similar result obviously holds for functions that are harmonic over just $\widehat{\Omega}_{({\sps})}$.  Also, given any function that is harmonic over $\widehat{\Omega}_{[{\sps}]}$ it can be extended to a harmonic function over a slightly larger domain, so the issue of whether it is stated to include the boundary or not is not, in itself, especially germane (this is one of several reasons why it is assumed that admissible functions for the II norm are harmonic over $\widehat{\Omega}_{[{\sps}]}$ and not simply $\widehat{\Omega}_{({\sps})}$).

   Obviously, Theorem~\ref{T:f-independence} strongly suggests a proof by construction, but as just noted this does no mean that consideration must be given to practical numerical implementation issues.  In particular no weight is given to, or needs to be given to, the consequences of ill-conditioning.  

  Before proceeding, some additional mathematical apparatus and associated terminology is required.  First, recall the usual concept of a star-shaped region, $\mathring{\Omega}$, 
with respect to some specified interior point ${\vec{X}}_0$: $\mathring{\Omega}$ is said to be a star-shaped region with respect to ${\vec{X}}_0$ if every straight-line segment connecting ${\vec{X}}_0$ and  another arbritrary point of $\mathring{\Omega}$ is constrained to lie wholly inside $\mathring{\Omega}$ (i.e., such line segments contain no points that are not in $\mathring{\Omega}$).  Here the primary interest is not in star-shaped regions themselves, but in the boundary surface of such regions; accordingly the term closed star surface will be used to denote the boundary of a star-shaped region.  Moreover, since not only are the boundaries of star-shaped regions of interest needed, but also the hull-like counterparts of finite sets of points are needed, the required generalization will be labeled a star-surfaced set or a star-surface.  Thus with this understanding, a general definition can be given that includes not only closed surfaces, but also finite sets of points.  The criteria that a set of points ${\Sigma} \subset \mR{n}$ (which can be either a finite or infinite) is star-surfaced with respect to some given point ${\vec{X}}_0$ is thus defined as follows:

\vskip 7pt

\noindent
{\bf{Definition B.1}}\ \ 
The criteria for a region to be \emph{star-surfaced with respect to some given point ${\vec{X}}_0$} is that for all $\vec{P} \in \Sigma$, the line segment joining ${\vec{X}}_0$ and $\vec{P}$ must contain no other point of $\Sigma$.  If ${\Sigma}$ is a finite set or is only part of a closed surface it will be called an \emph{open star surface}; alternatively, if ${\Sigma}$ is a closed surface it will be called a \emph{closed star surface}.

\vskip 7pt

\noindent
Next, observe that this concept can be used to prove a theorem that will come in useful later (which is stated and proved for \R{n} for $n > 1$, rather than \R{3}):
\begin{theorem}\label{T:Star-Surface}
Let $\Sigma \subset \mR{n}$, for $n \geq 2$ be a finite set of distinct points, then $\Sigma$ is star-surfaced with respect to some point ${t}_0$ or other, where ${t}_0$ is constrained to be inside some given specified neighborhood of $\mR{n}$.
\end{theorem}
\begin{proof}
Here it is obviously assumed that neither the neighborhood nor any of the points of $\Sigma$ is at infinity.  Let $N_k$ be the number of points of $\Sigma$, then consider the set of extended straight lines $L$ going through each pair of distinct points $t_k$ and $t_{k'}$ of $\Sigma$, which are indexed by the subscripts $k$ and $k'$: $\{L_{k\,k'}\}$.  Let $d(s,\,k\,k')$ be the (minimum) distance between a point $s$ and $L_{k\,k'}$ for all points of $L_{k\,k'} \in \{L_{k\,k'}\}$.  Then consider some given fixed line $L_u$ that is not in $\{L_{k\,k'}\}$ and let $u \in L_u$, so that the function
\begin{equation}\notag
d_{\pi}(u) \eq \prod\limits_{k,k'\neq k}d(u,\,k\,k')
\end{equation}
is zero when the point $u$ and some points $t_k$ and $t_{k'}$ are all common points of some line or other.  (Here, while it is assumed that $L_{u} \notin \{L_{k\,k'}\}$, $L_u$ may be parallel to some element of $\{L_{k\,k'}\}$ and in this case $d_{\pi}(u) = const. \neq 0$ may occur.) 
Since there are at most $N_k(N_k - 1)/2$ zeros of $d_{\pi}(u)$ corresponding to the possible $N_k(N_k - 1)/2$ intersections of the $L_{k\,k'}\,$'s with $L_u$ there are at most a like number of collinear points; hence, for any specified neighborhood it is obviously possible to find a point $u \in L_u$ where $d_{\pi}(u) \neq 0$ and when $t_0$ is taken to be such a point, $\Sigma$ must be star-surfaced with respect to $t_0$.
\end{proof}

   Given this result, the general strategy will be to first of all show that a closed star surface with respect to ${\vec{X}}_0$ can always be constructed from the given $N_k$ points, where ${\vec{X}}_0$ can be taken to be a point $t_0$ guaranteed to exist by Theorem~\ref{T:Star-Surface}.  (Notice that it is not necessary here that ${\vec{X}}_0$ be in the convex hull of the original specified points.)  Second, it will be shown that given such a surface, an interpolating function can be found that matches the prescribed point values on this surface; hence, this function also serves  as an interpolating function for the given finite number of specified points.  Third, it will be shown that this surface interpolating function can serve to specify a unique harmonic function over the interior of this region from the uniqueness of Dirichlet boundary conditions.  Leaving, fourth and finally, the issue of performing an associated harmonic extension from this subregion to the whole of ${\widehat{\Omega}}_{[{\sps}]}$.  Both of the first two steps here rely on the following lemma
\begin{lemma}\label{T:SH0}
Consider a set of distinct points $\{{\hat{U}}_k\}_{k=1}^{N_k}$ that lie on a unit \R{3} sphere that has a center at $\vec{X}_0$.  If a set of bounded real (function) values $\{F_k\}_{k=1}^{N_k}$ are specified that correspond to the locations $\{{\hat{U}}_k\}_{k=1}^{N_k}$, then  a smooth interpolating function $F({\hat{X}})$ (where $|\hat{X}| = 1$) can always be generated that matches the prescribed values:  $F({\hat{U}}_k) = F_k$.
\end{lemma}
\begin{proof}
The proof will be by construction. Let $(r_k,\,\theta_k,\,\phi_k) = (1,\,\theta_k,\,\phi_k)$ be the polar coordinates of ${\hat{U}}_k$.  Let the minimum angular separation of the ${\hat{U}}_k$'s be $\beta_{\text{min}}$. Also let the specified values of $F$ be $F_k \eq F(\theta_k,\,\phi_k)$.   Further let $F_0 \eq F_{k'}$ be the minimum value of $F_k$, which is associated with the point ${\hat{U}}_{k'}$. This value will be the default value of $F({\hat{X}})$ (i.e., the interpolating values at all points will take on this value unless otherwise specified).  Notice that all that is really required is that around each interpolating point a smooth extrusion, which can be visualized as a local ``function surface'' deformation [where a function surface here is in the sense of ordinary calculus where $z \eq Z(x,y)$ often occurs], be defined and added to $F_0$ so that $F_k$ will be matched.  Next consider how this will be done for each point ${\vec{U}}_{k}$ in turn. To avoid confusion let $k''$ be the index for the specific point under current consideration that is temporarily being singled out.  First, temporarily erect a Cartesian coordinate system with a $z$-axis through the point ${\vec{U}}_{k''}$.  Reexpress the other $\theta_k$ and $\phi_k$'s in the new corresponding polar coordinates.  Finally, for $\theta < \beta_{\text{min}}/2$ let $F(\theta_{k''},\,\phi_{k''}) = F_0 + (F_k - F_0)(\cos\,\alpha(\theta)) + 1)/2$, where $\alpha(\theta) = \pi(\theta/\beta_{\text{min}})$, and  $F$ is otherwise as previously specified.  After considering each point transcribe the results back to the original coordinates and repeat the procedure until it has been applied to all the points reindexed to $k''$.  Although he interpolating function that results is clearly a smooth (not only $C^1$, but even $C^{\infty}$) interpolating function, it is clearly not generally a practical one.
\end{proof}
\noindent
The transition function here \{i.e., $H_{.5}(X) = [\cos\,(\pi X) + 1]/2$, for $X \in \mathbb{R}$ for $1 \geq X \geq 0$\} is known as a fifty-percent cosine taper.

Notice that this lemma (and the following lemmas) can immediately be generalized to handle complex or vector valued functions by simply applying the lemma to each required component in turn.  This lemma can also be used to show that a smooth interpolating surface always exists:
\begin{lemma}\label{T:SurfInterp}
Given a finite set of distinct bounded points $\{{\vec{P}}_k\}_{k=1}^{N_k}$ in \R{3}, a smooth interpolating surface exists that goes through these points; moreover, this interpolating surface is a star-shaped surface with respect to some point $\vec{X}_0$ so that the interior of this interpolating surface is a bounded star-shaped region.  Moreover, $\vec{X}_0$ can be constrained to lie in any specified neighborhood.  
\end{lemma}
\begin{proof}
From Theorem~\ref{T:Star-Surface} the set of points $\{{\vec{P}}_k\}_{k=1}^{N_k}$ is star-surfaced with respect to some \R{3} point ${\vec{X}}_0$, which, in turn, is inside some specified neighborhood.  Place the origin at $\vec{X}_0$, and label all the coordinates accordingly so that, for example, $\theta_k$ and $\phi_k$ will be defined as usual in this new coordinate system.  These angles can then be used to define the unit sphere required by Lemma~\ref{T:SH0}: ${\hat{U}}_k = {\vec{P}}_k/{\vec{P}}_k$, where $P_k \eq |{\vec{P}}_k|$, as before.
Next let $F_k = P_k$ in Lemma~\ref{T:SH0}, which guarantees an interpolating function $F(\theta,\,\phi)$ exists. Finally set $r = F(\theta,\,\phi)$, which yields a smooth interpolating function with all of the desired properties.
\end{proof}

  Lemma~\ref{T:SH0} can be used again to show the following useful intermediate theorem:
\begin{theorem}\label{T:Interp-Surface}
Given a finite set of distinct bounded points $\{{\vec{P}}_k\}_{k=1}^{N_k}$ in \R{3} and a set of associated (function) values at these points $\{F_k\}_{k=1}^{N_k}$, a smooth interpolating surface $\Sigma$ exists and a smooth interpolating function $F(\vec{X})$ also exists such that ${\vec{P}}_k \in \Sigma$ and $F{(\vec{P}}_k) = F_k$ for all $k$; moreover, $F{(\vec{X}})$ is smooth for all ${\vec{X}} \in \Sigma$ and $\Sigma$ is star-shaped surface with respect to some point $\vec{X}_0$, that lies inside a given neighborhood, which can be taken to be one that also contains the given coordinate origin.
\end{theorem}
\begin{proof}
 Lemma~\ref{T:SurfInterp} immediately yields an interpolating surface $\Sigma$ with all of the desired properties, but it remains to be shown that an appropriate $F(\vec{X})$ exists.  First reposition the origin over $\vec{X}_0$, then consider a function $G = G(\theta,\,\phi)$ defined on the unit sphere such that $G(\theta_k,\,\phi_k) = F_k$.  Lemma~\ref{T:SH0} guarantees that such a function exists.  Finally, since $\Sigma$ is a star-shaped surface $F(r,\,\theta,\,\phi) = G(\theta,\,\phi)$ is a well defined function that has all of the desired properties.  This function can then be transcribed back to the original coordinates.
\end{proof}

  Taking stock of where we stand, Theorem~\ref{T:Interp-Surface} shows immediately that given a set of points $\{{\vec{P}}_k\}_{k=1}^{N_k}$ in $\widehat{\Omega}_{(\sps)}$ a smooth interpolating surface $\Sigma$ and function $F(\vec{X)}$ exist for $\vec{X} \in \Sigma$.  From the uniqueness of Dirichlet boundary conditions, it follows that a harmonic function $f(\vec{X})$ exists that is defined throughout the whole interior of $\Sigma$ and which takes on the values $f(\vec{X}) = F(\vec{X})$ when $\vec{X} \in \Sigma$.  What remains to be examined is the possibility that the specified boundary values on $\Sigma$ also can be harmonically extended outward to cover the region between $\Sigma$  and $\partial{\Omega}_{[{\sps}]}$.  If this can be done, then it will show that a smooth closed interpolating surface $\Sigma$ exists with smooth specified boundary conditions $F(\vec{X})$ and that it can be uniquely extended to yield a function that is harmonic throughout all of ${\Omega}_{[{\sps}]}$.  Towards that end the following lemma is useful:
\begin{lemma}\label{T:SH1}
For $f \in {\mathcal{H}}_{[{\sps}]}$, if $f(\vec{X}) = 0$ throughout some sphere of diameter less than one that is centered over the origin of $\widehat{\Omega}_{[{\sps}]}$, then $f(\vec{X}) = 0$ throughout $\widehat{\Omega}_{[{\sps}]}$.
\end{lemma}
\begin{proof}
  Following the notation of Jackson \cite{Jackson}, let $Y_{l\,m}(\theta,\,\phi)$ denote the (complex) spherical harmonics specified by 
\begin{equation}\label{E:Ylm}
Y_{l\,m}(\theta,\,\phi) = \sqrt{\frac{2l + 1}{4\pi}\frac{(l - m)!}{(l + m)!}}P_l^m(\cos \theta)e^{im\phi}
\end{equation}
where $P_l^m(x)$ are the associated Legendre functions.  For complex constants $a_{l\,m}$ the Dirichlet boundary value problem for ${\mathcal{H}}_{[{\sps}]}$ in ${\Omega}_{[{\sps}]}$ has the solution
\begin{equation}\label{E:SphericlHarmonic}
f(r,\,\theta,\,\phi) = \sum\limits_{l=0}^{\infty} \sum\limits_{m=-l}^{l}a_{l\,m}\left[\left(\frac{r}{R_{\sps}}\right)^lY_{l\,m}(\theta,\,\phi)\right]
\end{equation}
and so all $f \in {\mathcal{H}}_{[{\sps}]}$ also have this form.
(Although the $a_{l\,m}$ and $Y_{l\,m}$ are complex valued they can obviously be used to represent the real boundary value problem here also.)

  Let $a$ denote the radius of the sphere that is centered over the origin, then evaluating (\ref{E:SphericlHarmonic}) at $r = a$ and using the ortho-normality conditions for the $Y_{l\,m}$, it follows immediately that $a_{l\,m} = 0$ for all allowed values of $l$ and $m$.  Hence from the form (\ref{E:SphericlHarmonic}) itself $f = 0$ throughout $\widehat{\Omega}_{[{\sps}]}$.
\end{proof}

This result can easily generalized:
\begin{lemma}\label{T:SH2}
For $f \in {\mathcal{H}}_{[{\sps}]}$, if $f(\vec{X}) = 0$ throughout some sphere of finite diameter enclosed wholly inside $\widehat{\Omega}_{[{\sps}]}$, then $f(\vec{X}) = 0$ throughout all of $\widehat{\Omega}_{[{\sps}]}$.
\end{lemma}
\begin{proof}
Again let $a$ be denote the radius of the sphere in question.  A sphere of radius $a/2$ can be situated wholly inside this given sphere as close to the origin as possible.  Consider a spherical harmonic expansion like (\ref{E:SphericlHarmonic}), except with spherical coordinates that have an origin at the center of this new sphere of radius $a/2$.  Evaluate this new expansion in exactly the same way as in Lemma~\ref{T:SH1}.  This harmonic expansion can then be used to show that $f = 0$ thorough a sphere of radius $(3/2)a$ centered over the sphere of radius $a/2$. Next again place a sphere of radius $a/2$ wholly inside this new sphere as close to the origin as possible and repeat the same argument.  After a finite number of iterations, clearly the origin will eventually be enclosed by this sequence of new spheres.  As soon as this happens the required result immediately follows from Lemma~\ref{T:SH1}. 
\end{proof}
Lemma~\ref{T:SH2} can be used to show that a harmonic function specified over an interpolating surface that is inside $\widehat{\Omega}_{({\sps})}$ has a unique harmonic extension throughout the whole of $\widehat{\Omega}_{[{\sps}]}$.  First Lemma~\ref{T:SH2} implies:
\begin{lemma}\label{T:SH3}
For $f \in {\mathcal{H}}_{[{\sps}]}$, if $f(\vec{X})$ is specified over some sphere of finite diameter enclosed wholly inside ${\Omega}_{({\sps})}$, then it has a unique harmonic extension throughout the whole of $\widehat{\Omega}_{[{\sps}]}$.
\end{lemma}
\begin{proof}
 Suppose to the contrary that this is not true and let $\Sigma$ denote the sphere in question, then $f$ and $g \in {\mathcal{H}}_{[{\sps}]}$ exist such that $g(\vec{P}) = f(\vec{P})$ for all $\vec{P} \in \Sigma$, but $g(\vec{X}) \neq f(\vec{X})$ for some $\vec{X} \in \widehat{\Omega}_{[{\sps}]}$.  Let $\chi(\vec{X}) = f(\vec{X}) - g(\vec{X})$, then from the fact that $\chi(\vec{P}) = 0$ for all $\vec{P} \in \Sigma$ and the uniqueness of Dirichlet boundary conditions Lemma~\ref{T:SH2} can be invoked to conclude that $\chi(\vec{X}) = 0$ throughout all of $\widehat{\Omega}_{[{\sps}]}$ and a proof by contradiction follows.
\end{proof}
Since the interpolation surfaces constructed earlier all enclose a sphere that, in turn, is inside $\widehat{\Omega}_{({\sps})}$, it might seem that the desired general result immediately follows, but it is at this stage that the proof falls apart because of a hidden assumption that was hinted at in the beginning of this appendix.  Before pointing it out and drawing the appropriate conclusions, it is useful to summarize the results so far. 

\begin{center}
\ \\
 \begin{large}\underline{\textbf{Summary and Conclusions}}\end{large}\\
\ \\
\end{center}

 In summary, for a specified interpolation procedure, it has been shown that a set of interior points can be considered to be star-shaped and that an interpolating function to the given $f({\vec{P}}_k)$ values can be constructed, which is unique over this surface.  The uniqueness of Dirichlet boundary conditions thus imply that the values of $f$ are uniquely determined throughout the interior of this star-shaped region by this interpolation scheme.  Clearly, the values at each of the  points $f({\vec{P}}_k)$ can be assigned differently and independently from each other by the very nature of Dirichlet boundary value problem itself.  Moreover, a sphere of finite radius can be wholly contained within this star-shaped region and $f$ can be uniquely specified over it.  In turn, by Lemma~\ref{T:SH3} this $f$ can then be harmonically extended to the whole of ${\Omega}_{[{\sps}]}$.   Thus our initial goal---an independent proof of Theorem~\ref{T:f-independence}---would seem to follow; however, there is a hidden assumption that has been made.  Namely that a smooth function (i.e., $C^{\infty}$) specified over a smooth boundary specifies not only a harmonic function, but one that is bounded and has bounded partials (i.e., a harmonic extension in ${\mathcal{H}}_{[{\sps}]}$.)  To see the distinction here consider let $\Sigma$ be a sphere of radius $b = 1/4$ and let $a_1 = 1.01$ and $a_2 = .99$ where $a_1$ and $a_2$ are specified $a$ parameter values for the function $U_a$ discussed at the first of this appendix.  Let $U_{a_1}$ and $U_{a_2}$ denote these two functions with $a = a_1$ and $a = a_2$ in $U_a(\vec{X})$, respectively. 
Then it is clear that $U_{a_1}(\vec{P}) \approx U_{a_2}(\vec{P})$ for all $\vec{P} \in \Sigma$ and that both functions are $C^{\infty}$ on $\Sigma$.   Furthermore, the values on $\Sigma$ have a unique harmonic extension to all of $\widehat{\Omega}_{[{\sps}]}$ (namely $U_{a_1}$ and $U_{a_2}$); however, $U_{a_2} \in {\mathcal{H}}_{[{\sps}]}$, but $U_{a_1} \notin {\mathcal{H}}_{[{\sps}]}$.  Finally, it seems to be hard to engineer a constructive interpolating function, with an associated interpolating surface, that is simultaneously guaranteed to match the specified values and be in ${\mathcal{H}}_{[{\sps}]}$---although this leads to some interesting research problems in itself.  [For example, what are the $N$ pointwise independence properties of truncated spherical harmonic series (including geographic factors) and how can they be proved?]

  These general sort of examples have been discussed elsewhere within a downward continuation context (see, for example, \cite{Moritz}) and they seem to indicate that some sort of direct regularization is called for when implementing harmonic extrapolation.  In particular, although 
Theorem~\ref{T:DIDACKSfit} guarantees that a weighted energy minimization principle underlies DIDACKS interpolation, which in itself is a very useful criteria that guarantees an answer in ${\mathcal{H}}_{[{\sps}]}$, if data errors are present or if the source locations are inappropriate then the implied harmonic extrapolations may be, at best, useless.  [Here even in the presence of exact data, at a minimum Nyquist frequency and other related digital signal processing issues (such as aliasing) probably should always be considered.]

  Nevertheless, several useful intermediate lemmas and theorems have been presented in this appendix.  Furthermore, as previously pointed out the Kelvin transformation can be used to obtain corresponding results for the exterior of a unit sphere and many of the lemmas and theorems clearly apply to regions of a more general shape.  Also observe that these interpolating results also apply to analogous situations in $\mathbb{R}^2$, where the lines-of-proof followed are even easier to follow.  Many of the results here can also be readily generalized to  different shaped regions in $\mathbb{R}^n$ for $n > 3$; however, some technical issues must be dealt with to prove corresponding theorems in these general settings.

\newpage



\begin{thebibliography}{99}
\bibitem{Aronszajn}
N. Aronszajn,
\emph{Theory of Reproducing Kernels},
Am. Math. Soc. Trns. \textbf{68} (1950), 337--404.
\bibitem{AxlerEtAll}
Sheldon Axler,  Paul Bourdon and Wade Ramey,
\emph{Harmonic Function Theory},
 Second Edition,
Springer Verlag, New York, N.Y., 2001.
\bibitem{MFS1}
A. Bogomolny,
\emph{Fundamental solutions method for elliptic boundary value problems},
 SIAM J. Numer. Anal., \textbf{22}, (April 1985), 644--669.
\bibitem{MFS7}
C. S. Chen, Hokwon A. Cho and M. A. Goldberg,
\emph{Some comments on the ill-conditioning of the method of fundamental solutions},
 Engineering Analysis with Boundary Elements, \textbf{30}, (2006), 405--410.
\bibitem{BEMhistory}
Alexander H.-D. Cheng and Daisy T. Cheng,
\emph{Heritage and early history of the boundary element method},
 Engineering Analysis with Boundary Elements, \textbf{29}, (2005), 268--302.
\bibitem{Cohn}
Henry Cohn and Abhinav Kumar,
\emph{Universally Optimal Distributions of Point On Spheres},
arXiv:math.MG/0607446 v1, 19 Jul 2006. 
\bibitem{MFS5}
G. Fairweather and A. Karageorghis,
\emph{The method of fundamental solutions for elliptic boundary value problems},
Advances in Computational Mathematics, \textbf{9}, (1998), pp.~69--95.
\bibitem{Tdesign}
D. Gross, K. Audenaert and J. Eisert,
\emph{Evenly distributed unitaries: on the structure of unitary designs},
arXiv:quant-ph/0611002 v1, 31 Oct 2006.
\bibitem{Hille}
Einar Hille,
\emph{Analytic Function Theory Vol II},
Ginn and Company, Boston, MA., 1962.
\bibitem{PGII}
Bernhard Hofmann-Wellenhof and Helmut Moritz,
\emph{Physical Geodesy}, First Edition,
Springer-Verlag New York, New York, 2005.
\bibitem{Jackson}
John David Jackson,
\emph{Classical Electrodynamics},
Third Edition,
John Wiley \& Sons, Inc., New York, 1999.
\bibitem{Kellogg}
Oliver Dimon Kellogg,
\emph{Foundations of Potential Theory},
Dover Publications, New York, 1953.
\bibitem{Korns}
Granino A. Korn and Teresa M. Korn,
\emph{Mathematical Handbook for Scientists and Engineers: Definitions, Theorems, and Formulas for Reference and Review},
Second Edition,
McGraw-Hill Book Company (Reprinted by Dover Publishing Co. in 2003),
New York, 1968.
\bibitem{Krarup}
{\sc{Torben Krarup}},
\emph{A Contribution to the Mathematical Foundation of Physical Geodesy}, 
Pub. 44, Danish Geod. Inst., Copenhagen, Denmark 1969.
\bibitem{MFS6}
Y. J. Liu,  N. Nishimura and Z. H. Yao,
\emph{A fast multipole accelerated method of fundamental solutions for potential problems},
Engineering Analysis with Boundary Elements, \textbf{29}, (2005), pp.~1016--1024.
\bibitem{Mate}
L\'{a}szl\'{o} M\'{a}t\'{e},  
\emph{Hilbert Space Methods in Science and Engineering},
Adam Hilger imprint by IOP Publishing Ltd,
Bristol, England, 1989.
\bibitem{Moritz}
Helmut Moritz,
\emph{Advanced Physical Geodesy},
Abacus Press, Tunbridge Wells, Kent, England, 1980.
\bibitem{MFS4}
P. A. Ramachandran,
\emph{Method of fundamental solutions: singular value decomposition analysis},
Communication in Numerical Methods in Engineering, \textbf{18}, (2002), pp.~789--801.
\bibitem{Ruf1}
Alan Rufty,
\emph{A Dirichlet Integral Based Dual-Access Collocation-Kernel Approach to Point-Source Gravity-Field Modeling}, [arxiv:physics/0612099] v1 11 Dec 2006. (available at http://arxiv.org).
\bibitem{DIDACKSI}
Alan Rufty,
\emph{Dirichlet integral dual-access collocation-kernel space analytic interpolation for unit disks: DIDACKS I}, [arxiv:math-ph/0702062].
\bibitem{MFS2}
I. Saavedra and H. Power,
\emph{Multipole fast algorithm for the least-squares approach of the method of fundamental solutions for three-dimensional harmonic problems},
Numerical Methods for Partial Differential Equations, \textbf{19}, No. 6, (2003) pp.~828--845.
\bibitem{MFS3}
I. Saavedra and H. Power,
\emph{Adaptive refinement scheme for the least-squares approach of the method of fundamental solution for three-dimensional harmonic problems},
Engineering Analysis with Boundary Elements, \textbf{28}, (2004), pp.~1123--1133.
\bibitem{SandW}
{\sc{Robert Schaback and Holger Wendland}}, 
\emph{Kernel Techniques: From Machine Learning to Meshless Methods}, 
Acta Numerica, \textbf{15} (2006), pp.~1--97.
\bibitem{Schmeidler}
Werner Schmeidler,
\emph{Linear Operators in Hilbert Space},
Academic Press, New York, 1965.
\bibitem{geoPMuniq}
D. Stromeyer and L. Ballani,
\emph{Uniqueness of the Inverse Gravimetric Problem for Point Mass Models},
Manuscripta Geodetica, \textbf{9} (1984), 125--136.
\bibitem{RKHSbook}
Howard L. Weinert (editor),
\emph{Reproducing Kernel Hilbert Spaces Applications in Statistical Signal Processing},
Benchmark Papers in Electrical Engineering and computer Science / 25,
Hutchinson Ross Publishing Company, Stroudsburg, Pennsylvania, 1982.
\end{thebibliography}
\end{document}